\pdfoutput=1

\documentclass[preprint,fleqn,5p,numbers,sort&compress]{elsarticle}

\usepackage{graphicx}
\usepackage{amssymb,amsmath}
\usepackage{natbib}
\usepackage{hyperref}
\hypersetup{pdftitle = The title of my PDF, pdfauthor = My name, pdfsubject= The subject, pdfkeywords = keyword1 keyword2 keyword3}
\hypersetup{colorlinks = true, linkcolor = blue, anchorcolor = red, citecolor = blue, filecolor = red, pagecolor = red, urlcolor = blue}

\journal{International Journal of Non-Linear Mechanics}

\begin{document}

\begin{frontmatter}

\title{Fractal basins of convergence of libration points in the planar Copenhagen problem with a repulsive quasi-homogeneous Manev-type potential}

\author[mss]{Md Sanam Suraj\corref{cor1}}
\ead{mdsanamsuraj@gmail.com}

\author[eez]{Euaggelos E. Zotos}

\author[ck]{Charanpreet Kaur}

\author[mss]{Rajiv Aggarwal}

\author[am]{Amit Mittal}

\cortext[cor1]{Corresponding author}

\address[mss]{Department of Mathematics, Sri Aurobindo College,
University of Delhi, Delhi, India}

\address[eez]{Department of Physics, School of Science,
Aristotle University of Thessaloniki, GR-541 24, Thessaloniki, Greece}

\address[ck]{Department of Mathematics, SGTB Khalsa College,
University of Delhi, North Campus, New Delhi, India}

\address[am]{Department of Mathematics, ARSD College,
University of Delhi, Delhi, India}

\begin{abstract}
The Newton-Raphson basins of convergence, corresponding to the coplanar libration points (which act as attractors), are unveiled in the Copenhagen problem, where instead of the Newtonian potential and forces, a quasi-homogeneous potential created by two primaries is considered. The multivariate version of the Newton-Raphson iterative scheme is used to reveal the attracting domain associated with the libration points on various type of two-dimensional configuration planes. The correlations between the basins of convergence and the corresponding required number of iterations are also presented and discussed in detail. The present numerical analysis reveals that the evolution of the attracting domains in this dynamical system is very complicated, however, it is a worth studying issue.
\end{abstract}

\end{frontmatter}

\section{Introduction}
\label{intro:1}
The few-body problem, especially the restricted three-body problem, is the most celebrated problem in Celestial Mechanics. The history of this problem starts since the epoch of Lagrange and Euler. The restricted problem of three bodies describes the motion of a body with infinitesimal mass under the Newtonian gravitational attraction of two bodies, known as primaries, which move around their common center of mass in circular orbits under their mutual Newtonian attraction. In the special case of the so-called Copenhagen problem, the masses of the primaries are equal (\cite{sze67}).

Recently, various extrasolar planetary systems have been discovered which consist of two major bodies that in some cases can be presumed to have almost equal masses. Therefore, the basic configuration of the Copenhagen problem can have a real application in which the third body may be considered as a "Trojan" asteroid, moving in the vicinity of two main bodies. There are two versions of the Copenhagen problem: (i) the original version of the Newtonian gravitational attraction (e.g., \cite{ben96}; \cite{per96}; \cite{bro01}; \cite{kal08, kal12}; \cite{pap09}; \cite{zot15}) and (ii) the modified version in which the effective potential have been modified by introducing other perturbing parameters. More precisely, the pseudo-Newtonian planar restricted three-body problem deals with the modified version, where additional general relativistic corrections have been included to the effective potential (\cite{zot17a}).\\
Newton in his famous work Philosophiae Naturalis Principia Mathematica considered the quasi-homogeneous potential of the form -$\Big(\frac{a}{r}+\frac{e}{r^2}\Big)$, where $a, e$ are real constants, while the distances between the particles are denoted by $r$. In the framework of an inverse-square force law, the impossibility to explain the Moon's apsidal was the main reason to add the term $\frac{e}{r^2}$. An improvement in the universal law of gravity, by adding the corrective term of the form $\frac{e}{r^2}$ so as the gravitational potential to fall into a more general class of potential, always referred as the quasi-homogeneous potential (\cite{dia96}).\\
In a recent paper \cite{fak13}, the numerical aspects of the dynamics of the test particle under the action of a Maxwell-type $N-$body problem, by considering the central body as a spheroidal, were revealed. In the presented system, the non-sphericity is modeled by taking a corrective term that coincides with a Manev type potential. Moreover, in \cite{fak17} they studied the restricted three-body problem by taking primaries with equal masses and a quasi-homogeneous potential. The locations of the equilibrium points, the zero-velocity curves and the evolution of the regions where the motion of the test particle is allowed, were investigated. \\
In the present paper, we use the quasi-homogeneous potential to reveal the fractal basins of convergence by using the multivariate version of the Newton-Raphson iterative scheme in the Copenhagen problem. By taking the correction term into consideration, the effective potential becomes mono-parametric.\\
The structure of the present article is as follows: The most important properties of the dynamical model are discussed in Section \ref{sec:2}. The following section contains the parametric evolution of the libration points and their stability. The evolution of the Newton-Raphson basins of convergence, associated with the libration points, are discussed in Section \ref{sec:4}. The conclusions and all the main results of this work are presented in Section \ref{sec:5}.
\section{Description of mathematical model}
\label{sec:2}
According to the theory of the classical circular restricted three-body problem (\cite{sze67}), two bodies $P_1$ and $P_2$, with masses $m_1$ and $m_2$ respectively, called primaries, move on circular orbits around their common center of mass. The third body, also known as test particle, with mass $m$ moves under the gravitational influence of the two primaries and does not disturb, in any manner, their circular motion. It is further assumed that the mass of the third body is significantly smaller in comparison to the masses of the primaries $(m \ll m_1, m_2)$. Furthermore, we consider an inertial coordinate system $OXYZ$, where the plane $OXY$ coincides with the plane of the motion of the primaries $P_{1,2}$ (see Fig.1).
\begin{figure}
\label{Fig01}
\begin{center}
\resizebox{0.8\hsize}{!}{\includegraphics*{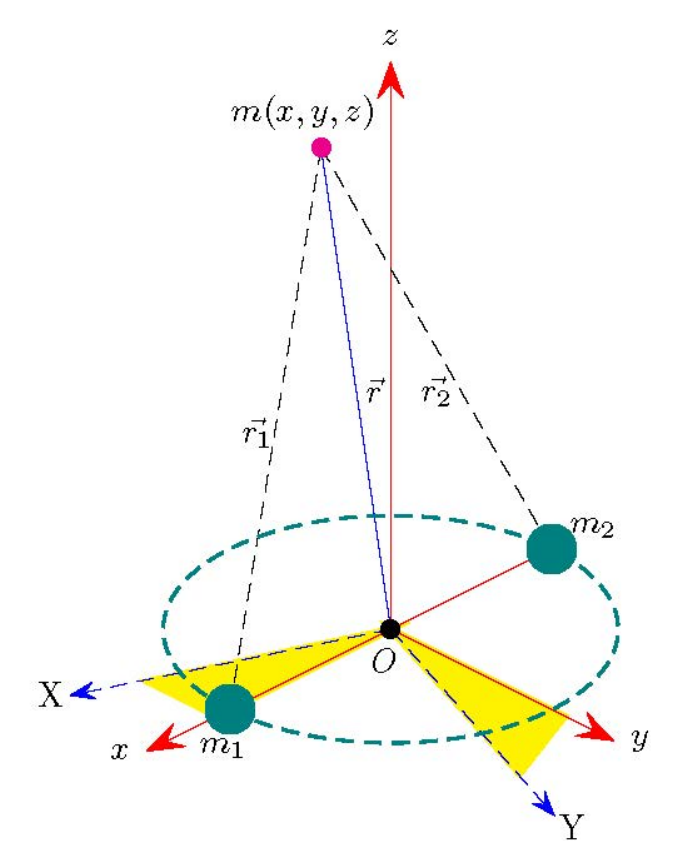}}
\caption{The Copenhagen problem: the configuration of the problem with the synodic coordinate system $Oxyz$ and the inertial frame $OXYZ$.}
\end{center}
\end{figure}

According to \cite{fak17}, the expression of the time-independent effective potential in a synodic coordinates system $Oxyz$ is
\begin{equation}\label{Eq:1}
\Omega(x, y, z)=\frac{1}{\Delta}\Big[\sum_{i=1}^{2}\Big(\frac{1}{r_i}+\frac{e}{r_i^2}\Big)\Big]+\frac{1}{2}(x^2+y^2),
\end{equation}
where $(x,y,z)$ are the coordinates of the test particle, while $r_i, i=1,2$ are the distances of the test particle from the primaries $P_i$, respectively,
\begin{equation}
  r_1 = \sqrt{\Big(x-\frac{1}{2}\Big)^2+y^2+z^2},\quad
  r_2 = \sqrt{\Big(x+\frac{1}{2}\Big)^2+y^2+z^2}.\nonumber
\end{equation}
Moreover, the quantity $\Delta$ is a proper number which depends on all the parameters and the geometrical characteristics of the configuration. In this study, the Copenhagen case is considered as a special case where $\Delta=M(\Lambda+2e\Lambda_1),$ with $M=2, \Lambda=1,$ and $\Lambda_1=1$. Since $\Delta$ must be positive, the parameter $e$ must satisfy the condition $e>-0.5$ (more details on the involved parameters and their meaning can be found in \cite{fak17}).\\
Using the transformation from the inertial to the synodic coordinate system and scaling the physical quantities, where we have considered the constant angular velocity of the primaries equal to unity, the equations of motion of the test particle $m$ in the rotating frame of reference are
\begin{subequations}
\begin{eqnarray}
\label{Eq:2a}
\ddot{x}-2\dot{y}&=&\Omega_x,\\
\label{Eq:2b}
\ddot{y}+2\dot{x}&=&\Omega_y,\\
\label{Eq:2c}
\ddot{z}&=&\Omega_z,
\end{eqnarray}
\end{subequations}
where
\begin{subequations}
\begin{eqnarray}
\label{Eq:3a}
\Omega_x &=&  x-\frac{1}{\Delta}\Big(\sum_{i=1}^2\frac{x-x_i}{r_i^3}+2e\sum_{i=1}^2\frac{x-x_i}{r_i^4}\Big),\\
\label{Eq:3b}
\Omega_y &=&  y-\frac{y}{\Delta}\Big(\sum_{i=1}^2\frac{1}{r_i^3}+2e\sum_{i=1}^2\frac{1}{r_i^4}\Big),\\
\label{Eq:3c}
\Omega_z &=&  -\frac{z}{\Delta}\Big(\sum_{i=1}^2\frac{1}{r_i^3}+2e\sum_{i=1}^2\frac{1}{r_i^4}\Big),
\end{eqnarray}
\end{subequations}
while, $x_1=-x_2=\frac{1}{2}$.

Moreover, the partial derivatives of the second order, which will be used later for the  basins of convergence associated with the libration points, read as
\begin{subequations}
\begin{eqnarray}\label{Eq:4a}
\Omega_{xx} &=& 1-\frac{1}{\Delta}\Big(\sum_{i=1}^2\frac{1}{r_i^3}+\sum_{i=1}^2\frac{2e}{r_i^4}-\sum_{i=1}^2\frac{3(x-x_i)^2}{r_i^5}\nonumber\\
  &&-8e\sum_{i=1}^2\frac{(x-x_i)^2}{r_i^6}\Big),\\
\label{Eq:4b}
\Omega_{yy} &=& 1-\frac{1}{\Delta}\Big(\sum_{i=1}^2\frac{1}{r_i^3}+\sum_{i=1}^2\frac{2e}{r_i^4}-\sum_{i=1}^2\frac{3y^2}{r_i^5}-8e\sum_{i=1}^2\frac{y^2}{r_i^6}\Big),\\
\label{Eq:4c}
\Omega_{zz} &=&-\frac{1}{\Delta}\Big(\sum_{i=1}^2\frac{1}{r_i^3}+\sum_{i=1}^2\frac{2e}{r_i^4}-\sum_{i=1}^2\frac{3z^2}{r_i^5}-8e\sum_{i=1}^2\frac{z^2}{r_i^6}\Big),\\
\label{Eq:4d}
\Omega_{xy} &=&\Omega_{yx} =\frac{1}{\Delta}\Big(\sum_{i=1}^2\frac{3(x-x_i)y}{r_i^5}+\sum_{i=1}^2\frac{8e(x-x_i)y}{r_i^6}\Big),\\
\label{Eq:4e}
\Omega_{yz} &=& \Omega_{zy} =\frac{1}{\Delta}\Big(\sum_{i=1}^2\frac{3yz}{r_i^5}+\sum_{i=1}^2\frac{8eyz}{r_i^6}\Big),\\
\label{Eq:4f}
\Omega_{zx} &=&\Omega_{zx}= \frac{1}{\Delta}\Big(\sum_{i=1}^2\frac{3(x-x_i)z}{r_i^5}+\sum_{i=1}^2\frac{8e(x-x_i)z}{r_i^6}\Big).
\end{eqnarray}
\end{subequations}
The system of the equations (\ref{Eq:2a},...,\ref{Eq:2c}) admits the well known Jacobian-type integral of motion which corresponds to the total orbital energy
\begin{equation}\label{Eq:9}
J(x, y, z, \dot{x}, \dot{y}, \dot{z})=2\Omega(x, y, z)-(\dot{x}^2+\dot{y}^2+\dot{z}^2) =C,
\end{equation}
where $\dot{x}, \dot{y}$, and $\dot{z}$  represent the velocities, while the numerical value of the Jacobi constant is represented by $C$. This Jacobian constant is conserved and, in the canonical coordinates, the value of Hamiltonian corresponding to the Jacobian integral is known as the total orbital energy $h$, which is related to Jacobian constant by $C=-2h$.

The coplanar libration points on the configuration $(x,y)$ plane exist when $z=0$, while the out-of-plane libration points are those points which are located either on the $(x, z)$ plane (obtained by taking $y=0, x\neq0$) or on the $z-$axis. However, we restrict our analysis only to the coplanar libration points.
The position of the libration points is defined through the intersection of the equations $\Omega_x=0, \Omega_y=0$.

\begin{figure*}
  \centering
  (a)\resizebox{0.45\hsize}{!}{\includegraphics*{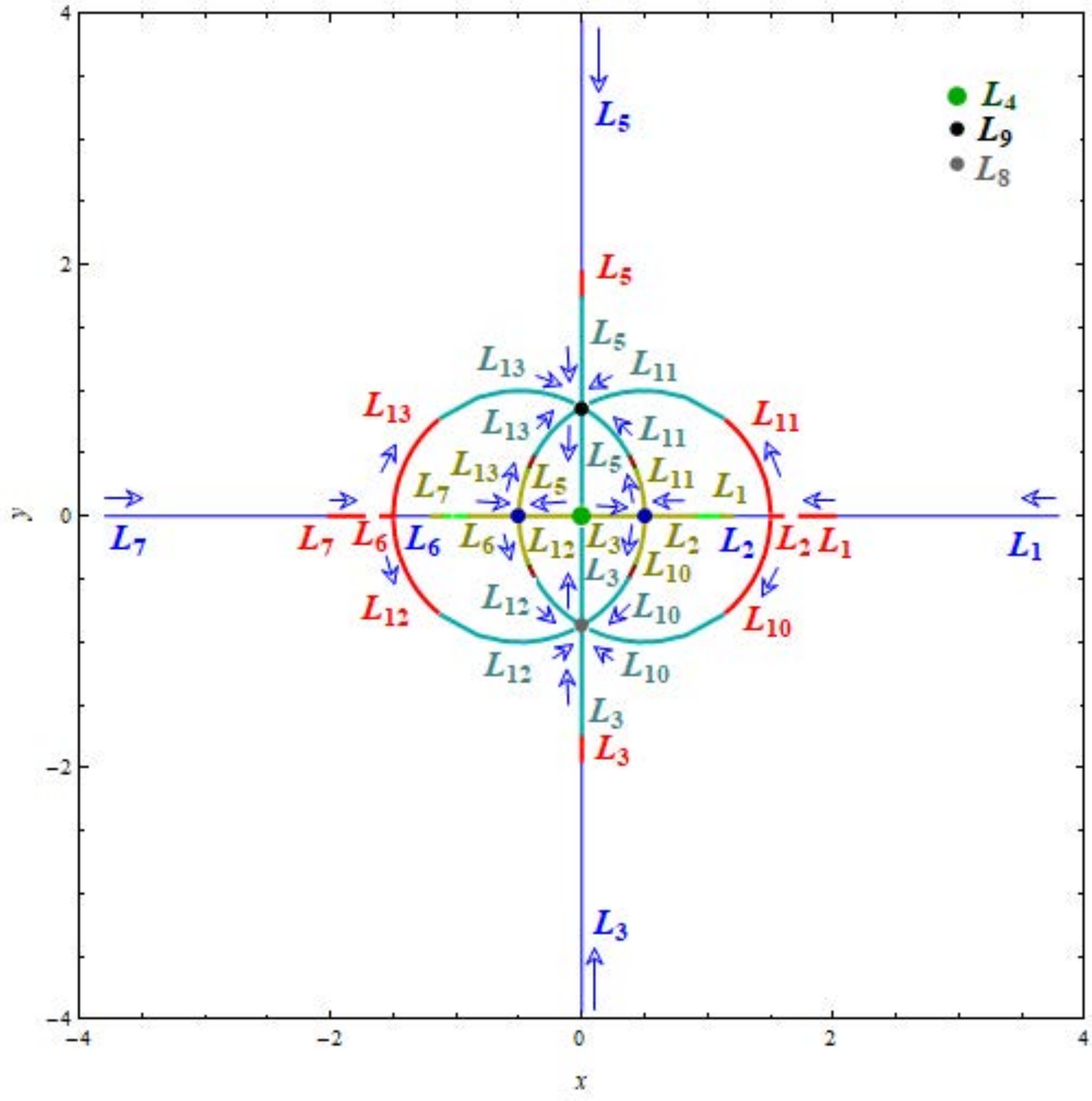}}
  (b)\resizebox{0.45\hsize}{!}{\includegraphics*{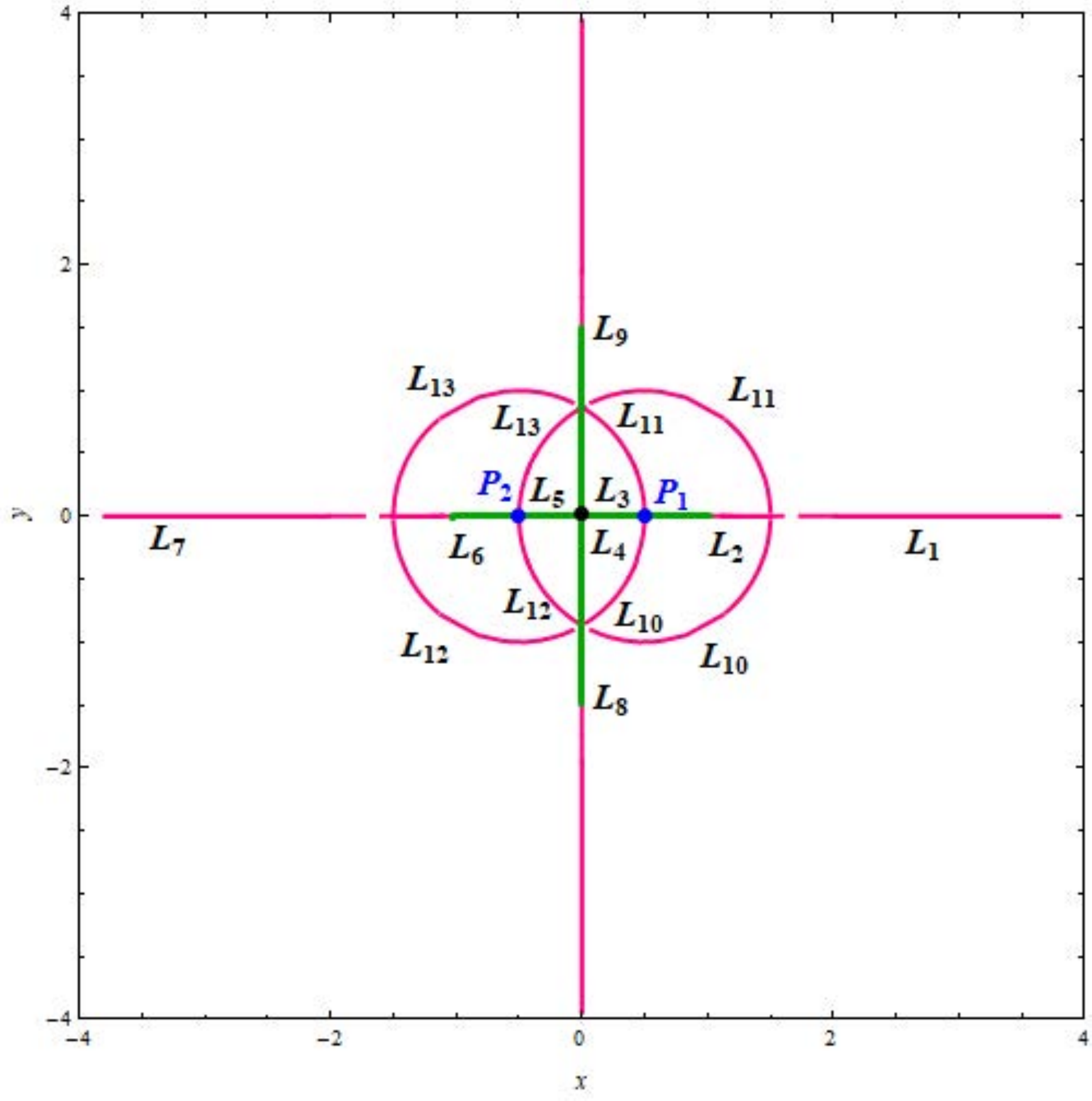}}
  \caption{The parametric evolution of (a-left): the positions and (b-right):
the linear stability (\emph{green}) or instability (\emph{rubine red}) of the libration points, $L_i, i=1,..., 13,$ in the Copenhagen restricted three-body problem with quasi-homogeneous potential, when $e\in(-0.5, 0)$. The arrows indicate the movement direction of the libration points as the value of the parameter $e$ decreases. The big blue dots pinpoint the fixed positions of the primaries, while the colour code is as follows: \emph{olive} for $e\in(-0.173395,0)$, \emph{green} for $e\in(-0.19526, -0.173395)$, \emph{crimson} for $e\in(-0.23334, -0.19526)$, \emph{teal} for $e\in(-0.457853, -0.2333
4)$, \emph{red} for $e\in(-0.46671, -0.457853)$, \emph{blue} for $e\in(-0.5, -0.46671)$. The libration points are depicted in the corresponding colour code. (Colour figure online).}
\label{Fig:2n}
\end{figure*}
The location as well as the existence of the libration points strongly depend on the parameter $e$ in the Copenhagen restricted three-body problem with a quasi-homogeneous potential. More precisely:
\begin{description}
 \item[$\bullet$] when $e\in (-0.23334, -0.19526)$ there are seven libration points: one collinear and six non-collinear libration points.
  \item[$\bullet$] when $e\in (-0.5, -0.46671)$  and $e\in (-0.457853, -0.23334)$ there are nine libration points: one or five collinear and eight or four non-collinear libration points respectively.
  \item[$\bullet$] when $e\in (-0.19526, -0.173395)$ there are eleven libration points: five collinear and six non-collinear libration points.
  \item[$\bullet$] when $e\in (-0.46671, -0.457853)$ and $(-0.173395, 0)$ there are thirteen libration points: five or seven collinear and eight or six non-collinear libration points respectively.
  \item[$\bullet$] when $e\geq0$ there are five libration points: three collinear and two non-collinear libration points.
\end{description}
It is worth studying the exact evolution of the positions of the libration points as a function of the parameter $e$, when $e\in (-0.5, 0)$. Figure \ref{Fig:2n}(a) illustrates the parametric evolution of the libration points, on the configuration $(x, y)$ plane.
We observe that as the parameter $e$ decreases (just below the zero), eight libration points, in two sets of four, come forth from the centers of the primaries $P_{1,2}$ (see Fig. \ref{Fig:2n}a, \emph{olive colour}). The interesting phenomenon occurs for the parameter $e$ , when two sets of four libration points existing on $(x, y)$ plane shrink to two single points on the $y-$axis (see $L_{8,9}$ in Fig. \ref{Fig:2n}a ) for $e\approx-0.377$ and for further decreasing value of the parameter $e$, they start expanding again (see Fig. \ref{Fig:2n}a, \emph{teal colour}). Therefore, we may argue that these sets of the libration points first collide to libration points $L_8$ and $L_9$ for the critical value of parameter $e\approx-0.377$ and then, again for decreasing values of $e$ they start originating from $L_8$ and $L_9$ and expand.

In panel: b of Fig. \ref{Fig:2n} we have depicted the evolution of stability of all the libration points, when the parameter $e \in (-0.5, 0)$. Our numerical computations reveal the following facts:
\begin{itemize}
  \item The libration points $L_{8,9}$ are stable when $e\in(-0.5, -0.445)$.
  \item When $e\in(-0.444, -0.23334)$, two more libration points $L_{3,5}$ are also stable along with the libration points $L_{8,9}$.
  \item The central libration point $L_4$ along with $L_{8,9}$ are stable if $e\in(-0.23334, -0.23)$, while only the central libration point $L_4$ remains stable if $e\in(-0.23, -0.19526)$.
  \item The libration points $L_{1,7}$ are stable if $e\in(-0.19526,$ $ -0.173395)$, while the libration points $L_{3,5}$ also become stable when $e\in(-0.173395, 0)$. It is further observed that the libration points $L_{3,5}$  originating from the vicinity of the origin, move towards the primaries $P_{1,2}$ from left and right to the respective primary, while the libration points $L_{1,7}$ also approach from right and left to the primaries $P_{1,2}$, respectively.
\end{itemize}
It may be concluded that the stable libration points are those which lie on either $x-$axis or $y-$axis while the rest of the coplanar libration points (i.e. $x\neq0$ , $y\neq0$) are always unstable for all values of the parameter $e$.

\section{Newton-Raphson basins of convergence}
\label{sec:4}
To solve the system of non-linear equations numerically, various iterative methods are available. Over the years, the Newton-Raphson method has become one of the most captivated as well as accurate iterative schemes to solve these type of equations. The associated multivariate iterative scheme is
\begin{equation}\label{Eq:401}
 \textbf{x}_{n+1}=\textbf{x}_n-J^{-1}f(\textbf{x}_n),
\end{equation}
where $f(\textbf{x}_n)$ represents the system of equations, while the $J^{-1}$  represents the corresponding inverse Jacobian matrix. In the present dynamical system, the system of the equations are:
\begin{subequations}
\begin{eqnarray}
\label{Eq:402a}
\Omega_{x}(x,y)&=&0,\\
\label{Eq:402b}
  \Omega_{y}(x,y)&=&0.
\end{eqnarray}
\end{subequations}
The multivariate version of the iterative scheme (\ref{Eq:401}) for each coordinate $(x, y)$, can be decomposed into two formulae as:
\begin{subequations}
\begin{eqnarray}
\label{Eq:403a}
x_{n+1} &= x_n - \left( \frac{\Omega_x \Omega_{yy} - \Omega_y \Omega_{xy}}{\Omega_{yy} \Omega_{xx} - \Omega^2_{xy}} \right)_{(x,y) = (x_n,y_n)}, \\
\label{Eq:403b}
y_{n+1} &= y_n + \left( \frac{\Omega_x \Omega_{yx} - \Omega_y \Omega_{xx}}{\Omega_{yy} \Omega_{xx} - \Omega^2_{xy}} \right)_{(x,y) = (x_n,y_n)},
\end{eqnarray}
\end{subequations}
where $x_n$ and $y_n$ are the values of $x$ and $y$ at the $n$-th step of the iterative process of the Newton-Raphson method. Here, the subscripts of $\Omega(x, y)$ represent the corresponding partial derivatives of the potential function.\\
The Newton-Raphson iterative scheme works on the following philosophy: the numerical code is activated with the given initial condition $(x_0, y_0)$, on the configuration plane, while the iterative procedure continues until a libration point is reached, with the predefined coveted accuracy. The iterative scheme converges for the particular initial condition, if this initial condition leads to one of the libration points, no matter what its state of stability is. The sets of all the initial conditions which converge to the same attractor, compile the attracting domains, also known as Newton-Raphson basins of convergence.\\

In dynamical system knowing the exact positions of the equilibrium points is an issue of paramount importance. Unfortunately, in many systems, such as those of the $N$-body problem (with $N \geq 3$), there are no explicit formulae for the positions of the libration points. Therefore, the locations of the equilibrium points can be obtained only by means of numerical methods. In other words, we need a multivariate iterative scheme for solving the system of the first order derivatives. It is well known that the results of any numerical method strongly depend on the initial conditions (staring points of the iterative procedure). Indeed, for some initial conditions the iterative formulae converge quickly, while for other starting points a considerable amount of iterations is required for reaching to a root (equilibrium point). Fast converging points usually belong to basins of convergence, while on the other hand slow converging points are located in fractal regions. On this basis, the knowledge of the basins of convergence of a dynamical system is very important because these basins reveal the optimal (regarding fast convergence) starting points for which the iterative formulae require the lowest amount of iterations, for leading to an equilibrium point. In addition, being aware of the fractal regions we know exactly which points should be avoided as initial conditions of the iterative formulae.\\

Over the years, many researchers and scientists have devoted their efforts to study the Newton-Raphson basins of convergence in various dynamical models, such as the restricted three-body problem (e.g., \cite{CK04b}, \cite{dou12}, \cite{zot16}, \cite{zot17a}, \cite{zot17c}, \cite{sur18a}), the restricted four-body problem (e.g.,  \cite{zot17b}, \cite{sur17a}, \cite{sur17b}, \cite{sur18b}), the restricted five-body problem ( e.g., \cite{ZS18}), the ring-type N-body problem (e.g., \cite{CK04a}, \cite{CK07}), or even the 2+2 body problem (e.g., \cite{CK13}).\\
The Newton-Raphson basins of convergence is determined by using the following algorithm: a double scan of the configuration $(x,y)$ plane is performed after classifying dense uniform grids of $1024\times1024$ initial conditions $(x_0, y_0)$. In the present computation, the predefined accuracy is set to $10^{-15}$ regarding the coordinates of the libration points, whereas the maximum number of iterations allowed is $N_{max}=500$.\\
The following subsections deal with the influence of the parameter $e$ on the topology of the Newton-Raphson basins of convergence, associated with the libration points, in the Copenhagen problem with a quasi-homogeneous potential. We will consider four cases, corresponding to the total number of the libration points which act as attractors.

\subsection{\emph{Case I: when seven libration points exist}}
\label{sec:401}
Our analysis starts with the first case, that is the case when $e\in (-0.23334, -0.19526)$, where seven libration points exist: $L_4$ at the center, four on the $(x, y)$ plane and two on the $y-$axis. In Fig. \textcolor[rgb]{1.00,0.00,0.50}{3}(a, d, g, j), we present the evolution of the topology of the Newton-Raphson basins of convergence for four different values the parameter $e$. It is observed that the configuration $(x, y)$ plane is covered by a profusion of well-formed basins of convergence in which the domain of the basins of convergence associated with the central libration point $L_4$ extent to infinity while for all other libration points the domains of convergence are finite. Moreover, the various local areas such as the vicinity of the basin boundaries composed by a highly fractal mixture of initial conditions. It is worth mentioning that the word "fractal" simply refers to the particular local region which displays a fractal-like geometry. This fact can be justified by the following philosophy: for an initial condition $(x_0, y_0)$ on the configuration plane it is observed that its final state is highly sensitive. More precisely, even a very small change in the initial conditions could result to a completely different attractor or final state. Therefore, it is almost impossible to predict from which of the libration points (attractors) each initial condition will be attracted by.\\
We can observe that the extent of the domain of the basins of the convergence associated with the libration points are almost unperturbed except their size (see panels: a, d). We may conclude that there is no significant change in the geometry of the Newton-Raphson basins of convergence as value of the parameter $e$ decreases. However, as the value of $e$ decreases, the topology of the attracting domain changes  (see panels: g, j). The most notable change is the appearance of figure-eight shaped tendrils at the outer parts of the colour coded diagrams. Moreover, these tendrils increase in size, mainly along the horizontal direction, while on the other hand, the other basins of convergence are being restricted to the central region of the colour coded diagram.\\
\begin{figure*}
\label{Fig:4}
\begin{center}
(a)\includegraphics[scale=.26]{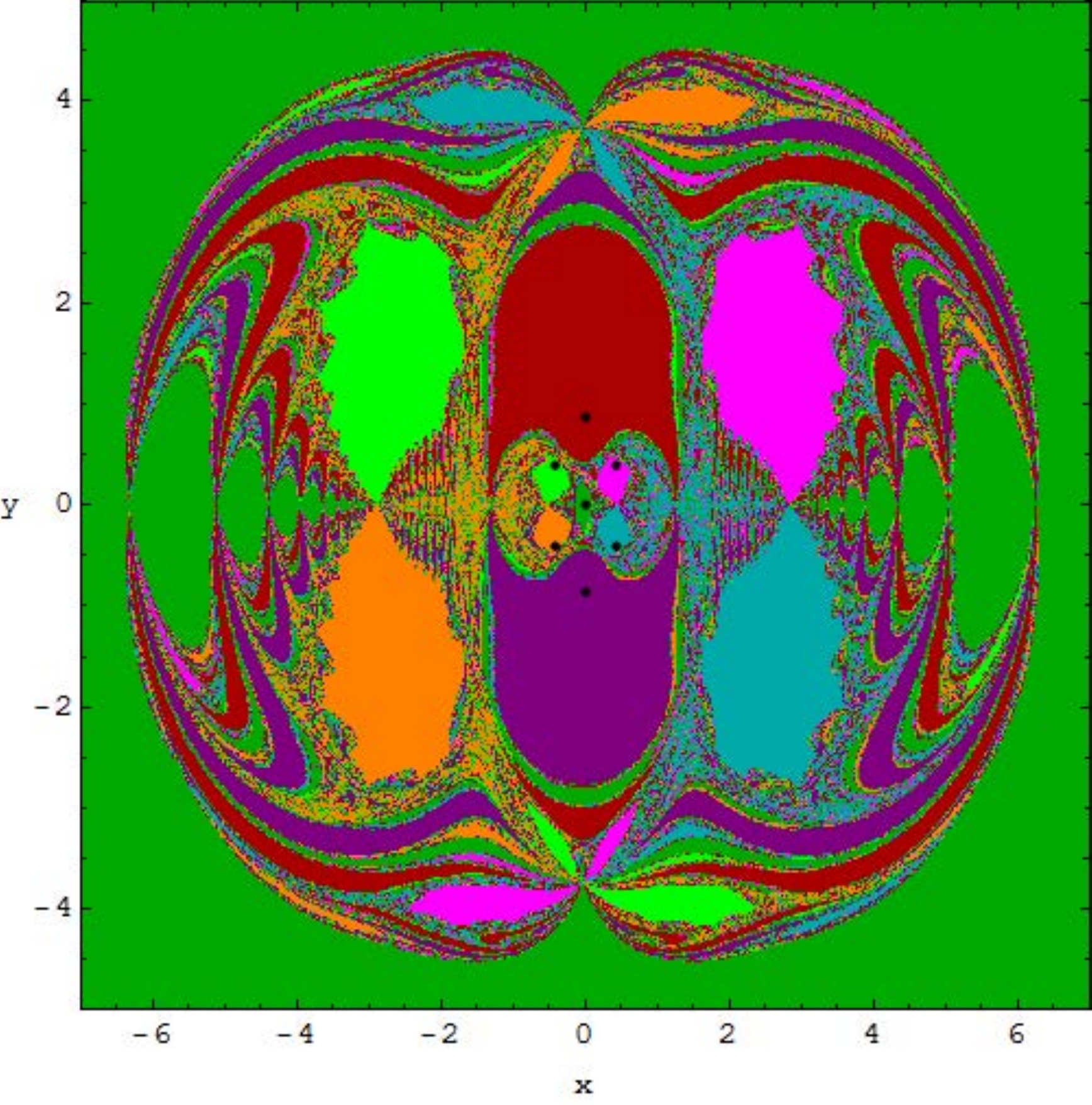}
(b)\includegraphics[scale=.26]{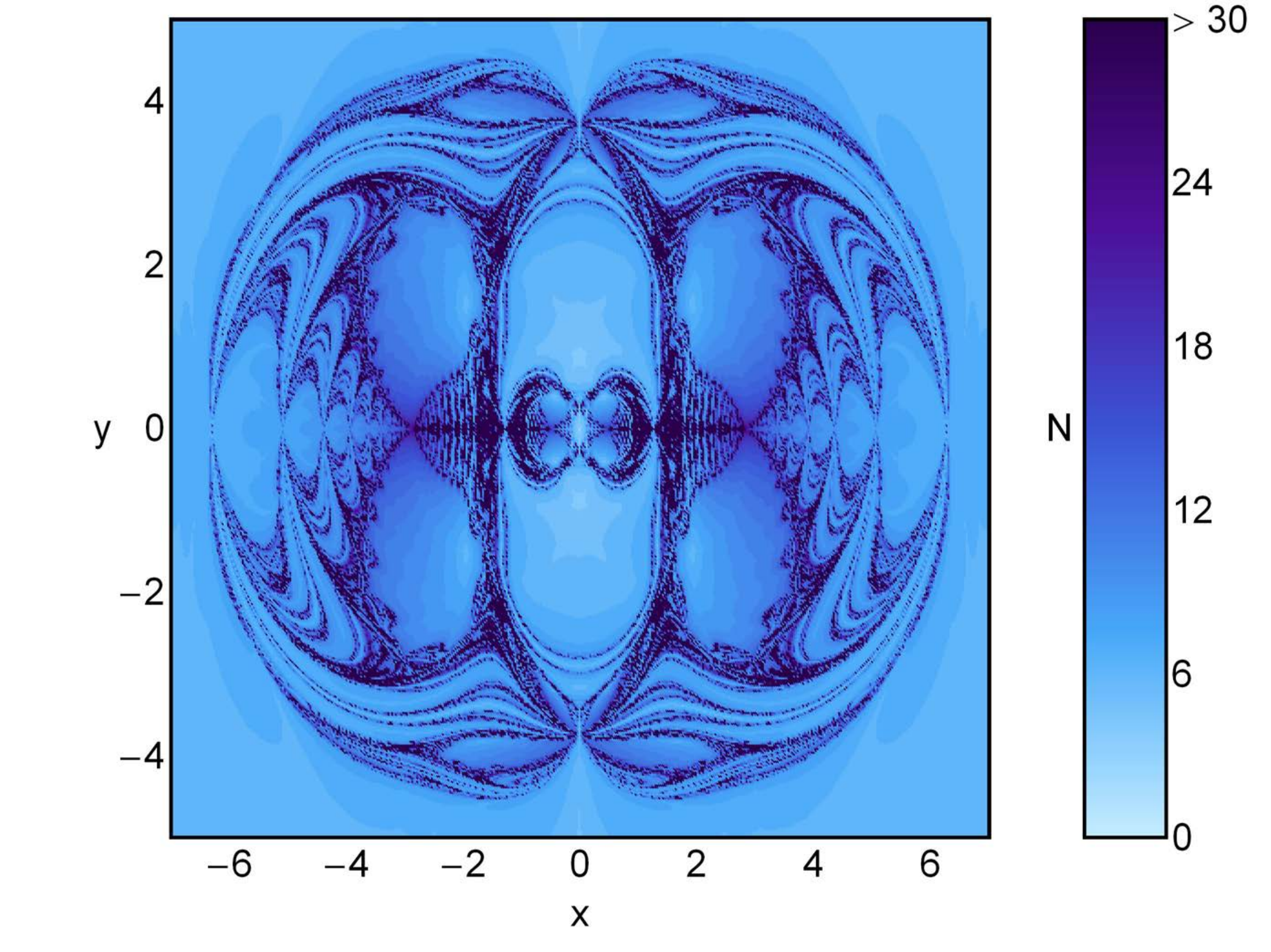}
(c)\includegraphics[scale=.26]{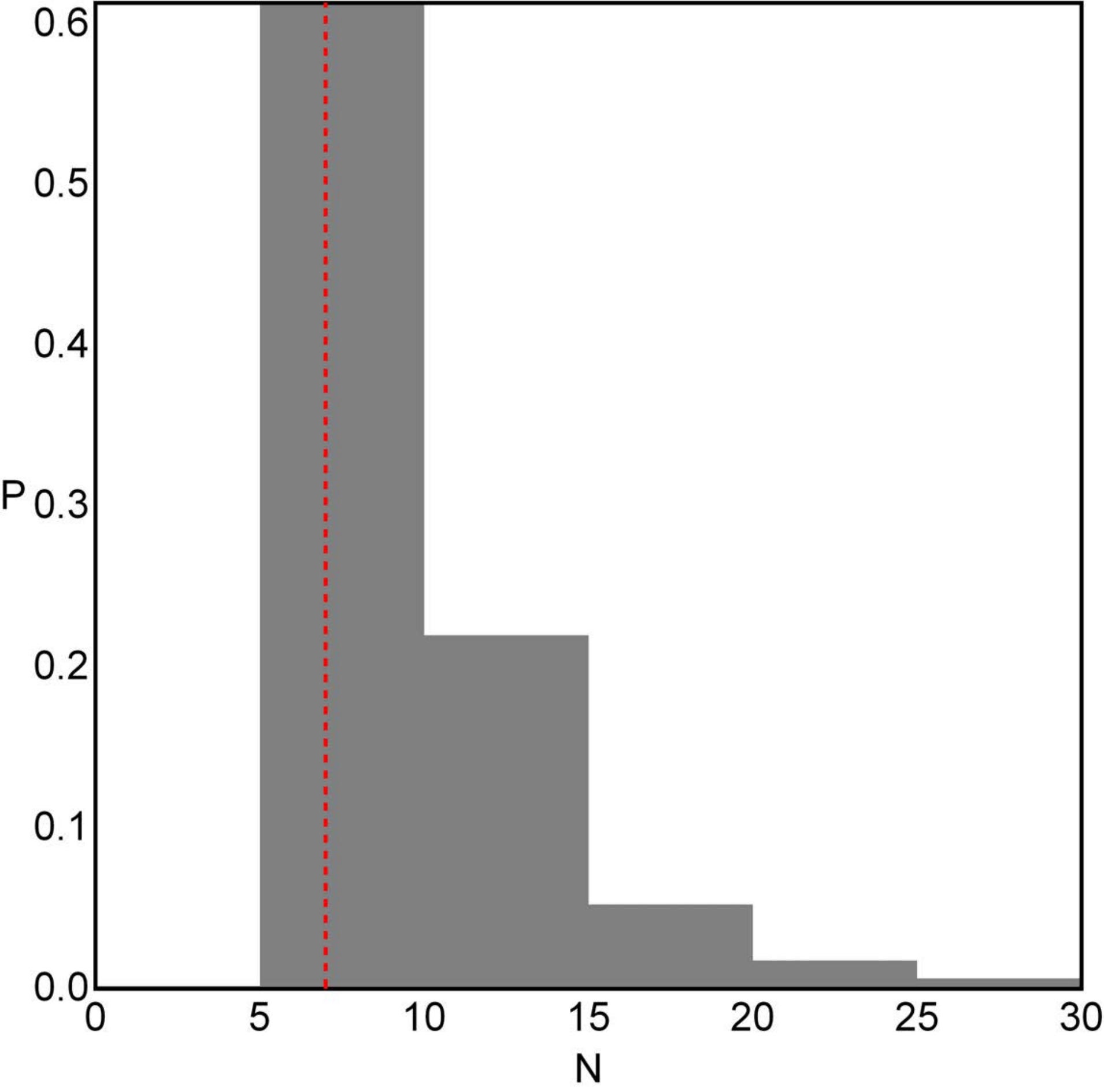}\\
(d)\includegraphics[scale=.26]{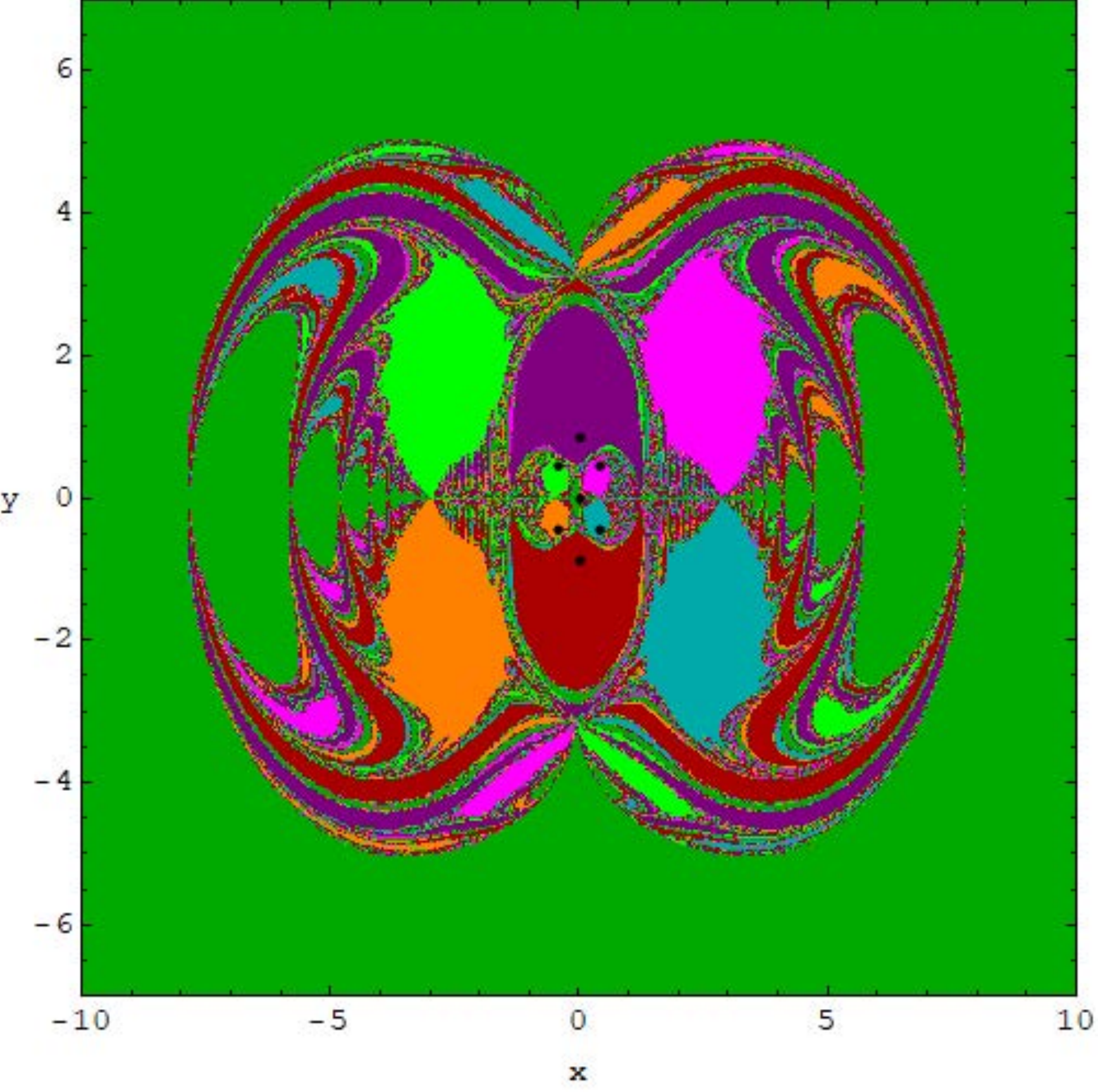}
(e)\includegraphics[scale=.26]{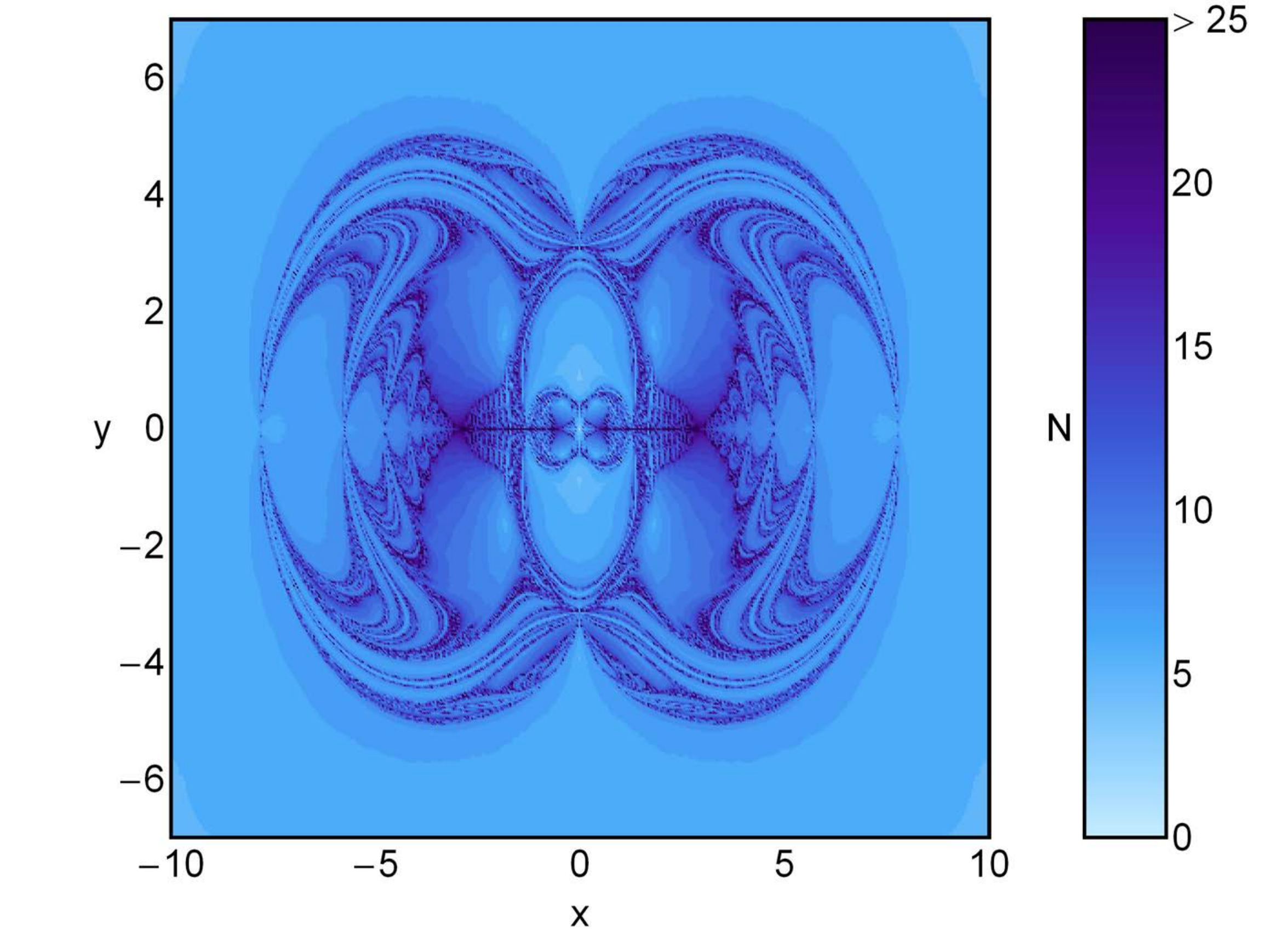}
(f)\includegraphics[scale=.26]{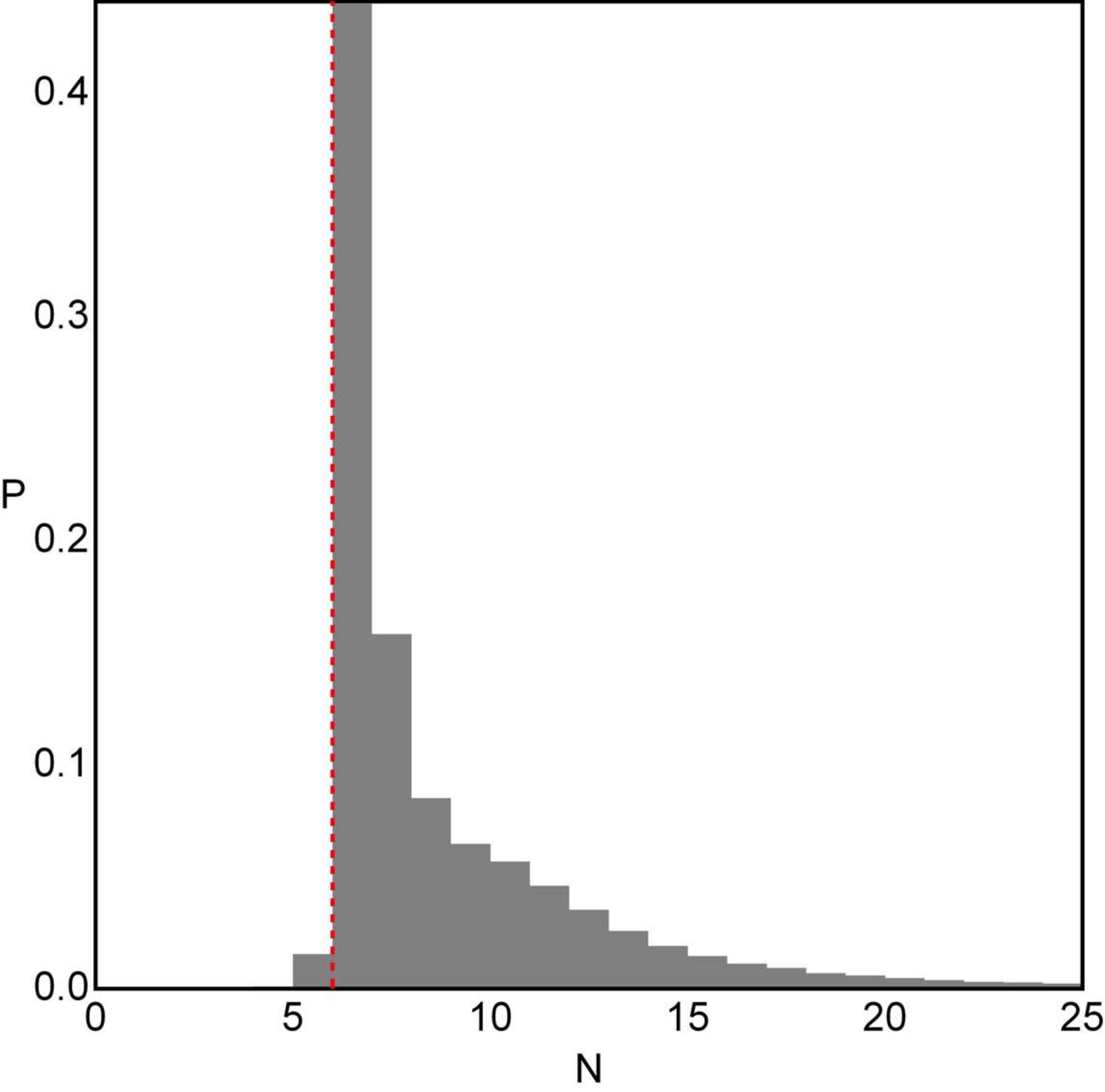}\\
(g)\includegraphics[scale=.26]{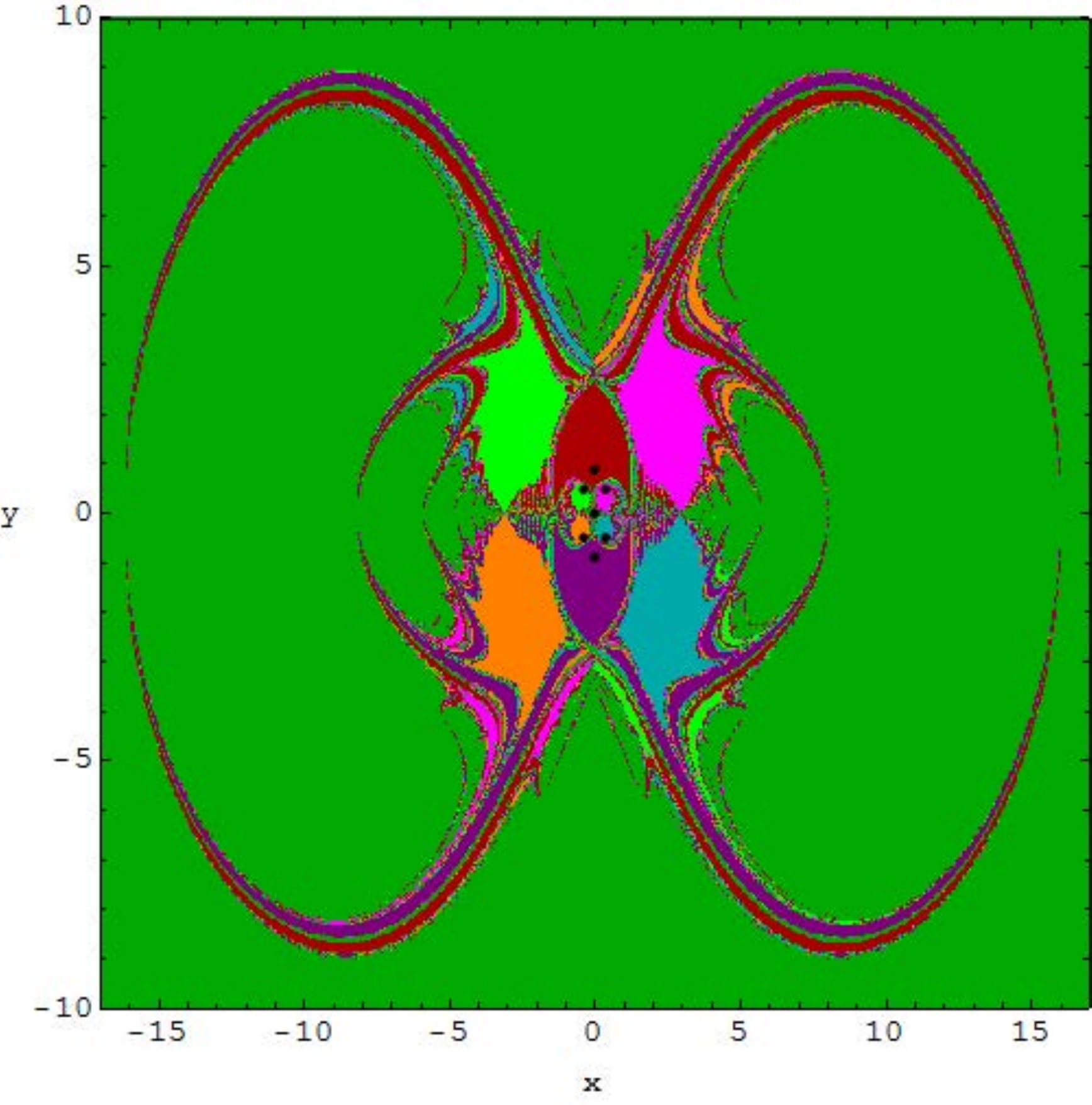}
(h)\includegraphics[scale=.26]{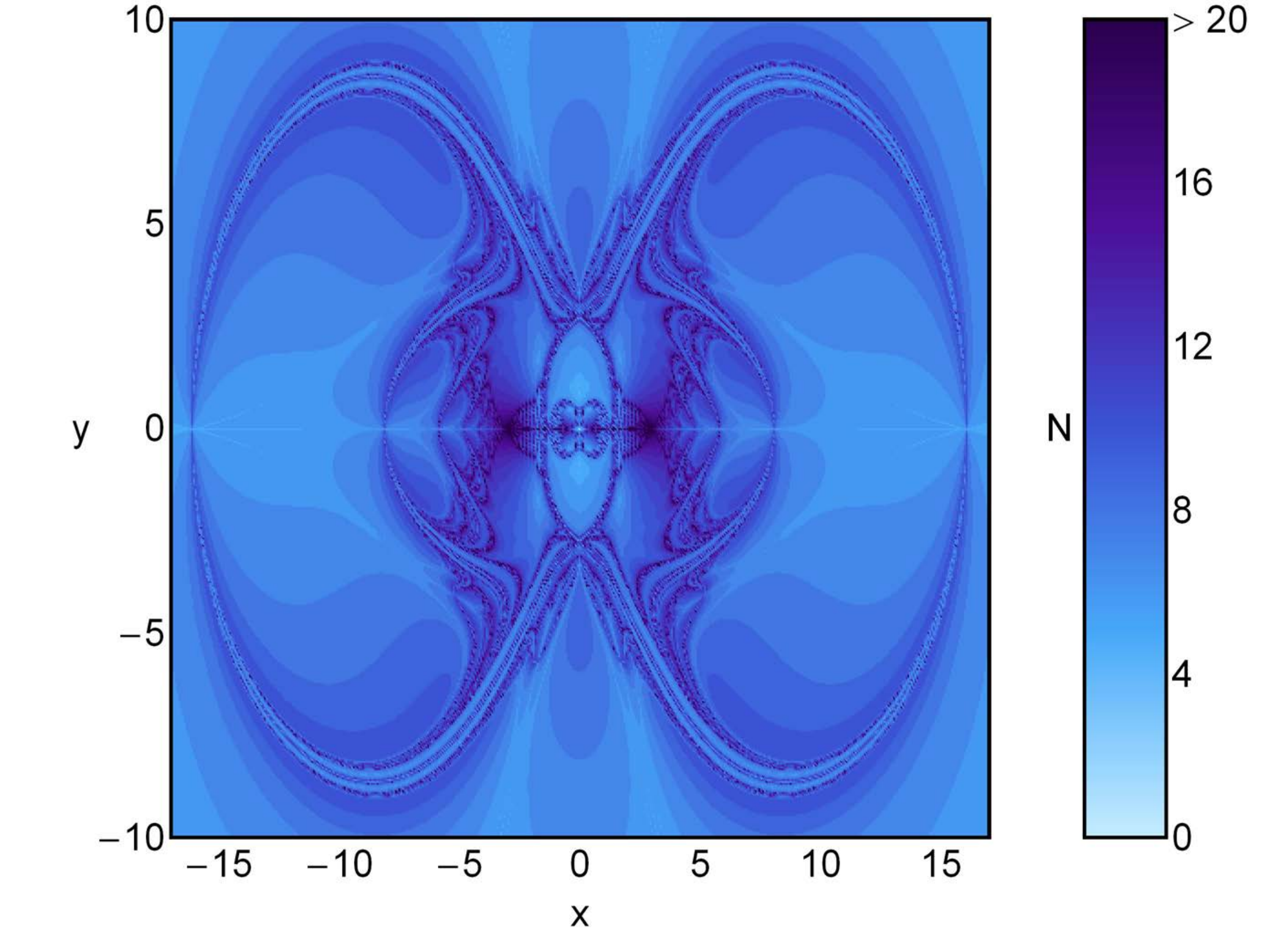}
(i)\includegraphics[scale=.26]{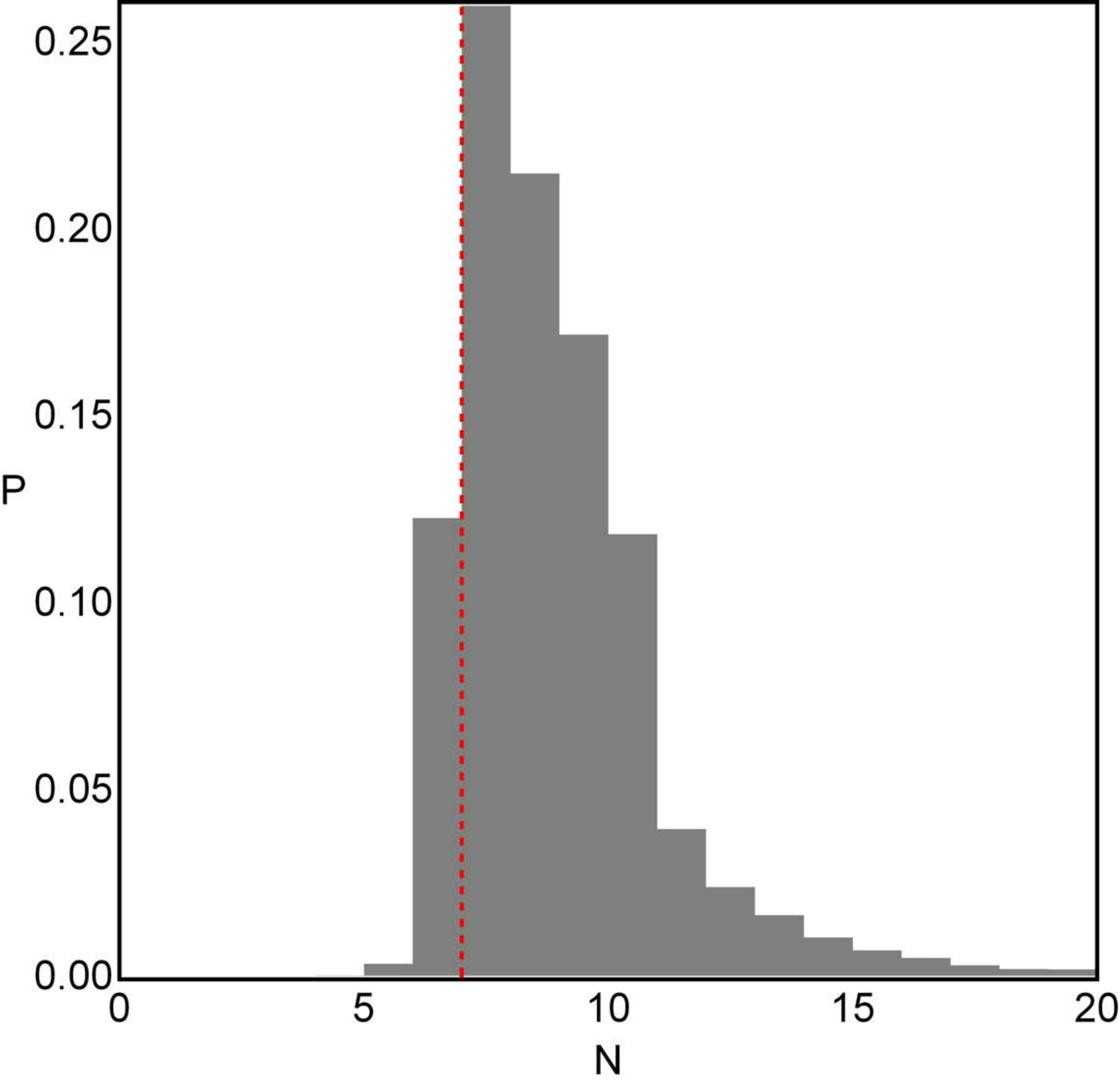}\\
(j)\includegraphics[scale=.26]{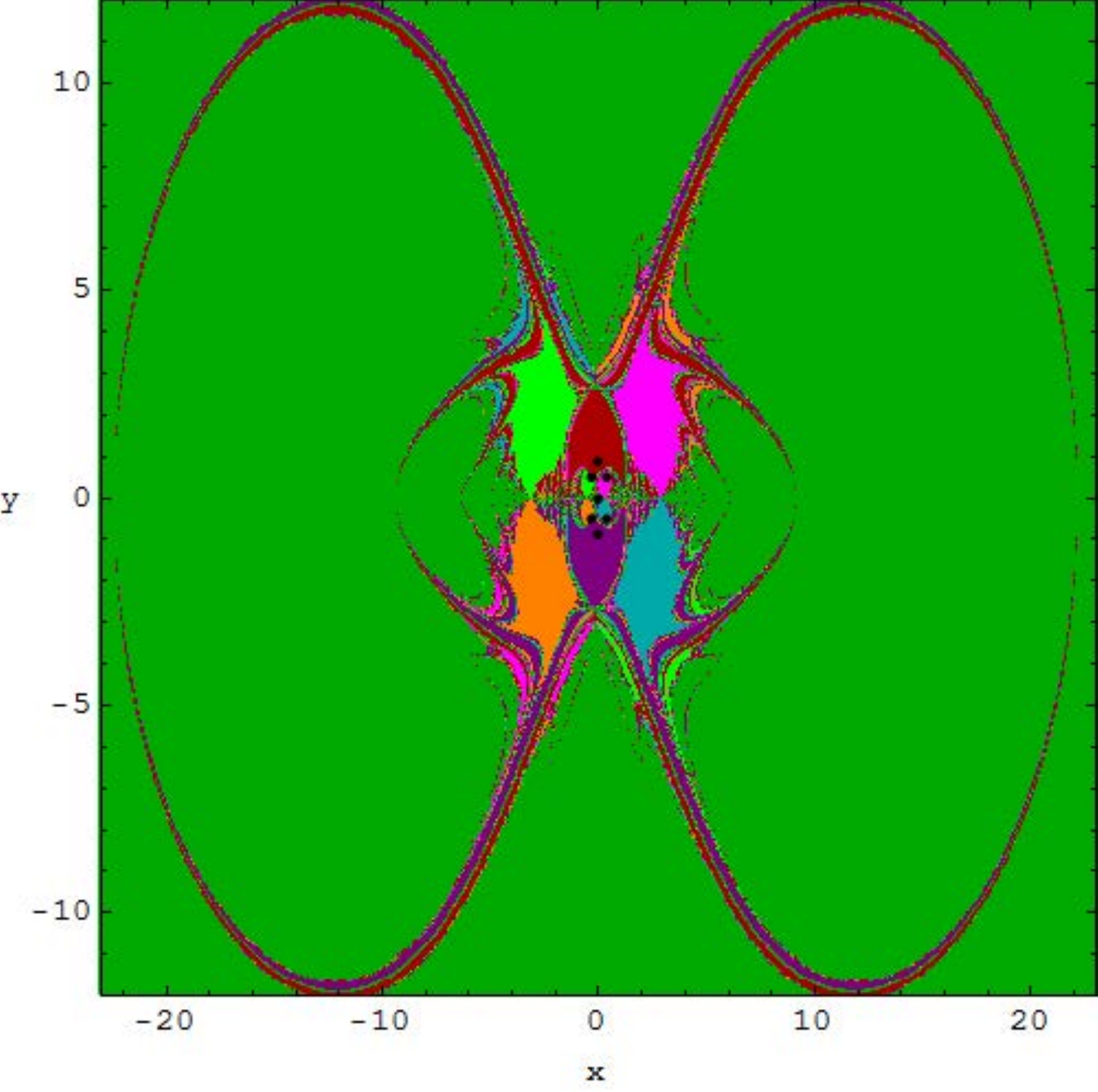}
(k)\includegraphics[scale=.26]{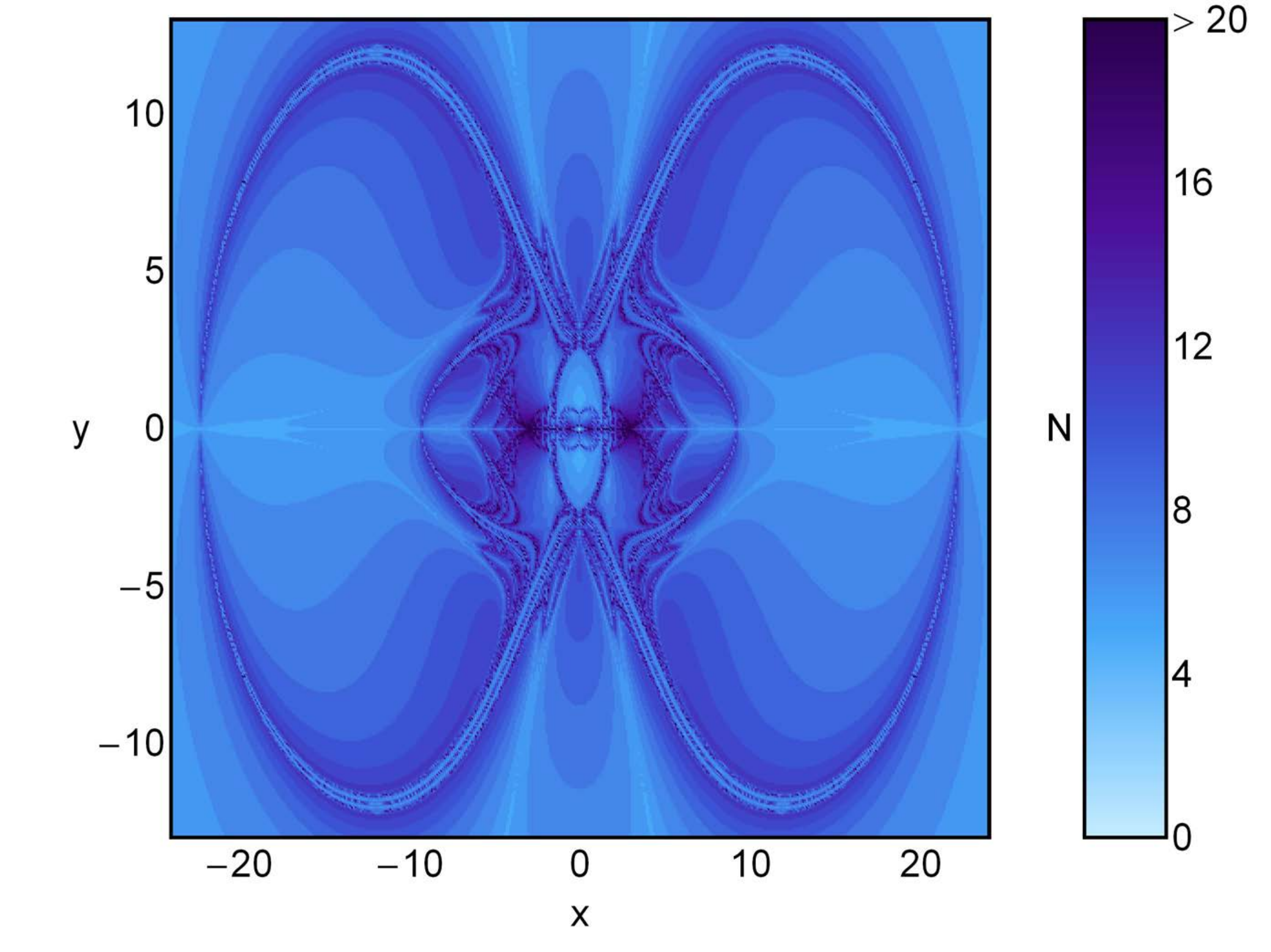}
(l)\includegraphics[scale=.26]{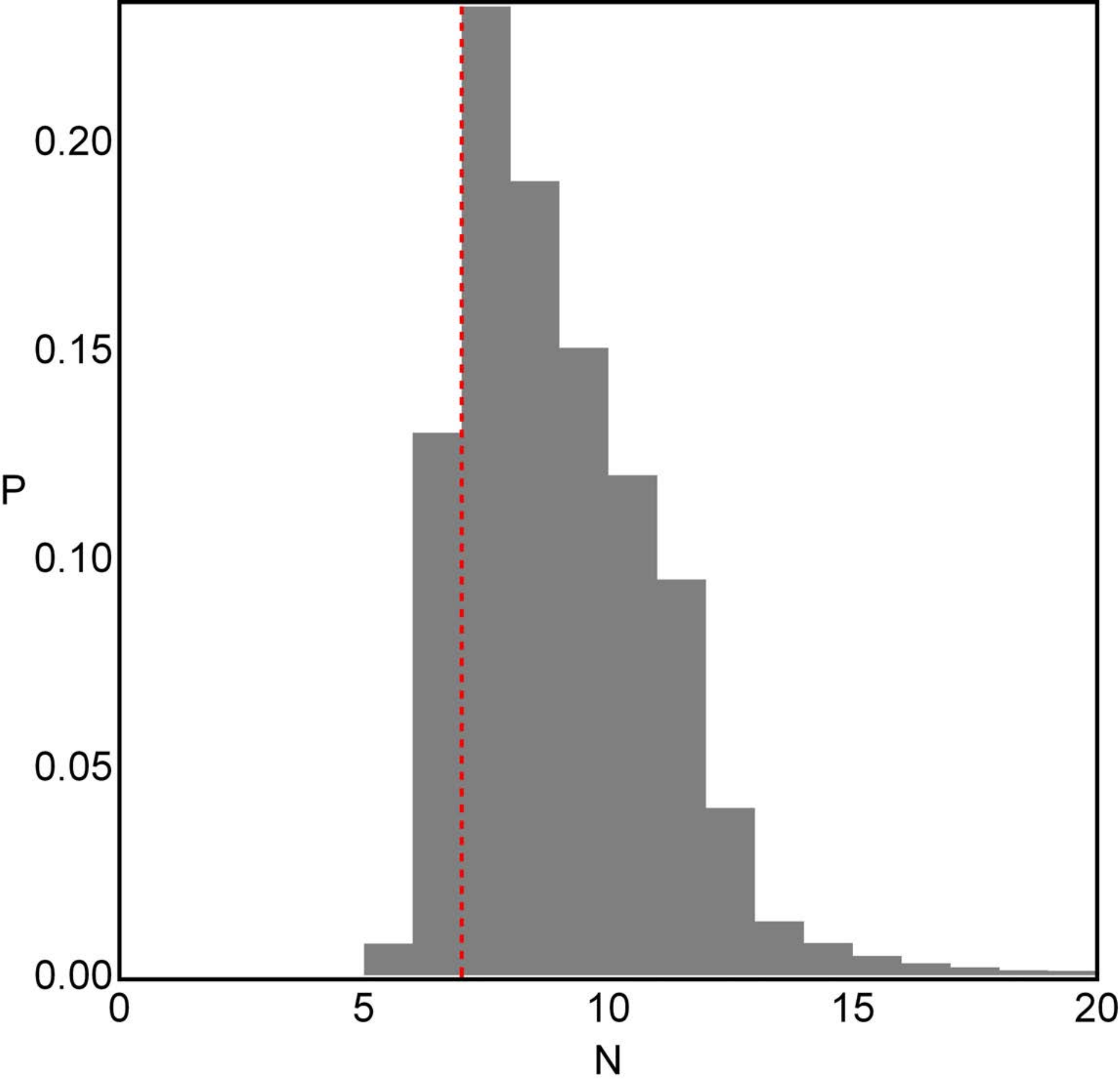}
\caption{The Newton-Raphson basins of attraction on the $(x,y)$ plane for
the case when seven libration points exist for:  (a) $e=-0.1956$; (d) $e=-0.2129$;
 (g) $e=-0.2318$; (j) $e=-0.2329$. The color code denoting the attractors is as follows:
  $L_4$ (\emph{green}); $L_8$ (\emph{purple}); $L_9$ (\emph{crimson}); $L_{10}$
  (\emph{teal}); $L_{11}$ (\emph{magenta}); $L_{12}$ (\emph{orange}); $L_{13}$
  (\emph{light green}) and non-converging points (\emph{white}). (b, e, h, k: the middle panels)
  the corresponding
  distribution of the number $N$ of required iterations for obtaining the attracting
  regions, (c, f, i, l: the right panels) the corresponding probability distributions of required number of iterations
  for obtaining the Newton-Raphson basins of convergence, shown in panels-(a, d, g, j) respectively. The vertical, dashed, red line indicates, in each case,
the most probable number $N^*$ of iterations. The black dots show the position of the libration points. (Color figure online).}
\end{center}
\end{figure*}
In Fig. \textcolor[rgb]{1.00,0.00,0.50}{3} (b, e, h, k), the distribution of the corresponding number $N$ of iterations requires to obtain the predefined accuracy are illustrated, using tones of blue. It is observed that the initial conditions located in the fractal regions are the slowest converging nodes, while the converging rate is relatively fast for all the initial conditions inside the attracting domains. The following Fig. \textcolor[rgb]{1.00,0.00,0.50}{3} (c, f, i, l) depicts the corresponding probability distribution of the iterations. In all the panels: (c, f, i, l), the histograms include almost $95\%$  of the corresponding distributions. The probability $P$ works according to the following philosophy: if $N_0$ initial conditions $(x_0, y_0)$ on the configuration plane converge, after $N$ iterations, to one of the libration points, then $P=\frac{N_0}{N_t}$, where $N_t$ correspond to total number of initial conditions in every colour coded diagram. Moreover, the most probable number of iterations is not constant throughout and it is equal to 6 for panels: (c, f) while for panels: (i, l) it is 7.
\subsection{\emph{Case II: when nine libration points exist}}
\label{sec:402}
We continue our exploration with the case when nine libration points exist, \textcolor[rgb]{0.00,0.00,0.00}{that is in subcase}-(i) when $e\in (-0.5, -0.46671)$,  there exist one central libration point $L_4$ and two sets of four libration points, one each on $x$-axis $(L_{1,2,6,7})$ and $y$-axis $(L_{3,5,8,9})$, respectively. Moreover, in subcase-(ii) when $e\in (-0.457853, -0.23334)$, there exist one central libration point $L_4$,   four non-collinear libration points $(L_{10,...,13})$ on $(x,y)$ plane and four libration points $(L_{3,5,8,9})$ on $y-$axis. The Newton-Raphson basins of attraction for six values of the parameter $e$ are presented in Fig. \textcolor[rgb]{1.00,0.00,0.50}{4}(a-f), in which the first three panels correspond to the subcase-(i), while the last three panels depict the subcase-(ii).\\
In Fig. \textcolor[rgb]{1.00,0.00,0.50}{4}(a-c), we observe that the topology of the basins of convergence regions, on the configuration $(x,y)$ plane, is composed by three patterns: (i) the central circular region, which is mainly occupied by well-defined basins of convergence associated with the libration points $L_1$ (in yellow), $L_7$ (in cyan), $L_3$ (in teal) and $L_5$ (in magenta), (ii) the middle region, where most of the convergence areas is covered by the shape of wings as well as by the exotic bugs shaped area and (iii) the exterior region,  which corresponds to the central libration point $L_4$ with infinite extent of the basins of convergence.

It is observed that the basin boundaries of the middle region are highly chaotic (more accurately highly fractal) which leads to the finding that the final state of the associated initial points is highly sensitive. Therefore, we may contend that for the majority of initial conditions in the middle region, it can not be predicted from which libration point these initial conditions they attracted by. The topology of the configuration $(x, y)$ plane changes drastically with the increase in the value of parameter $e$.\\
The most notable changes are as follows:
\begin{itemize}
  \item The area corresponding to libration points $L_{1,7}$  on $x-$axis and $L_{3,5}$ on $y-$axis decreases.
  \item The extent of the basins of convergence, associated with the libration points $L_{2,6}$, which look like exotic bugs with many legs and antennas, decreases.
  \item The basins of convergence, associated with the libration points $L_{2, 6}$ on $x-$axis and $L_{8, 9}$ on $y-$axis, in the central region increase.
\end{itemize}
\begin{figure*}\label{Fig:5}
\begin{center}
(a)\includegraphics[scale=.45]{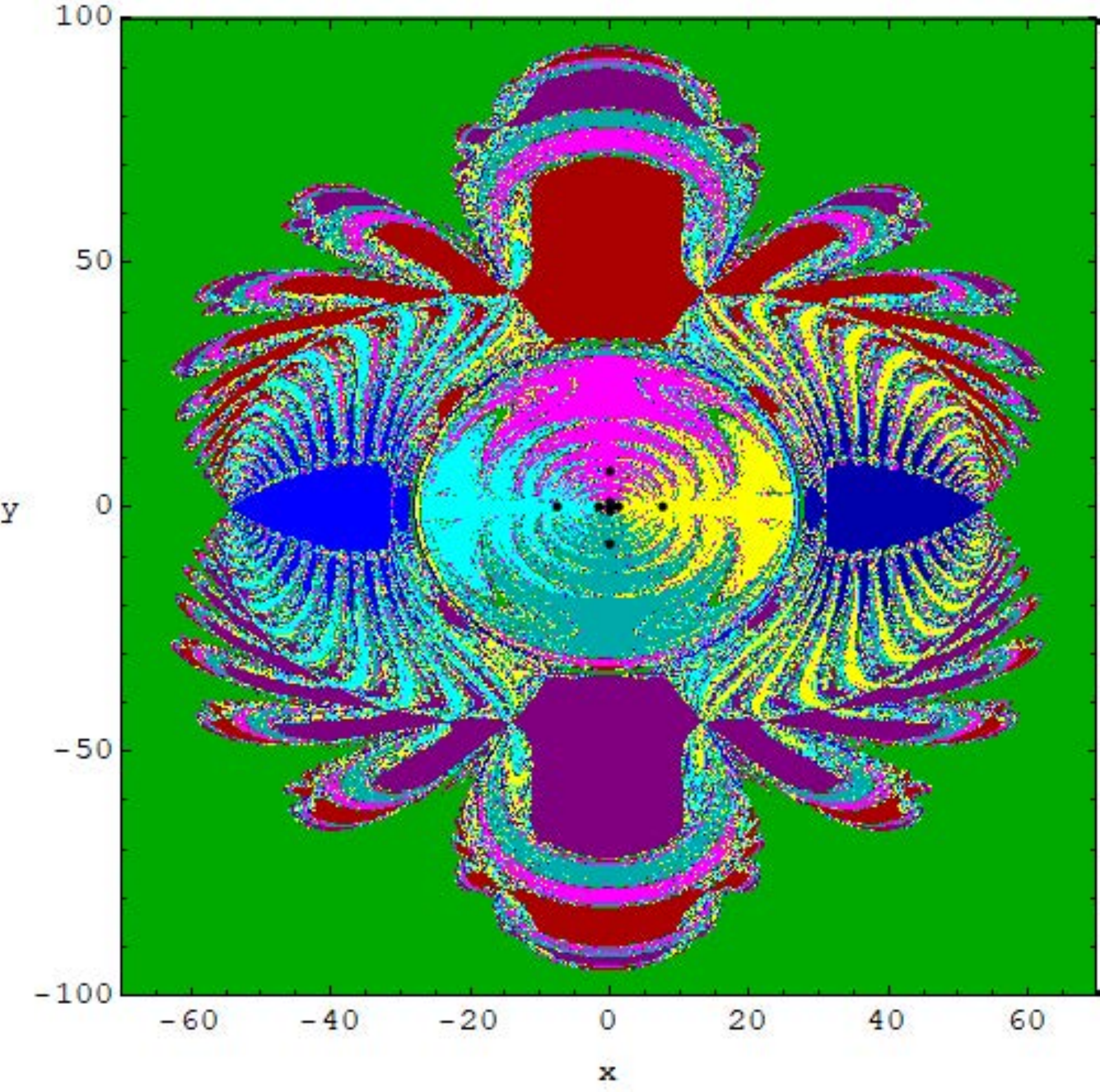}
(b)\includegraphics[scale=.45]{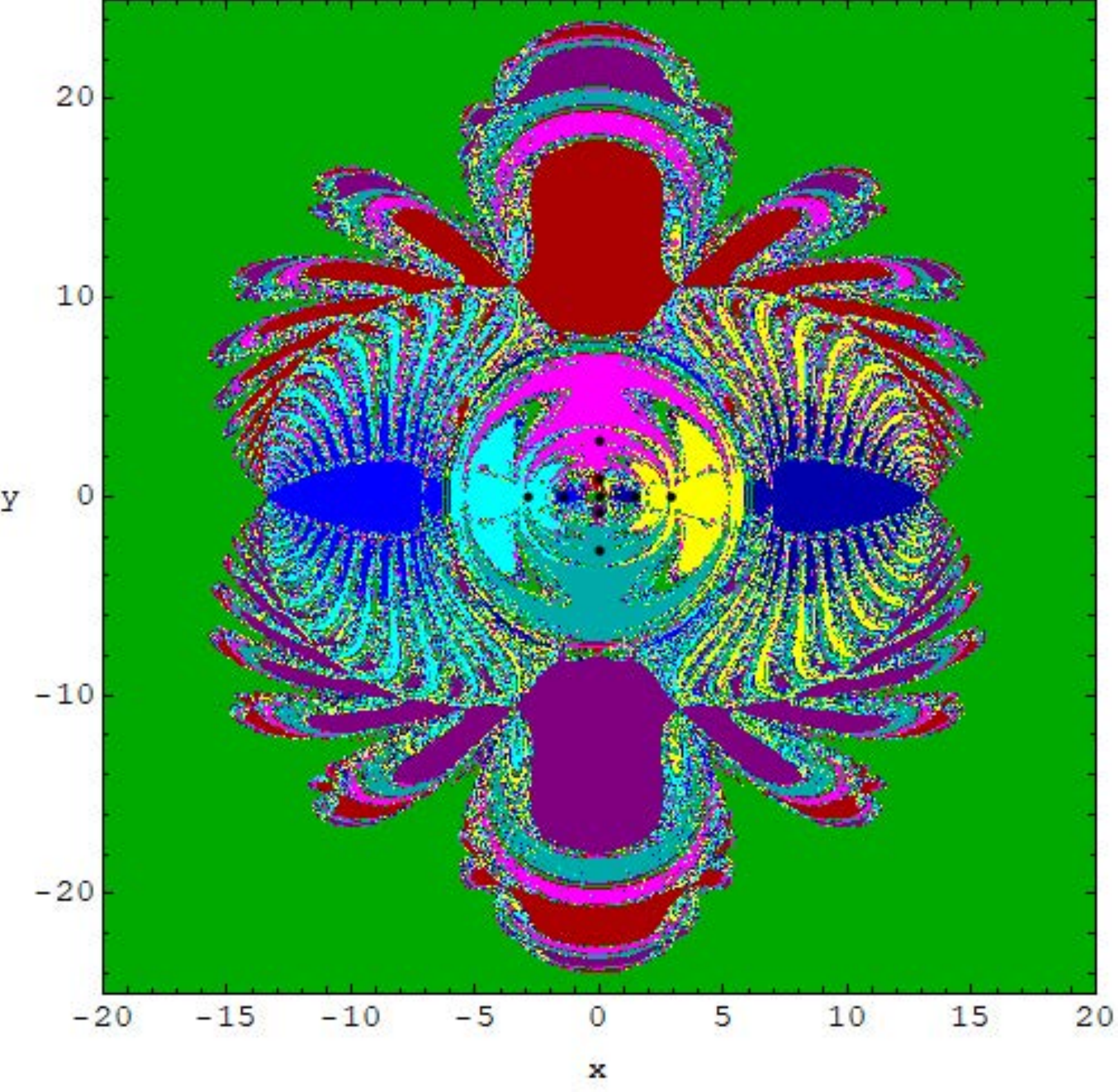}\\
(c)\includegraphics[scale=.45]{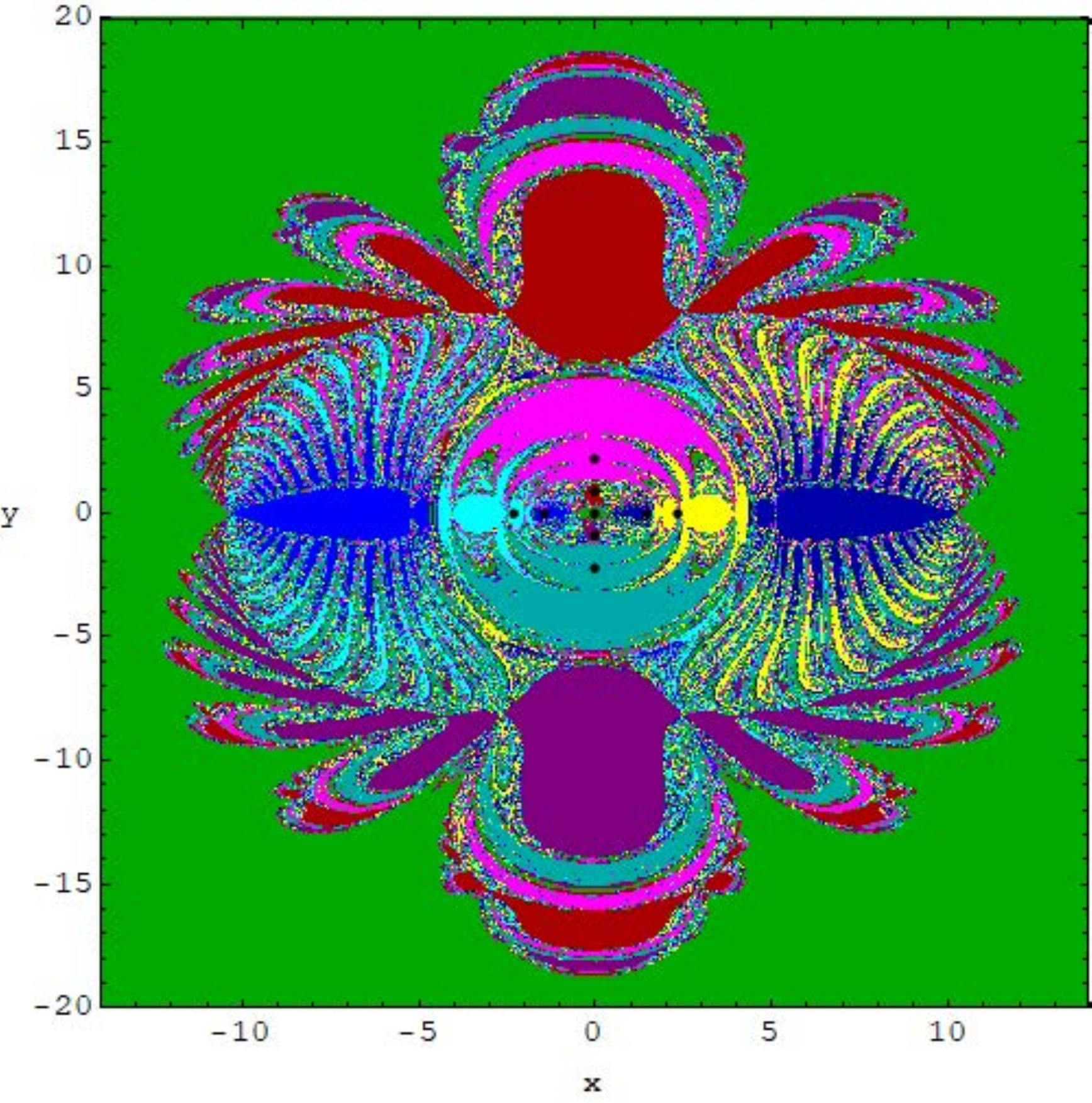}
(d)\includegraphics[scale=.45]{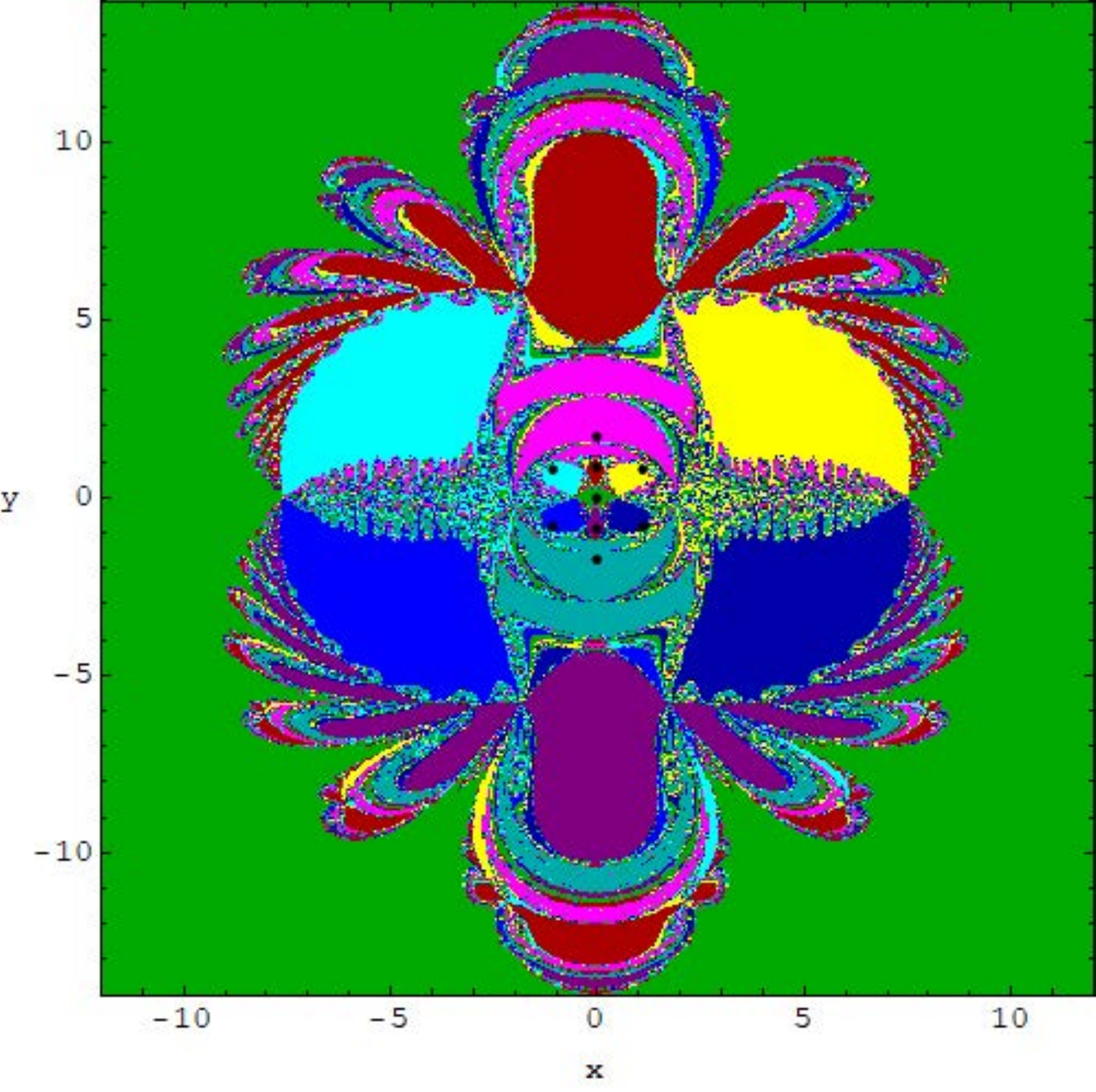}\\
(e)\includegraphics[scale=.45]{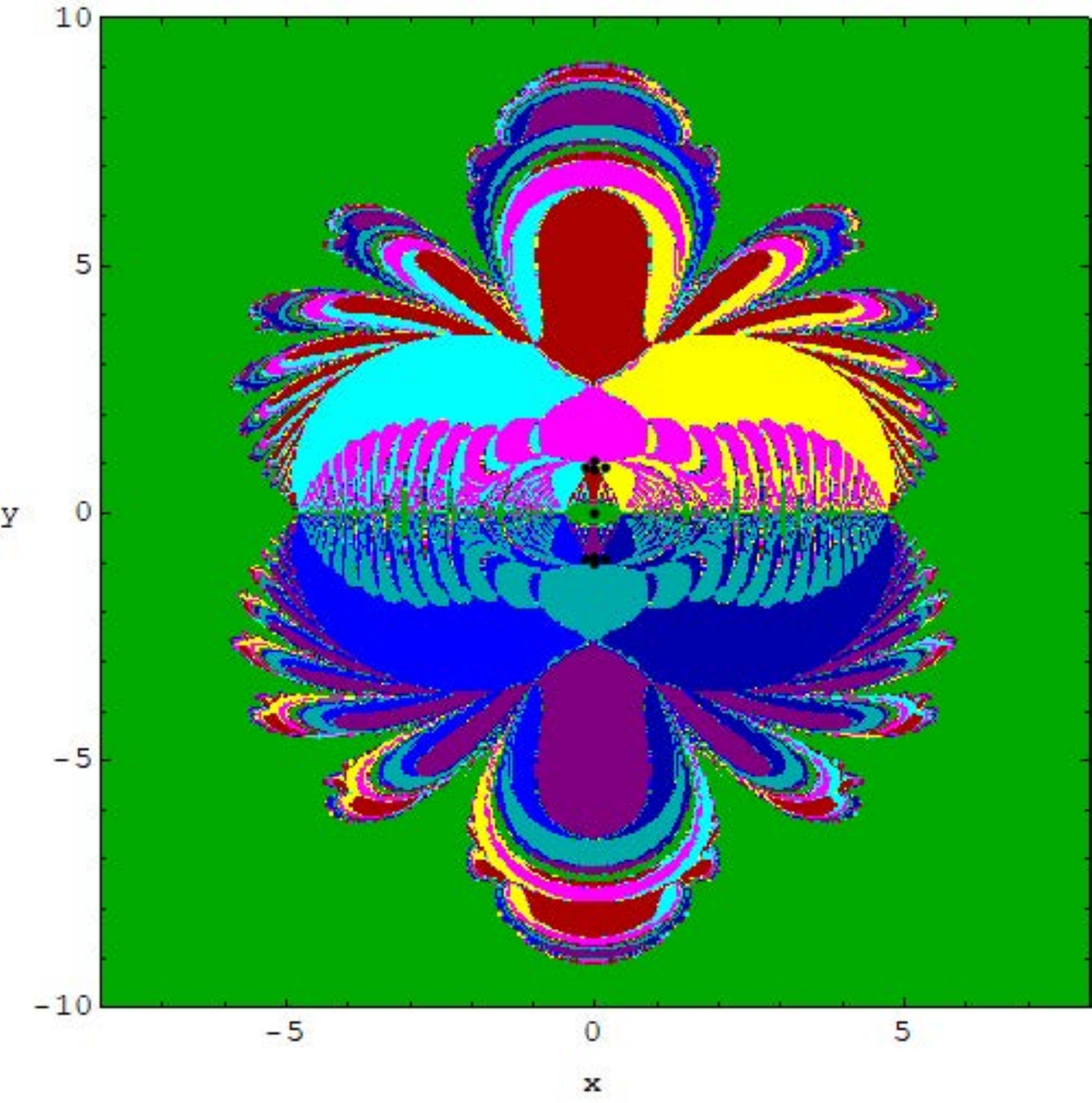}
(f)\includegraphics[scale=.45]{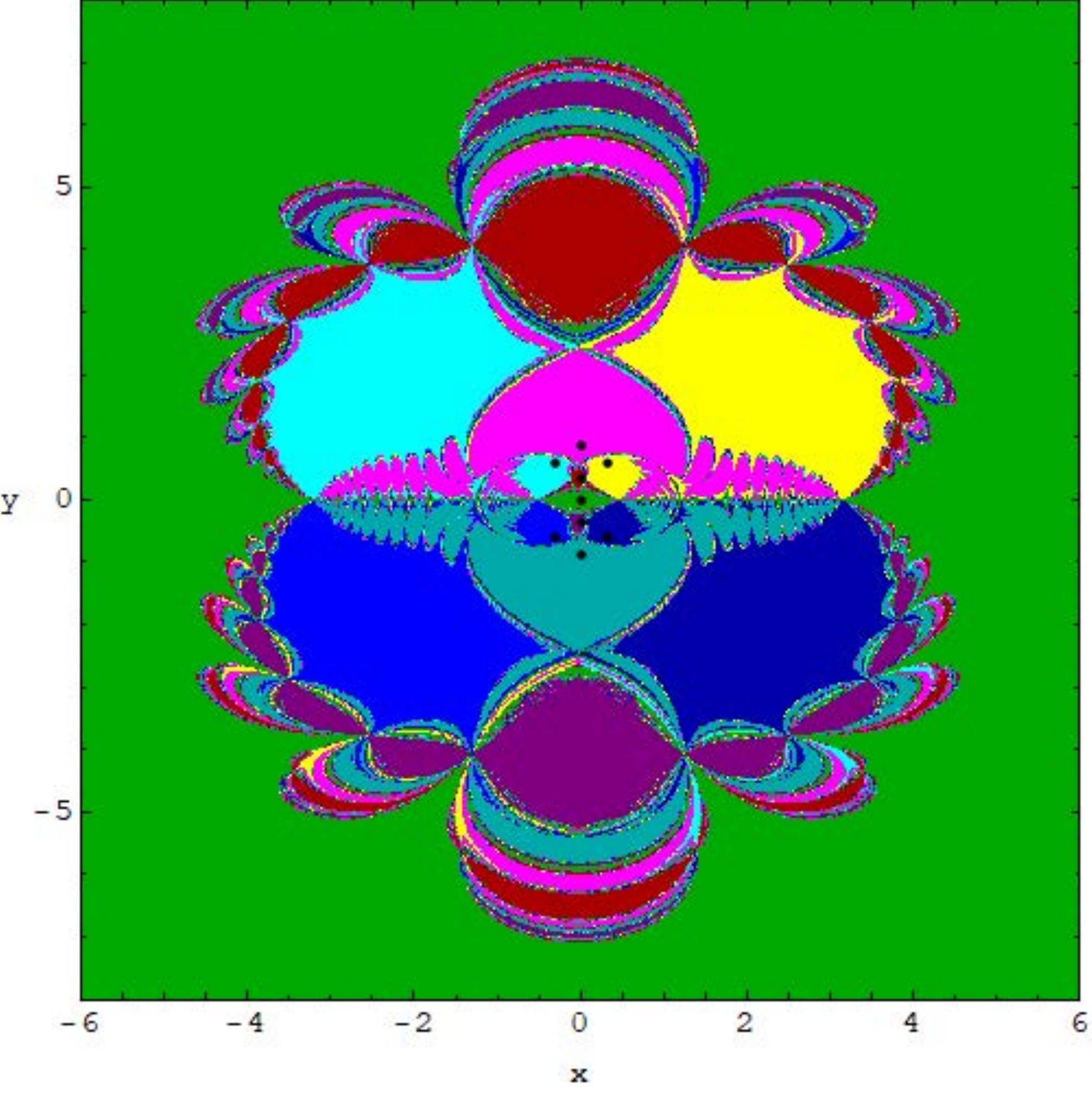}
\caption{The Newton-Raphson basins of attraction on the $xy$-plane for the case when nine libration points exist for:  (a) $e=-0.499$;  (b) $e=-0.485$; (c) $e=-0.4757$; (d) $e=-0.457$; (e) $e=-0.4$;  (f) $e=-0.275$. The color code denoting the attractors is as follows: $L_1$ (\emph{yellow}); $L_2$ (\emph{Darker blue}); $L_4$ (\emph{green}); $L_{6}$ (\emph{blue}); $L_{7}$ (\emph{cyan}); $L_{8}$ (\emph{purple}); $L_{9}$ (\emph{crimson}); $L_{10}$(\emph{teal}); $L_{11}$(\emph{magenta}) and non-converging points (\emph{white}). The black dots show the position of the libration points.}
\end{center}
\end{figure*}
\begin{figure*}\label{Fig:6}
\begin{center}
(a)\includegraphics[scale=.36]{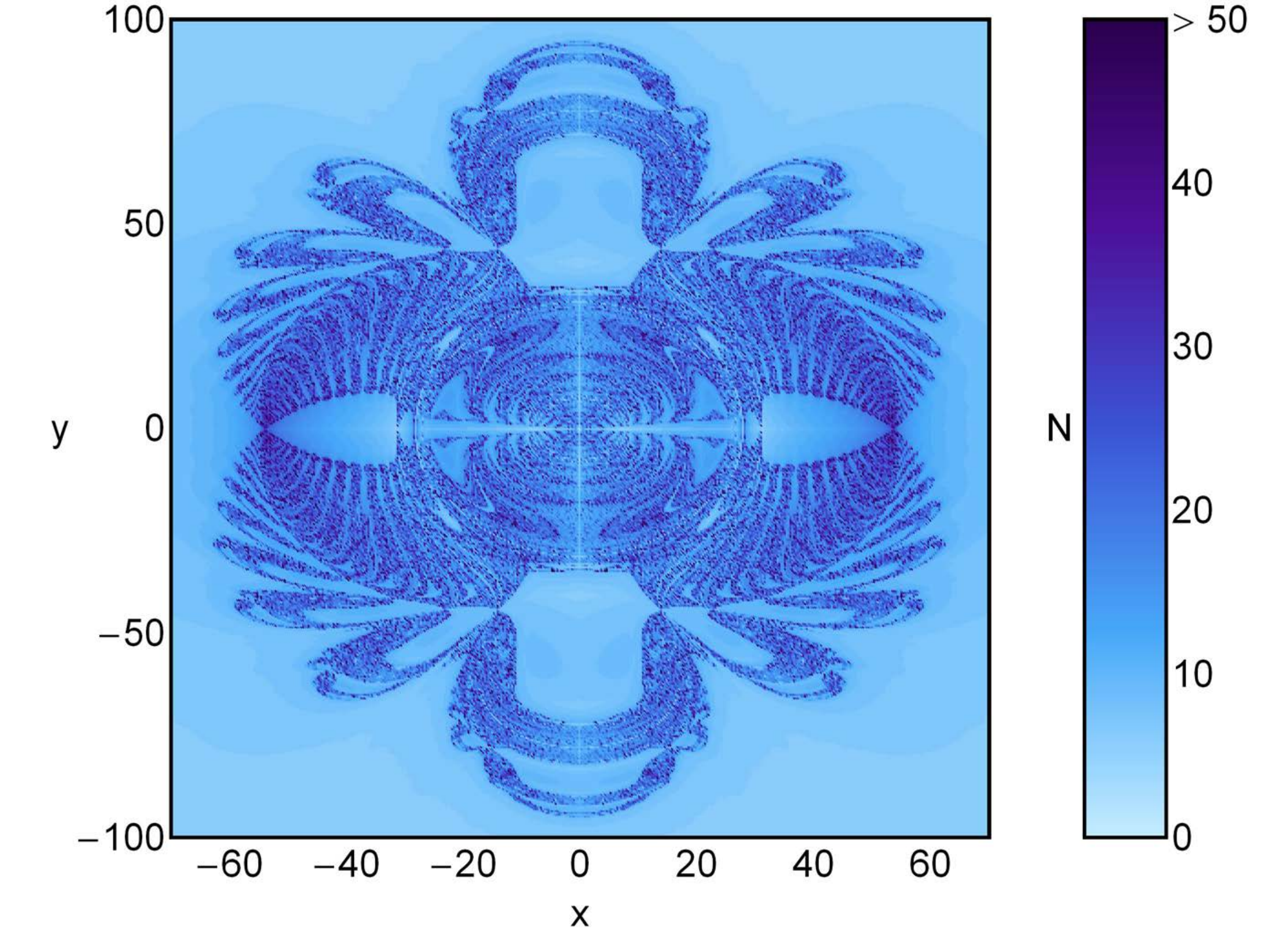}
(b)\includegraphics[scale=.36]{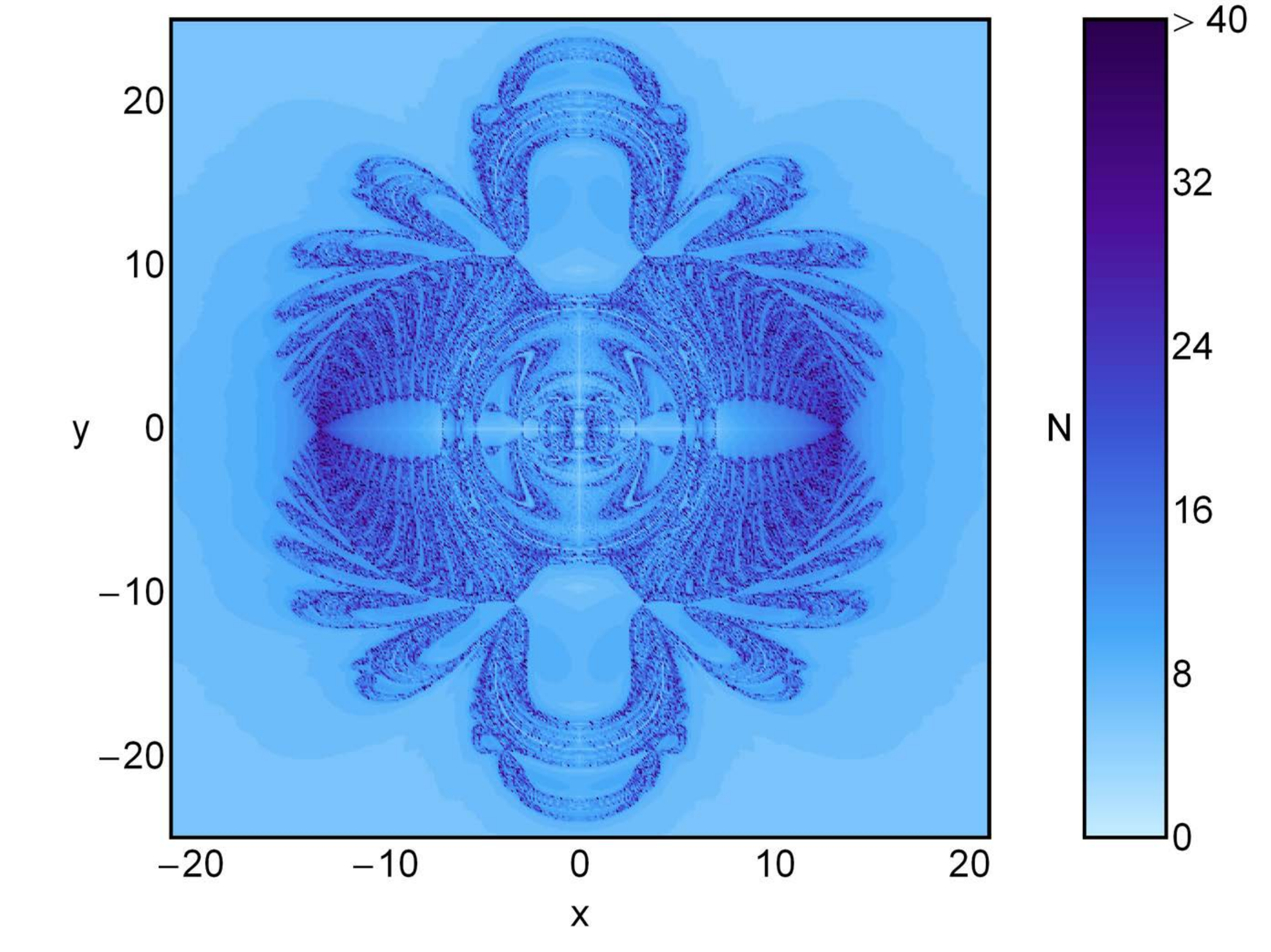}\\
(c)\includegraphics[scale=.36]{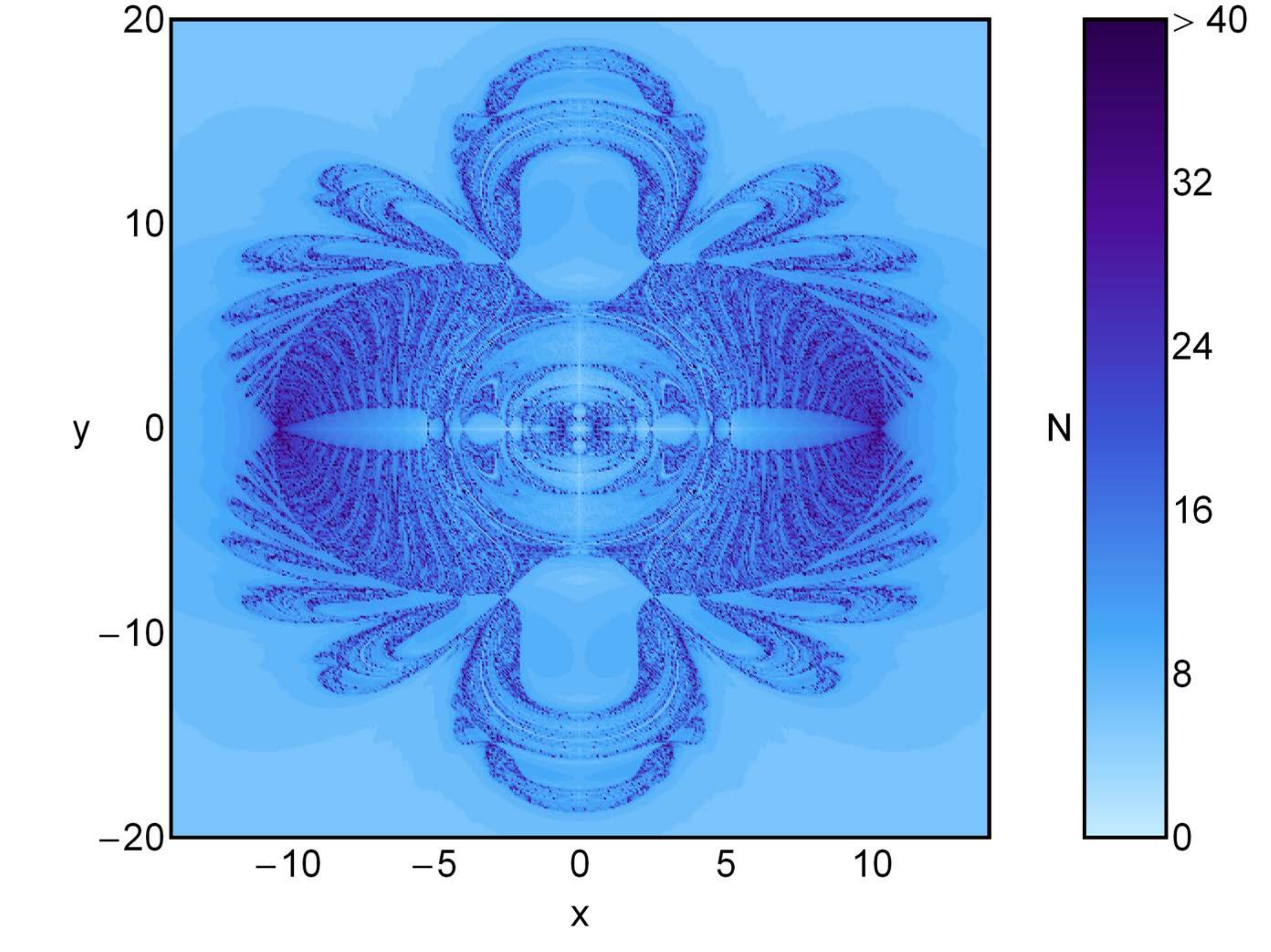}
(d)\includegraphics[scale=.36]{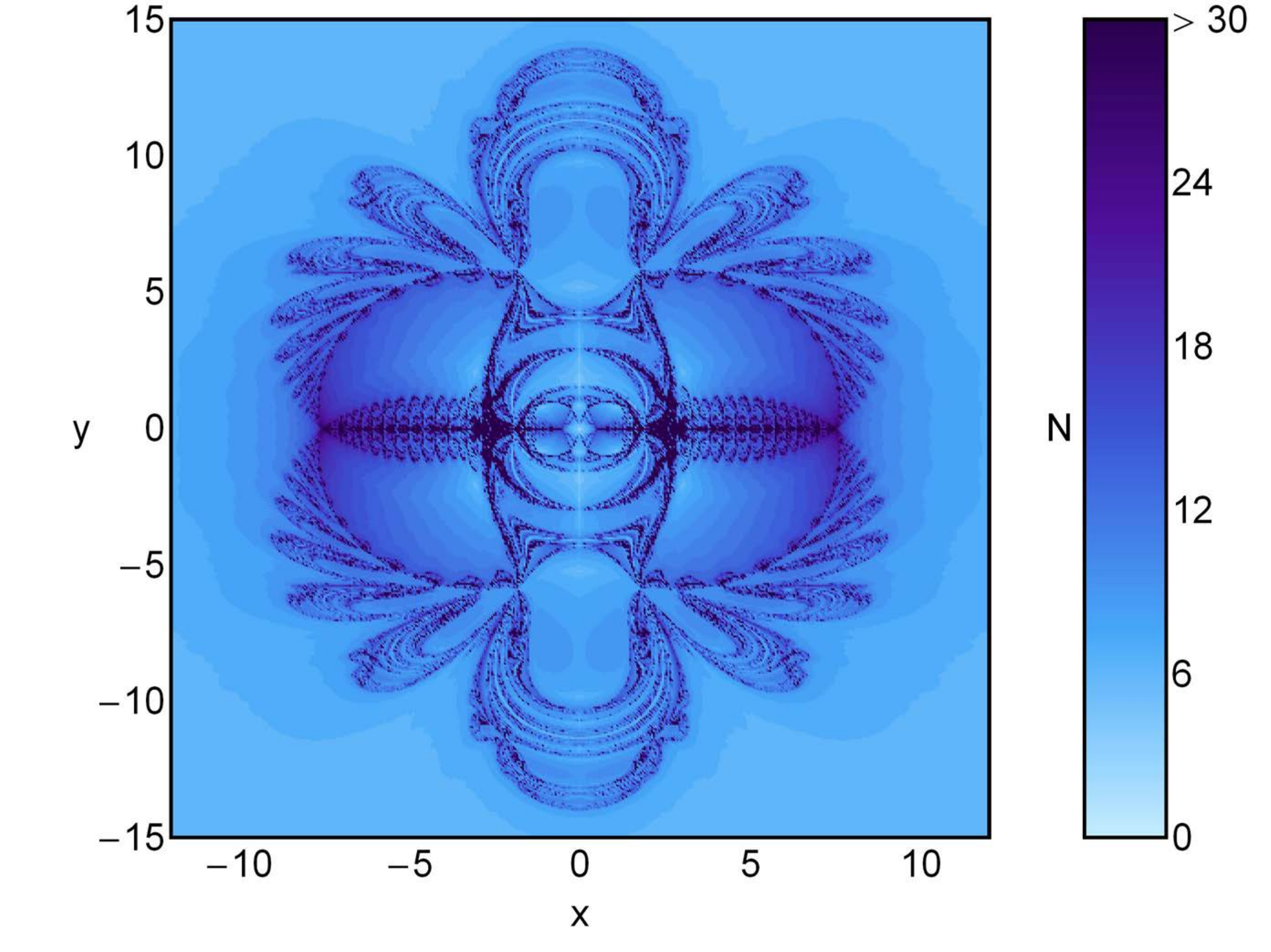}\\
(e)\includegraphics[scale=.36]{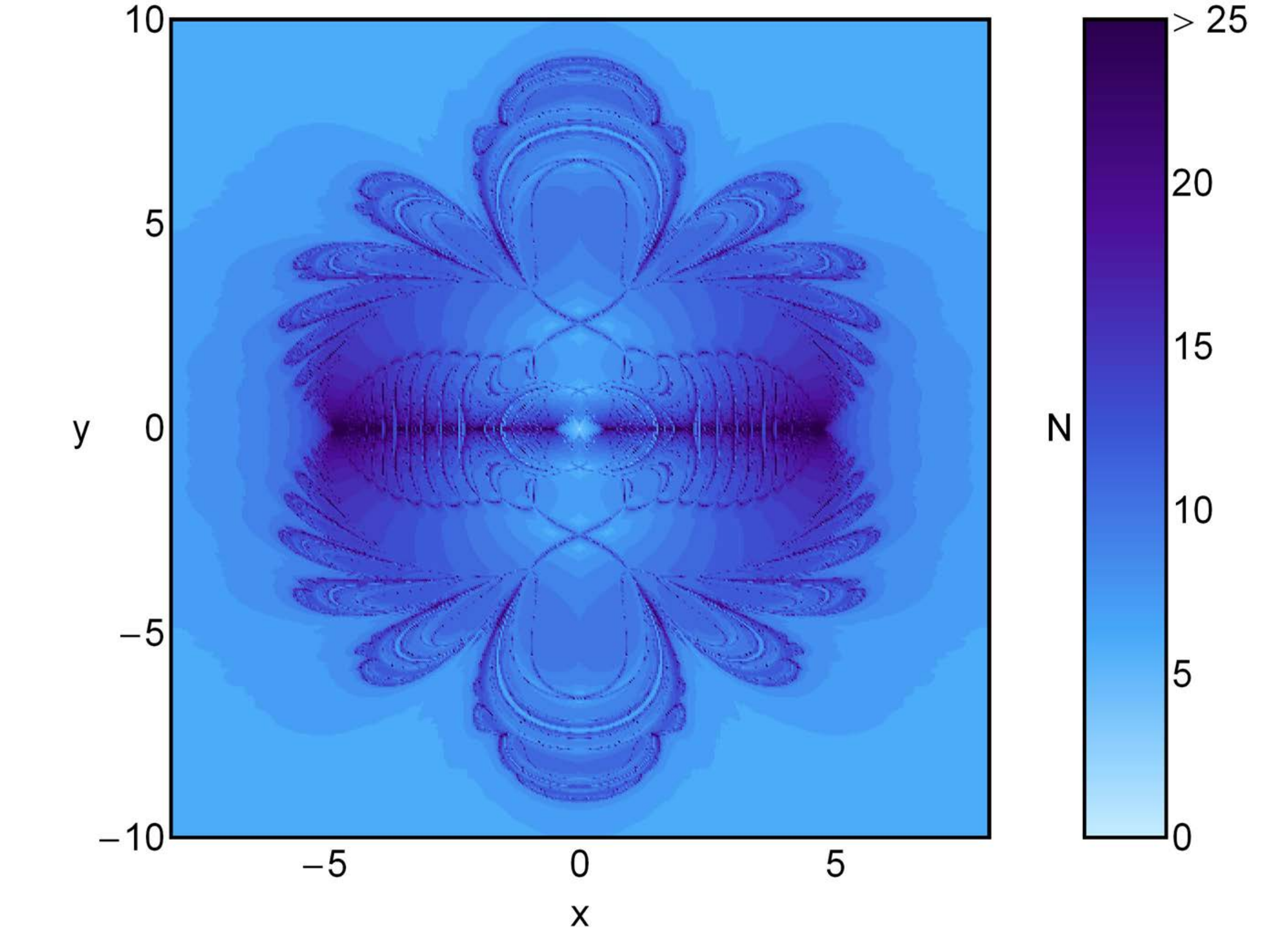}
(f)\includegraphics[scale=.36]{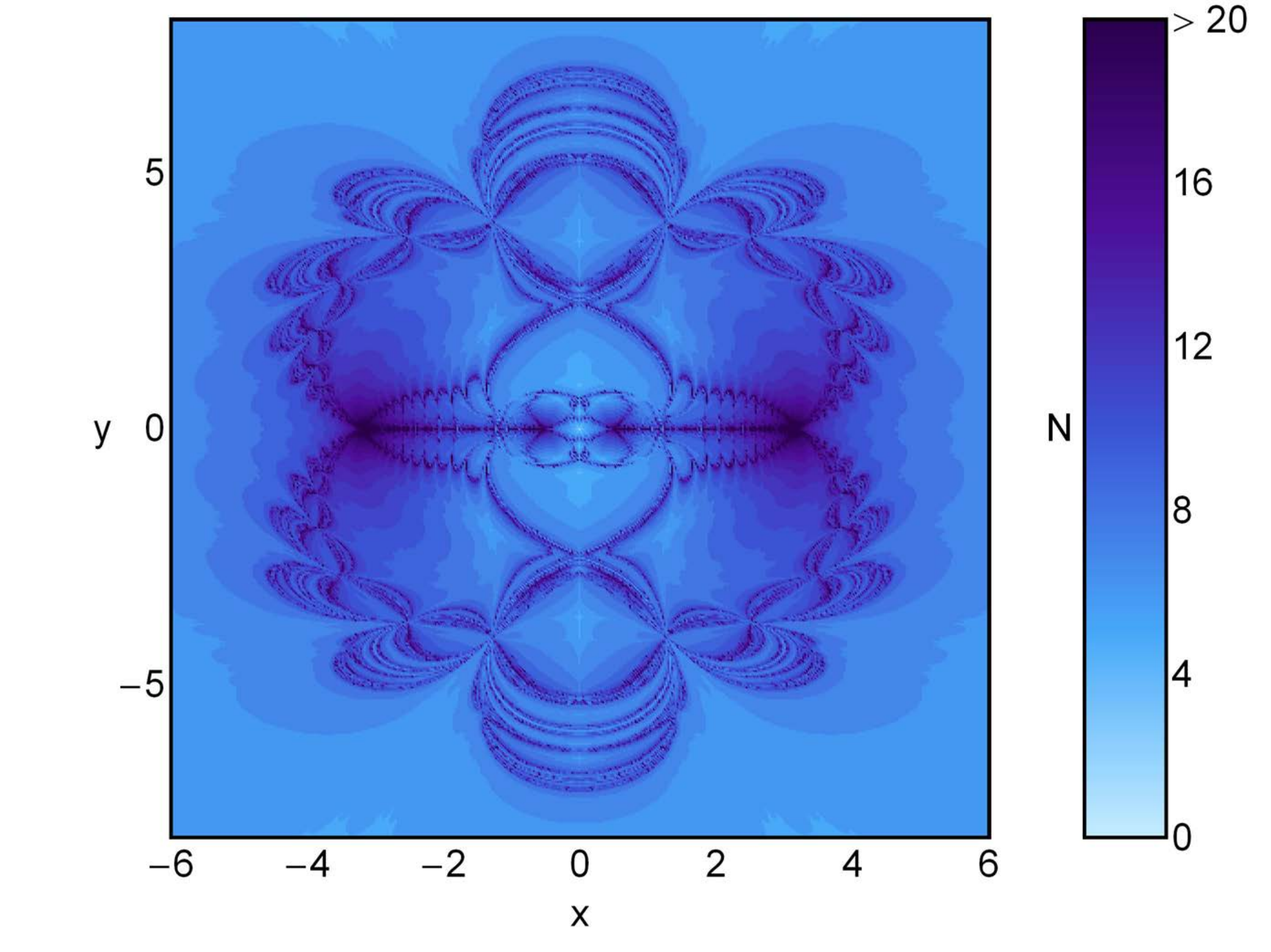}
\caption{The corresponding distributions of the number $N$ of the required iterations for obtaining the Newton-Raphson basins of convergence,
shown in Fig. 4(a-f). (Color figure online).}
\end{center}
\end{figure*}
\begin{figure*}\label{Fig:7}
\begin{center}
(a)\includegraphics[scale=.36]{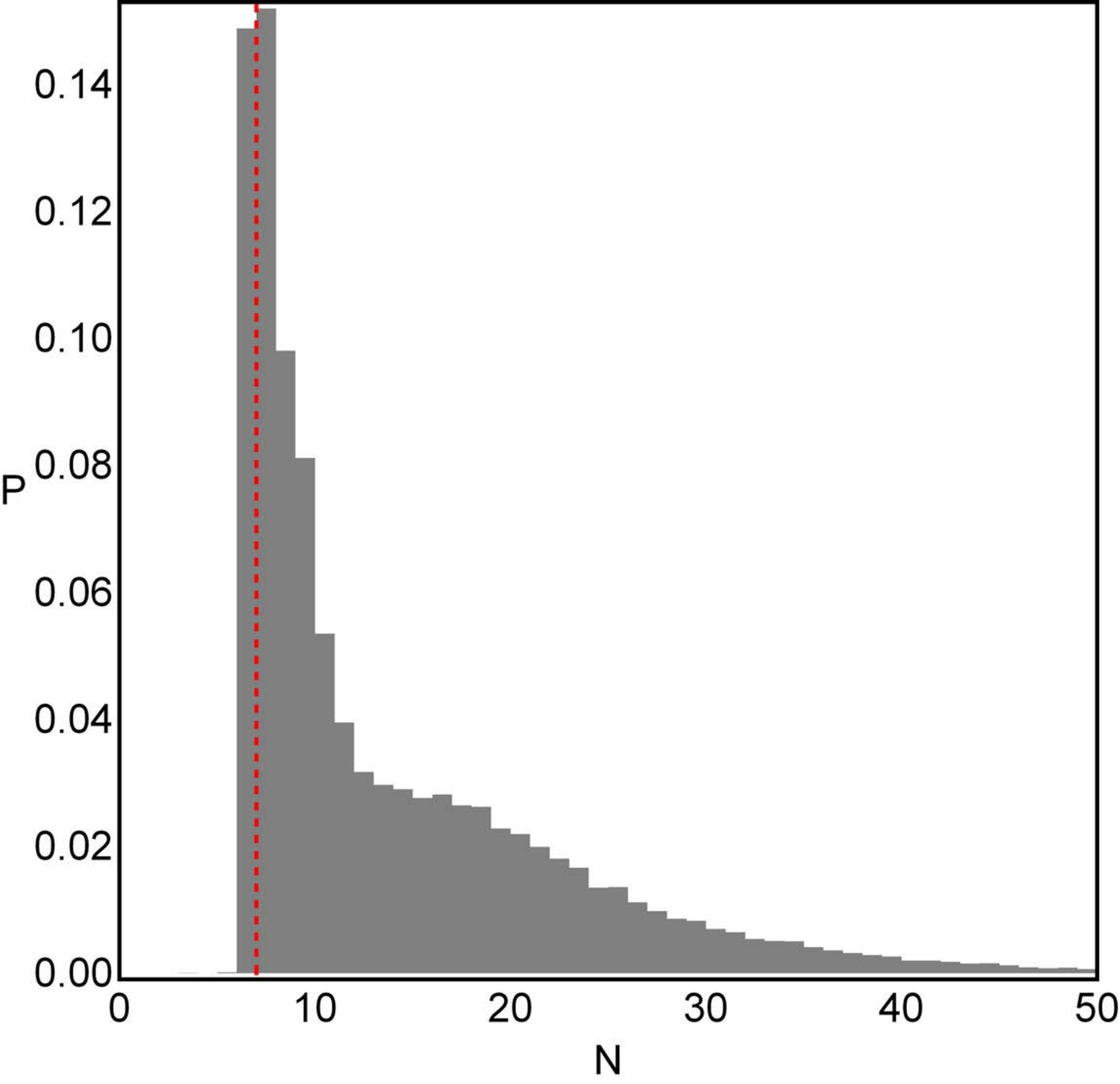}
(b)\includegraphics[scale=.36]{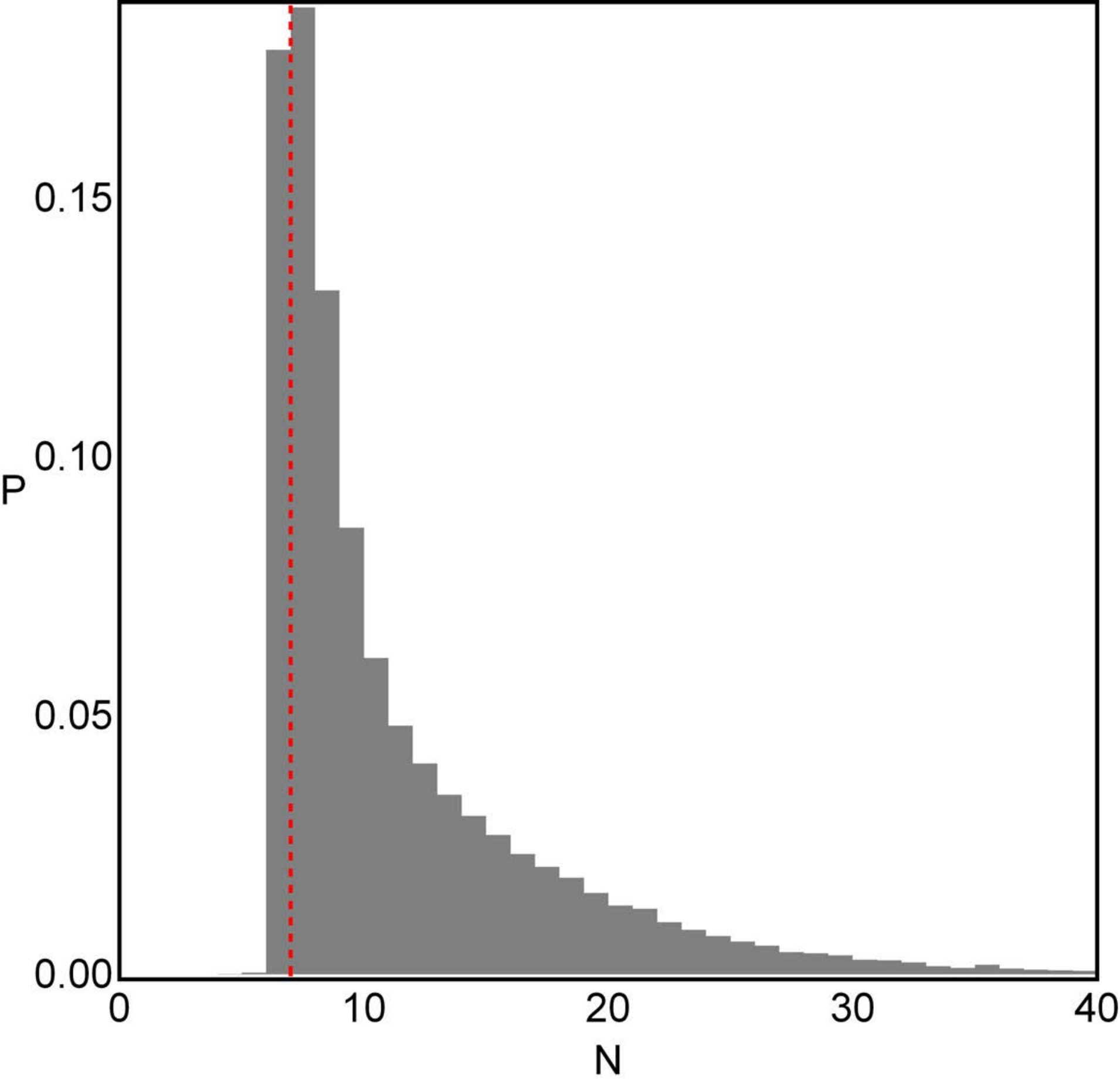}\\
(c)\includegraphics[scale=.36]{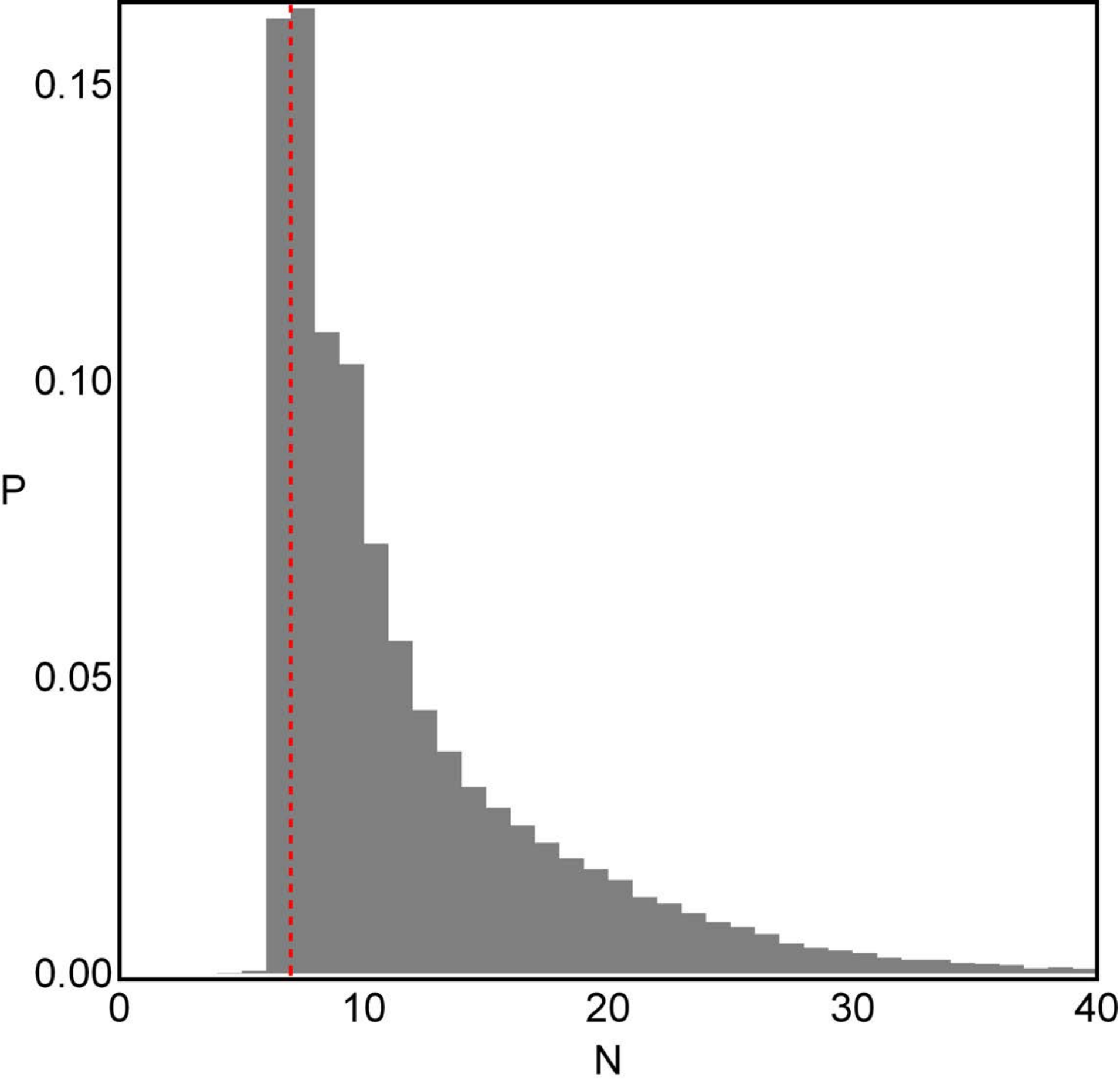}
(d)\includegraphics[scale=.36]{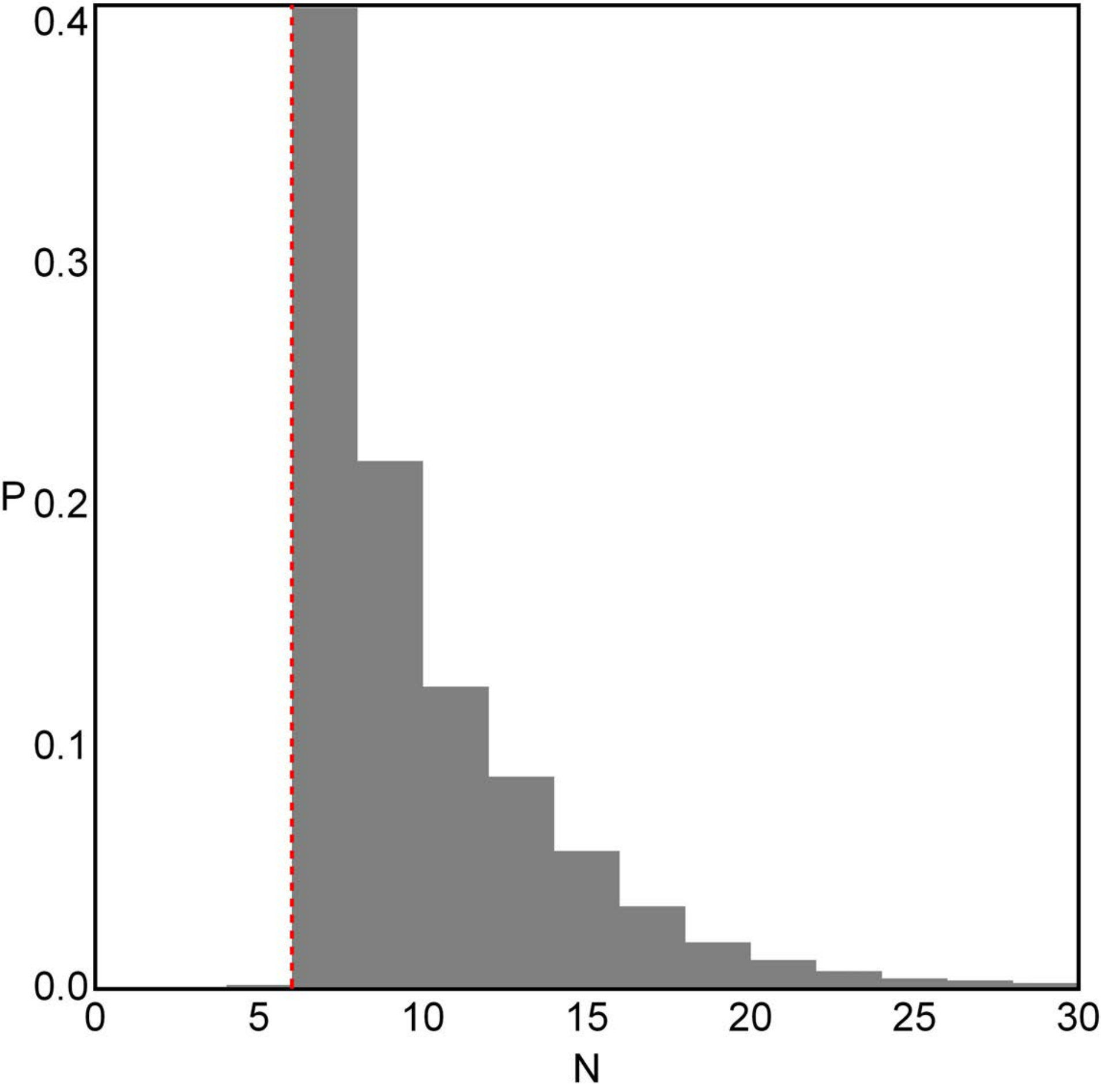}\\
(e)\includegraphics[scale=.36]{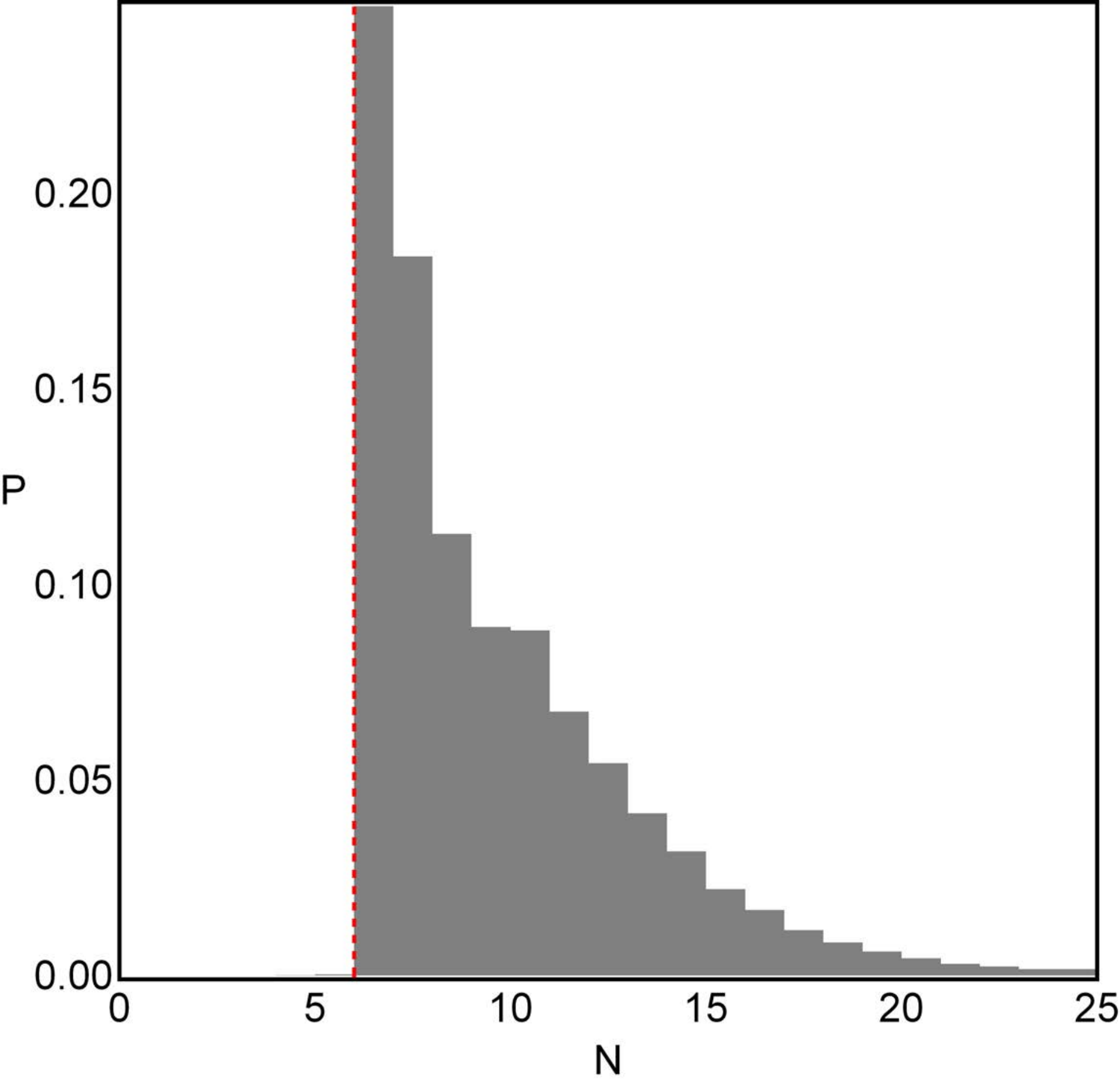}
(f)\includegraphics[scale=.36]{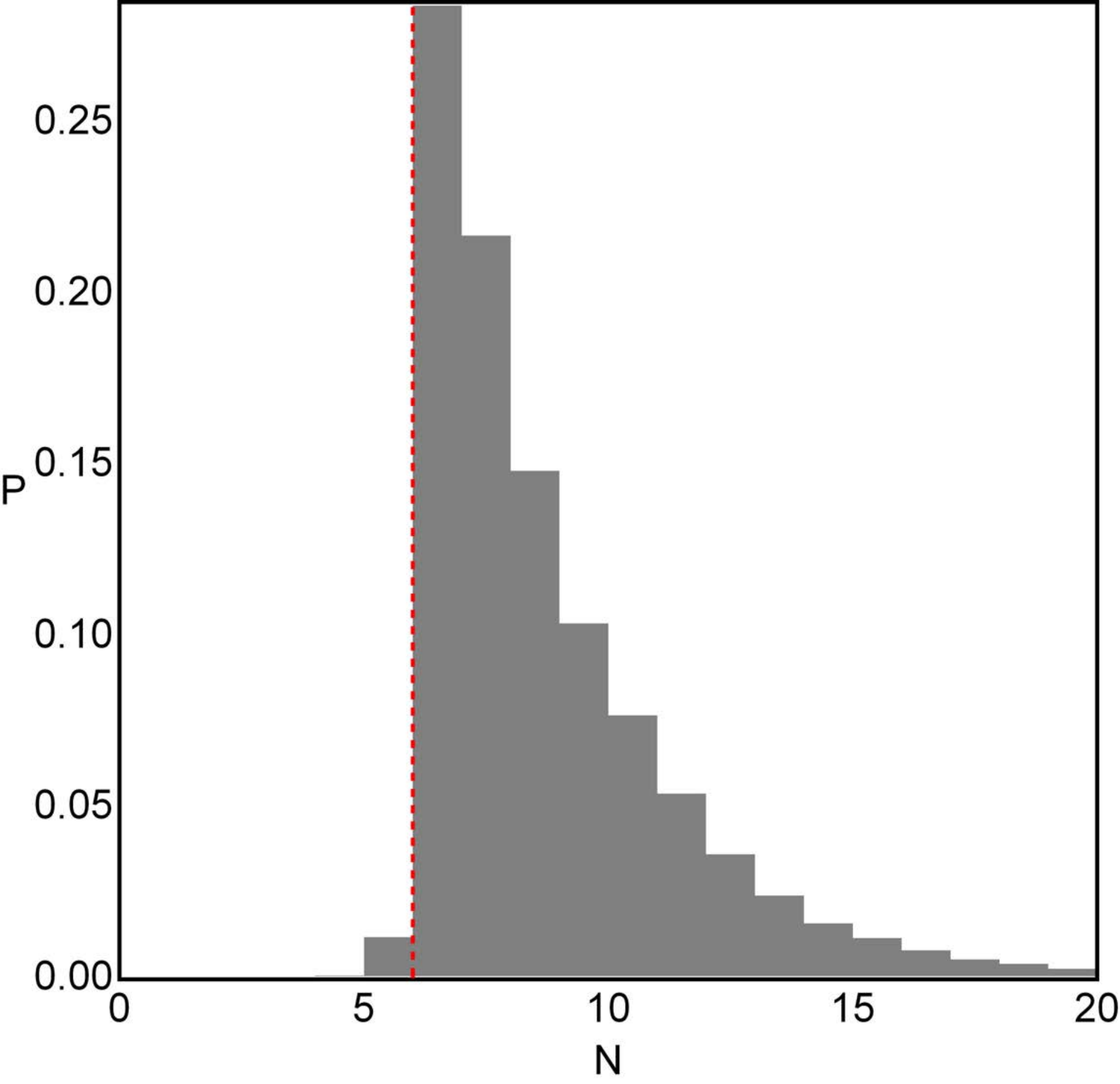}
\caption{The corresponding probability distributions of required number of iterations for obtaining the Newton-Raphson basins of convergence,
shown in Fig. \textcolor[rgb]{1.00,0.00,0.50}{4}(a-f). The vertical, dashed, red line indicates, in each case, the most probable number $N^*$ of iterations. (Color figure online).}
\end{center}
\end{figure*}
In panel-d, it is seen that a strip shaped region in the vicinity of the
$x-$axis is composed by a highly chaotic mixture of initial condition.
Therefore, it is impossible to predict from which libration point, the initial conditions
falling inside this strip are attracted by. This chaotic strip is converted into a
well shaped domain of basins of convergence, as the value of the parameter
$e$ grows.\\
The corresponding distributions of the number $N$ of the required iterations to obtain the
predefined accuracy is given in Fig. \textcolor[rgb]{1.00,0.00,0.50}{5}(a-f).
It can be observed that the initial conditions inside
the attracting domains converge relatively fast $(N < 25)$,
whereas the initial conditions located in
the fractal regions are the slowest converging nodes. In Fig. \textcolor[rgb]{1.00,0.00,0.50}{6}(a-f), the
corresponding probability distributions are illustrated. It is observed that with increasing value
of the parameter $e$, the most probable number $N$ of iterations decreases from 8 (in panel: a-c) to 6 (in panel: d-f).
\subsection{\emph{Case III: when eleven libration points exist}}
\label{sec:403}
This subsection is devoted to the case when $e \in (-0.19526,$ $ -0.173395)$, where eleven libration points exist:
five on the $x-$axis ($L_{1,2,4,6,7}$) in which $L_4$ is the central libration point, two on $y-$axis ($L_{8,9}$)
and four on $(x,y)$ plane ($L_{10, 11, 12, 13}$). The evolution of the basins of convergence, using
Newton-Raphson iterative scheme, for three values of parameter $e$ are illustrated in Fig. \textcolor[rgb]{1.00,0.00,0.50}{7}(a, d, g).

The following observations are made for increasing value of the parameter $e$ on the
configuration $(x, y)$ plane:
\begin{itemize}
  \item Four lobes corresponding to the libration points on configuration
  $(x, y)$ plane are present in central as well as in middle region, which look
  symmetrical with respect to both axes.
  \item Two hourglass shaped regions, associated with the libration points
  $L_{1,7}$ (grey and cyan colour respectively), emerge and shrink with increasing
  value of parameter $e$.
  \item Two thin figure-eight shaped tentacles elongated to the
  $y-$axis, originate from zenith and nadir of the middle region and lie at the $y-$axis
  with the increase in the value of parameter $e$. Moreover, these tentacles are composed of a
  chaotic mixture of initial conditions.
  \item The extent of the basins of convergence, corresponding to the central libration point, is always infinite while for all the other libration points, the domain of the basins of convergence is finite.
\end{itemize}
In Fig. \textcolor[rgb]{1.00,0.00,0.50}{7} (b, e, h), the corresponding distributions of the number $N$ of
required iterations to obtain the predefined accuracy are depicted. One can notice that almost every initial condition
on the configuration plane converges for $N< 20$ except for those points which lie on the boundaries of the basins
of convergence, i.e., the rate of convergence is much slower for those points.  From Fig. \textcolor[rgb]{1.00,0.00,0.50}{7} (c, f, i), we may argue that the most probable number
$N^*$ is not constant for each panel and it increases with increasing value of the parameter $e$.
\begin{figure*}\label{Fig:8}
\begin{center}
(a)\includegraphics[scale=.26]{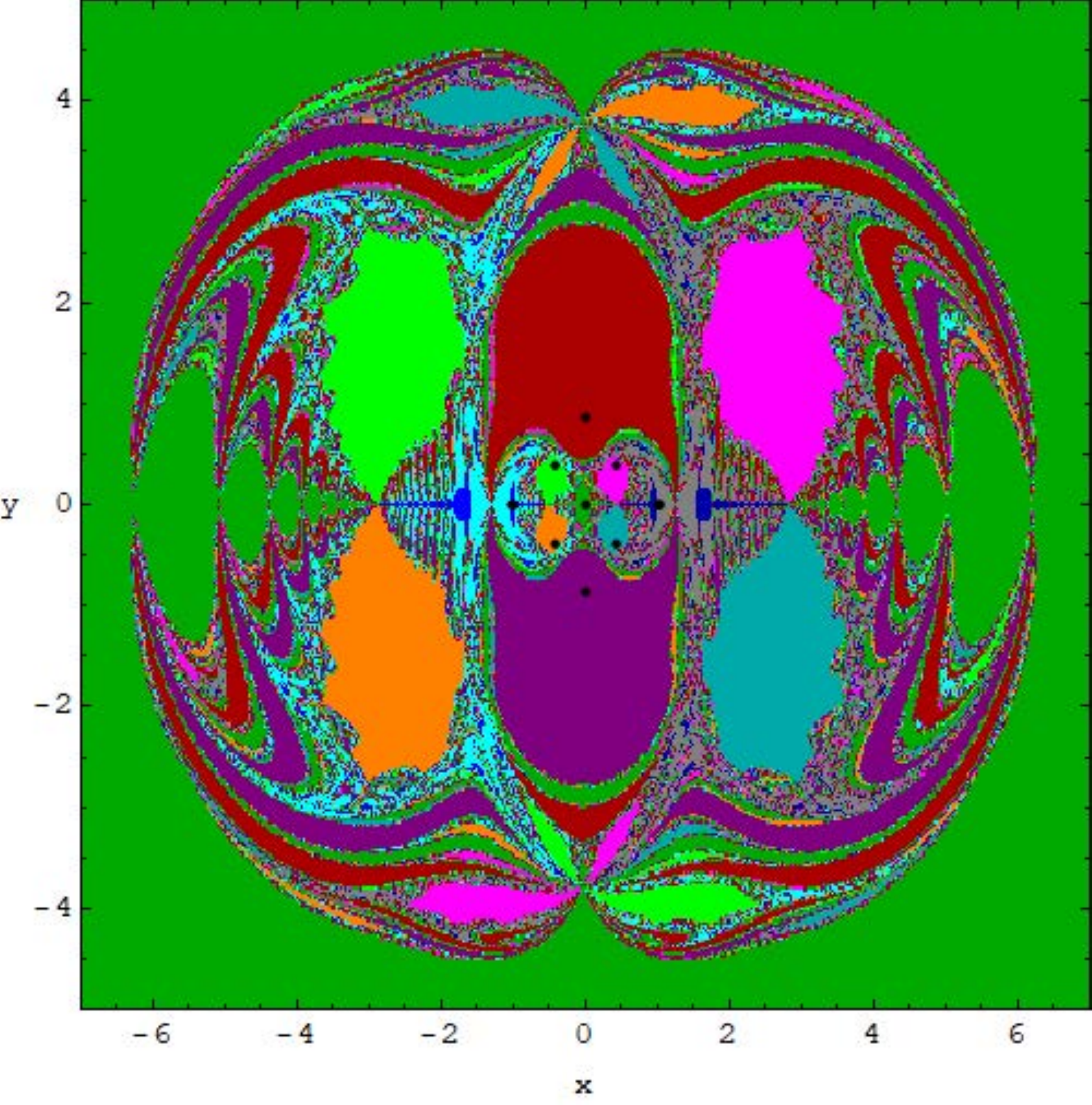}
(b)\includegraphics[scale=.26]{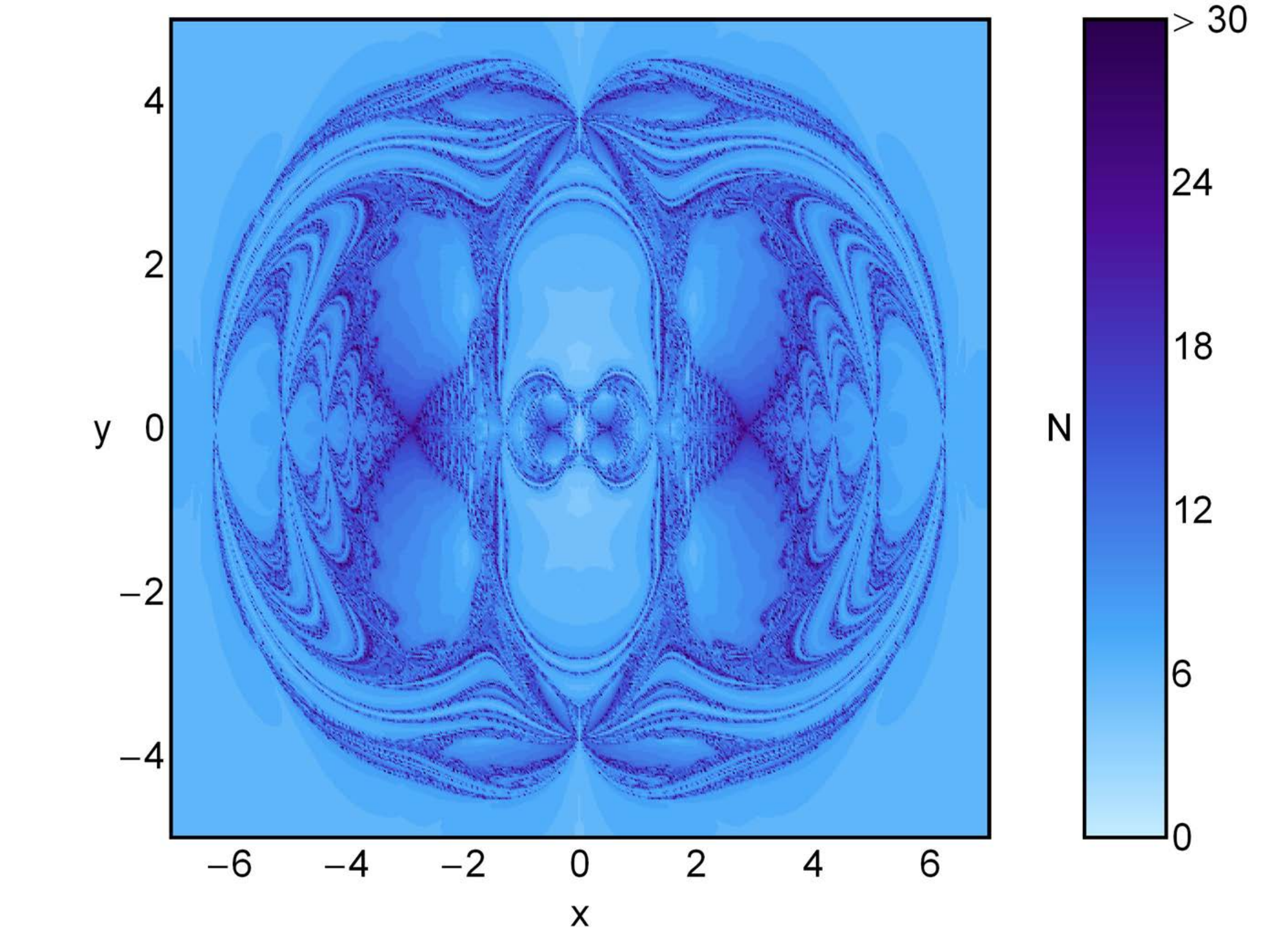}
(c)\includegraphics[scale=.26]{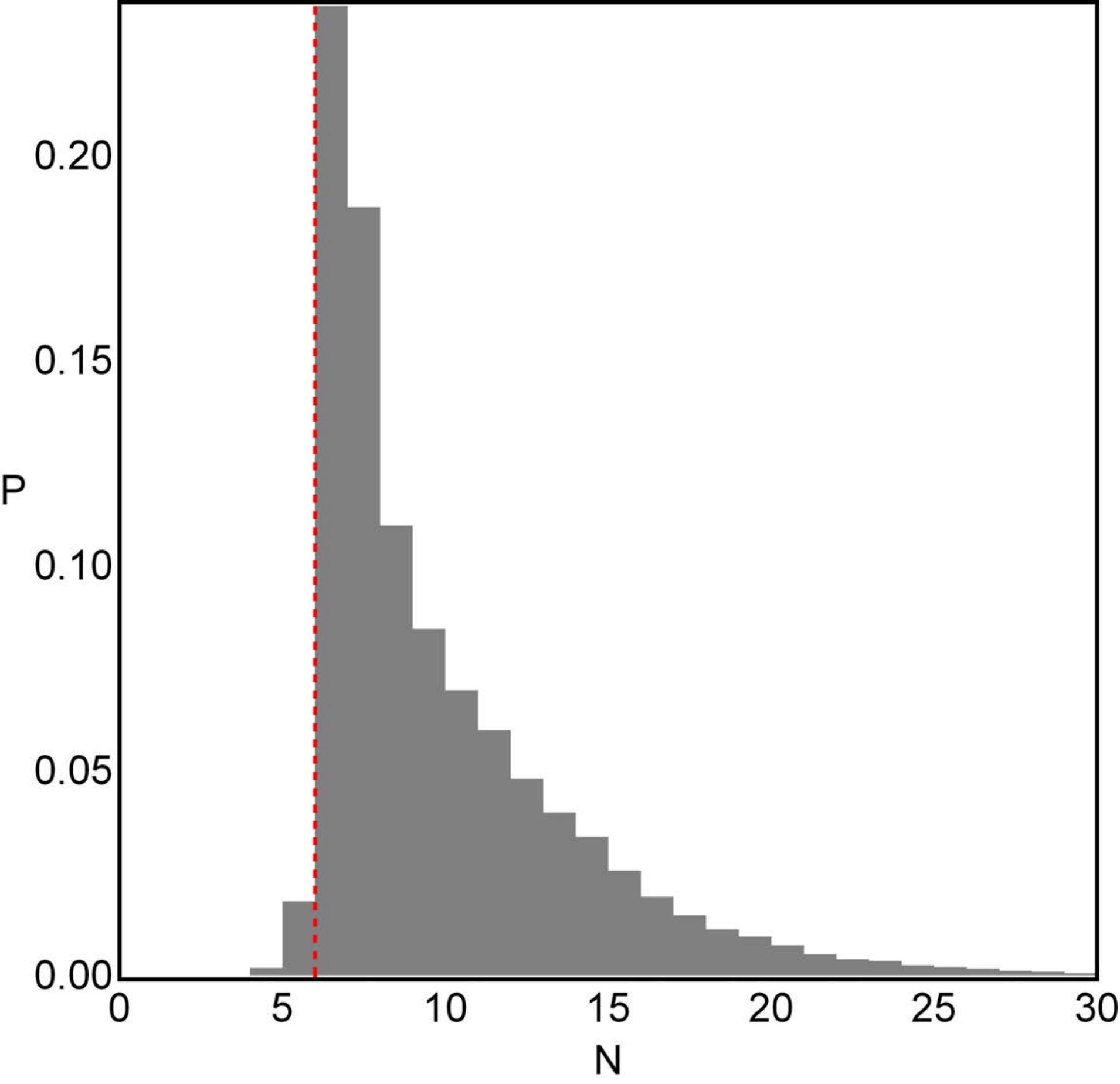}\\
(d)\includegraphics[scale=.26]{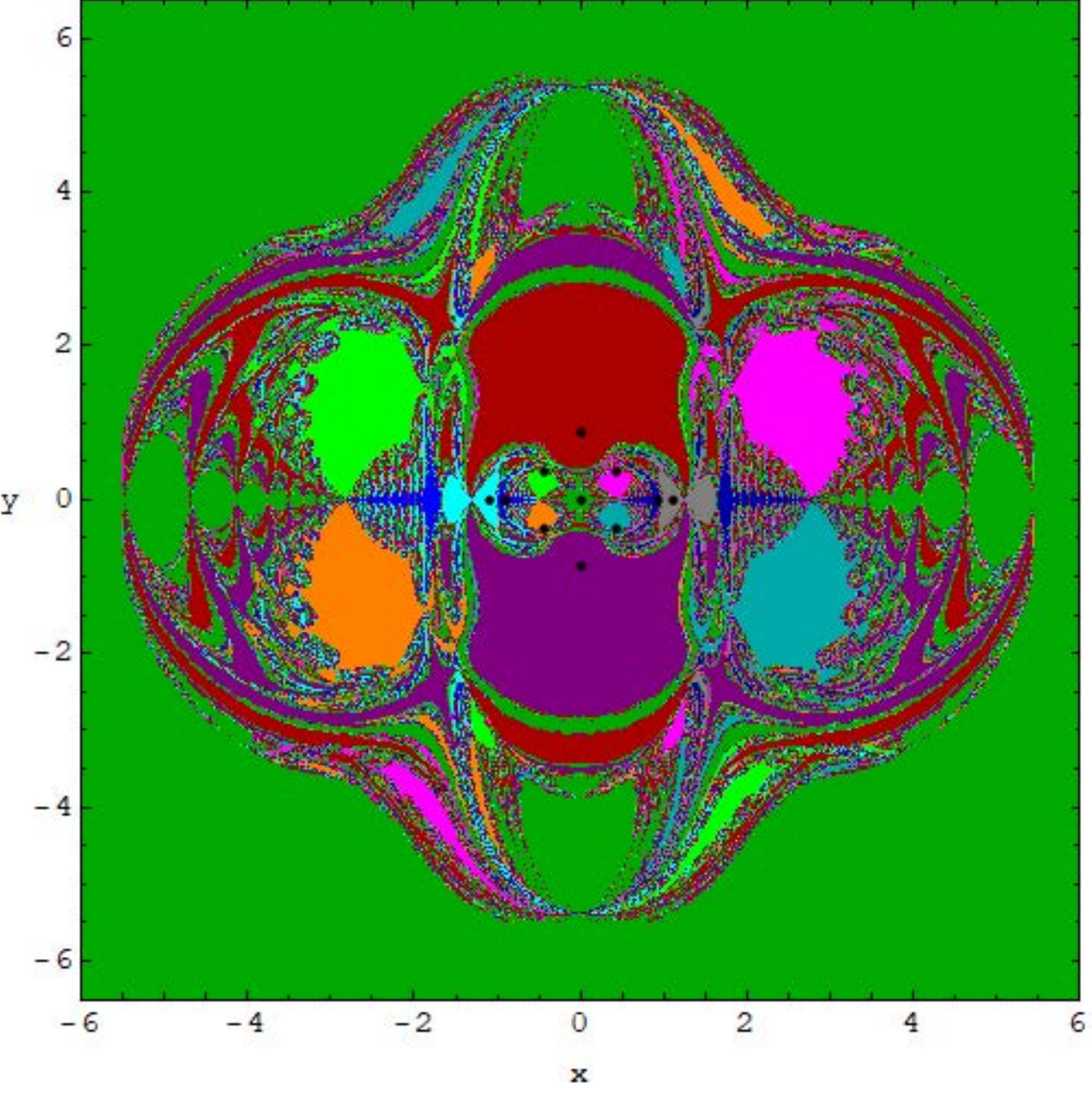}
(e)\includegraphics[scale=.26]{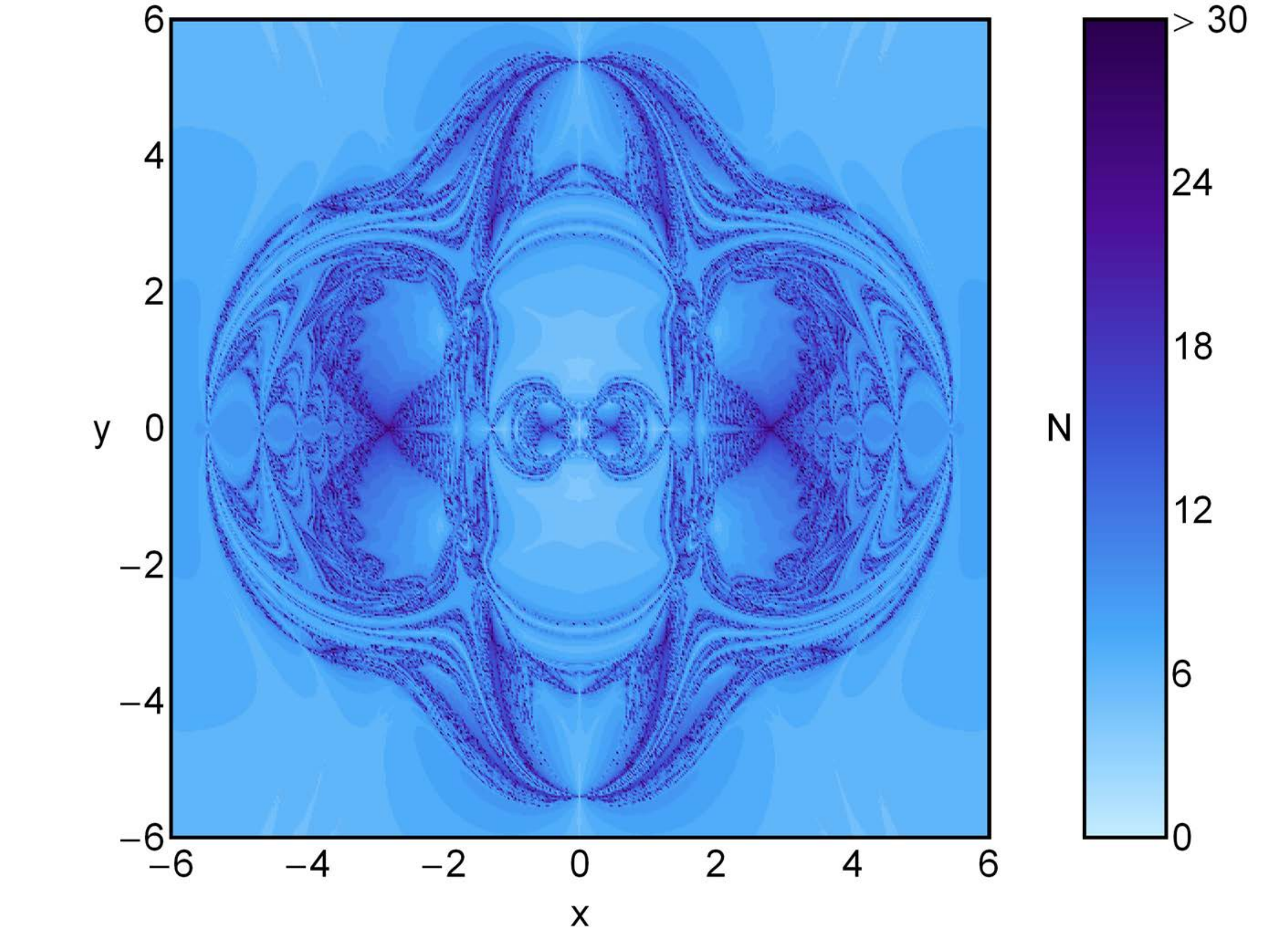}
(f)\includegraphics[scale=.26]{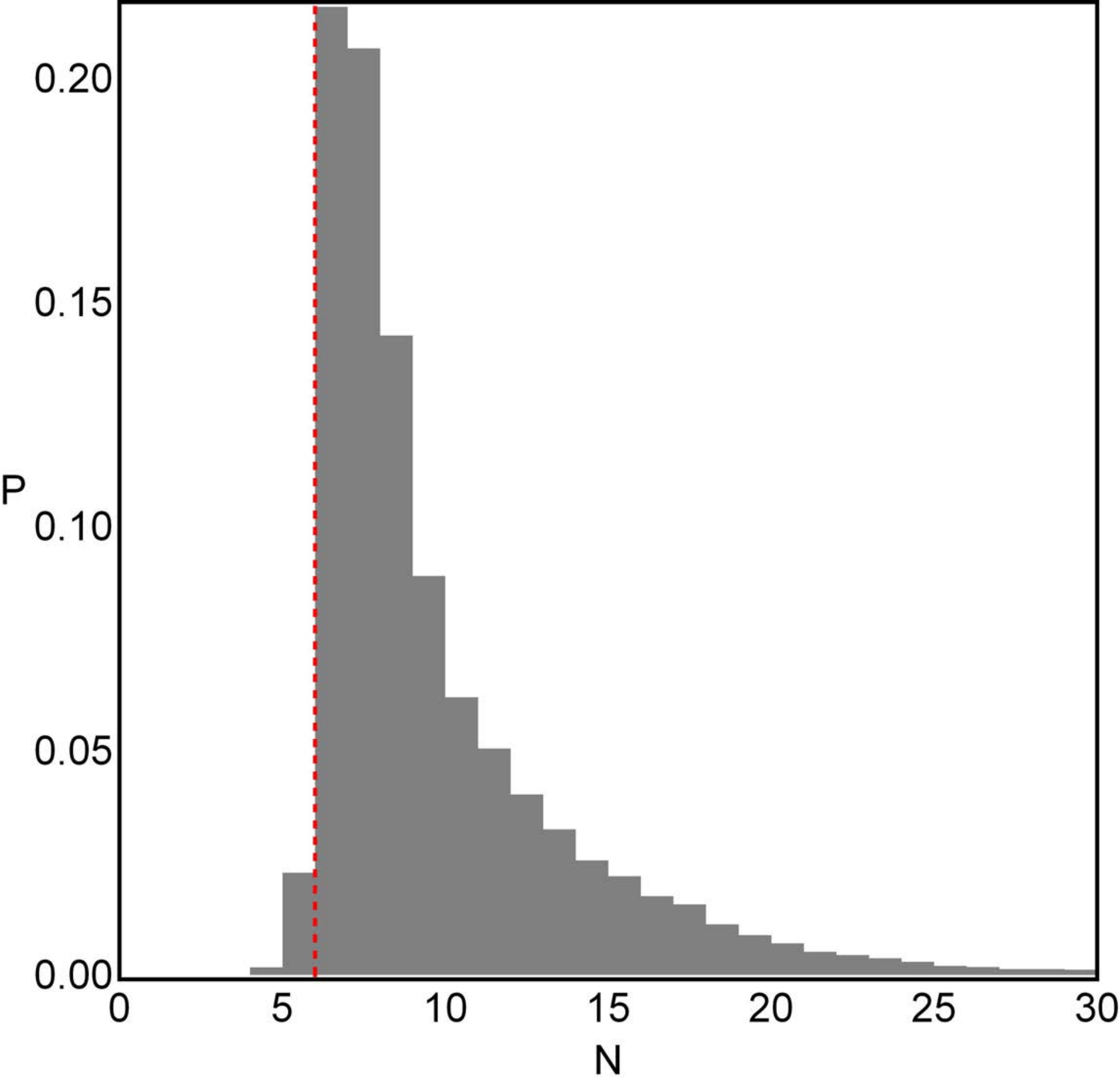}\\
(g)\includegraphics[scale=.26]{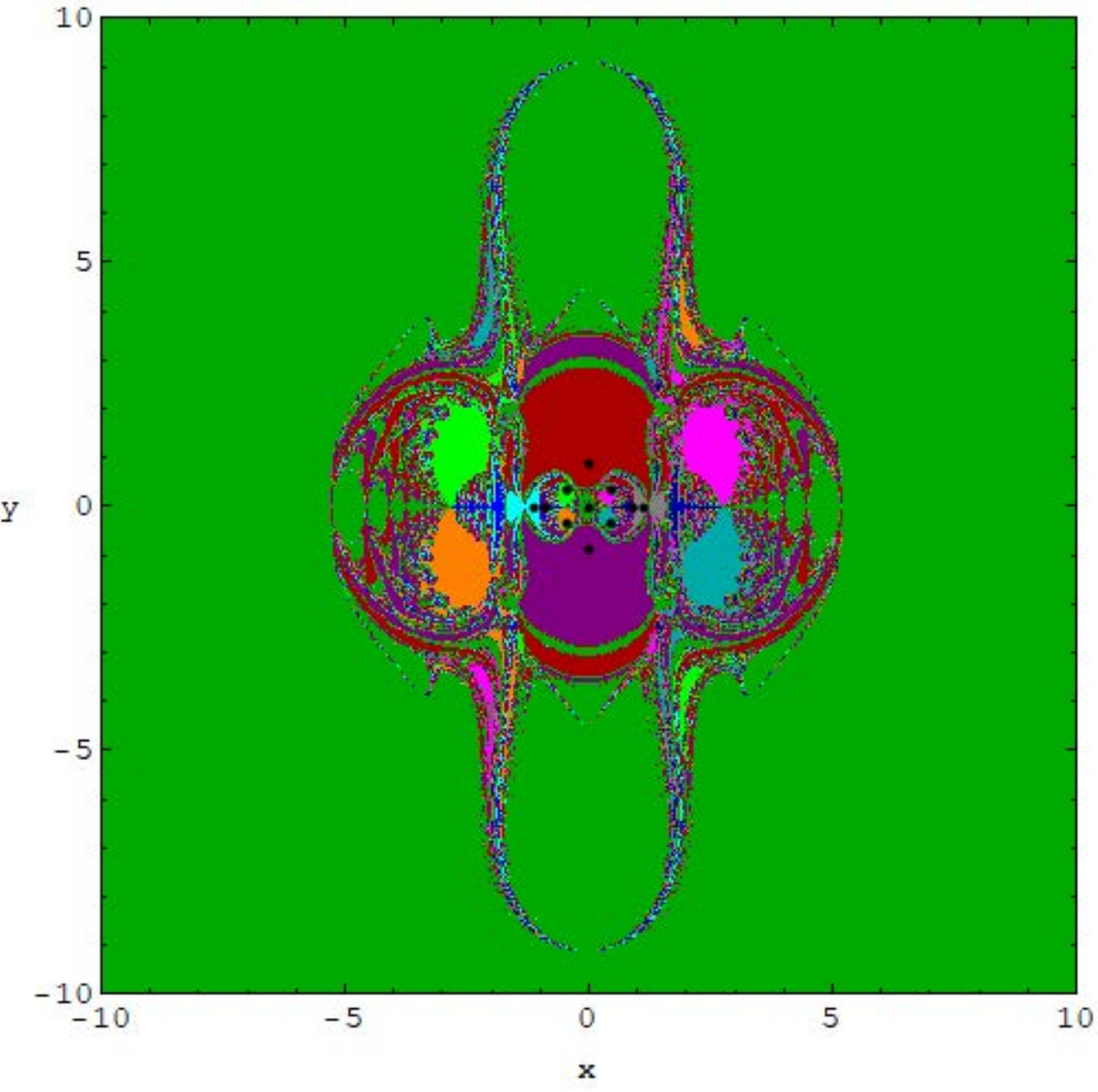}
(h)\includegraphics[scale=.26]{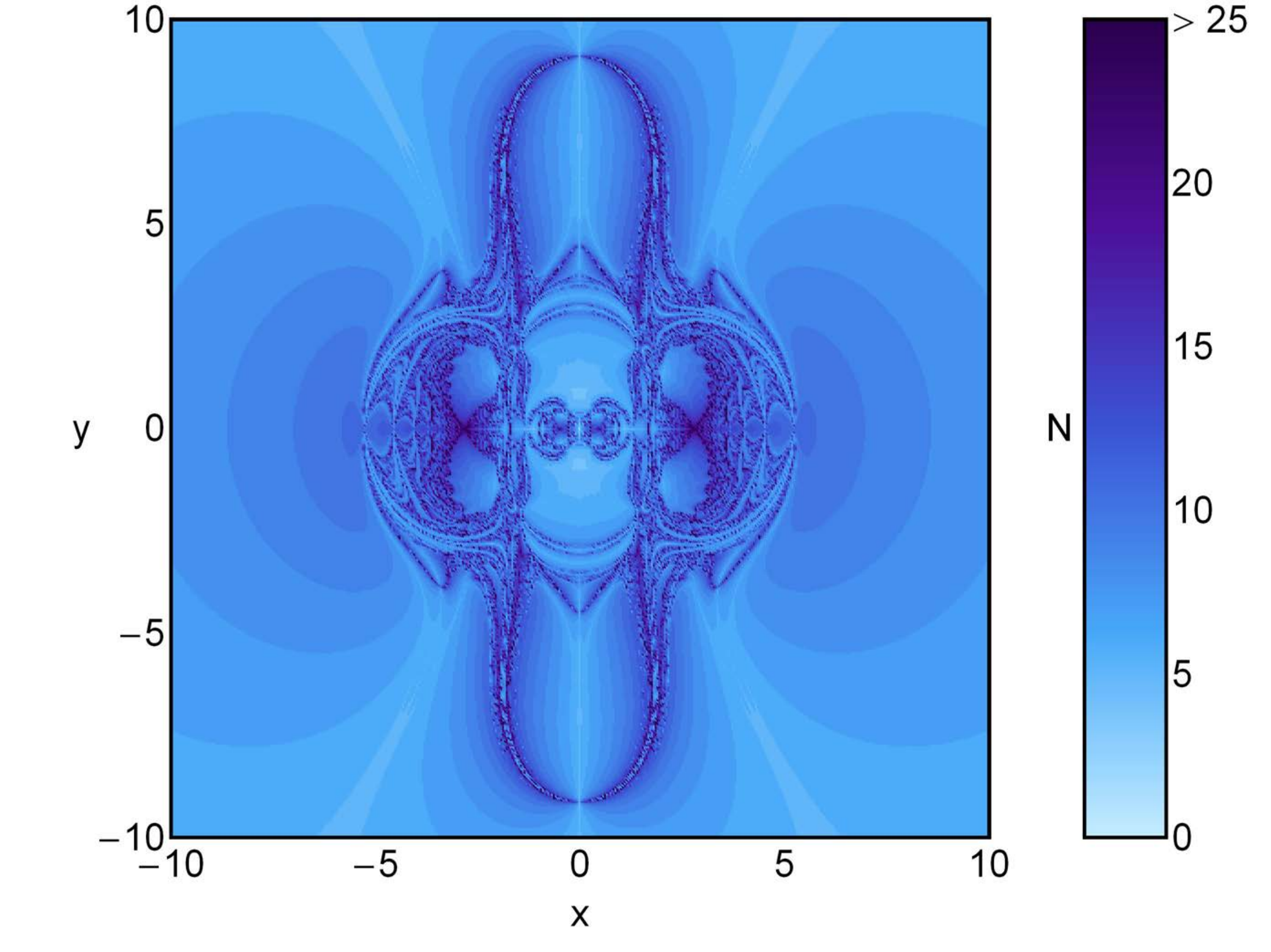}
(i)\includegraphics[scale=.26]{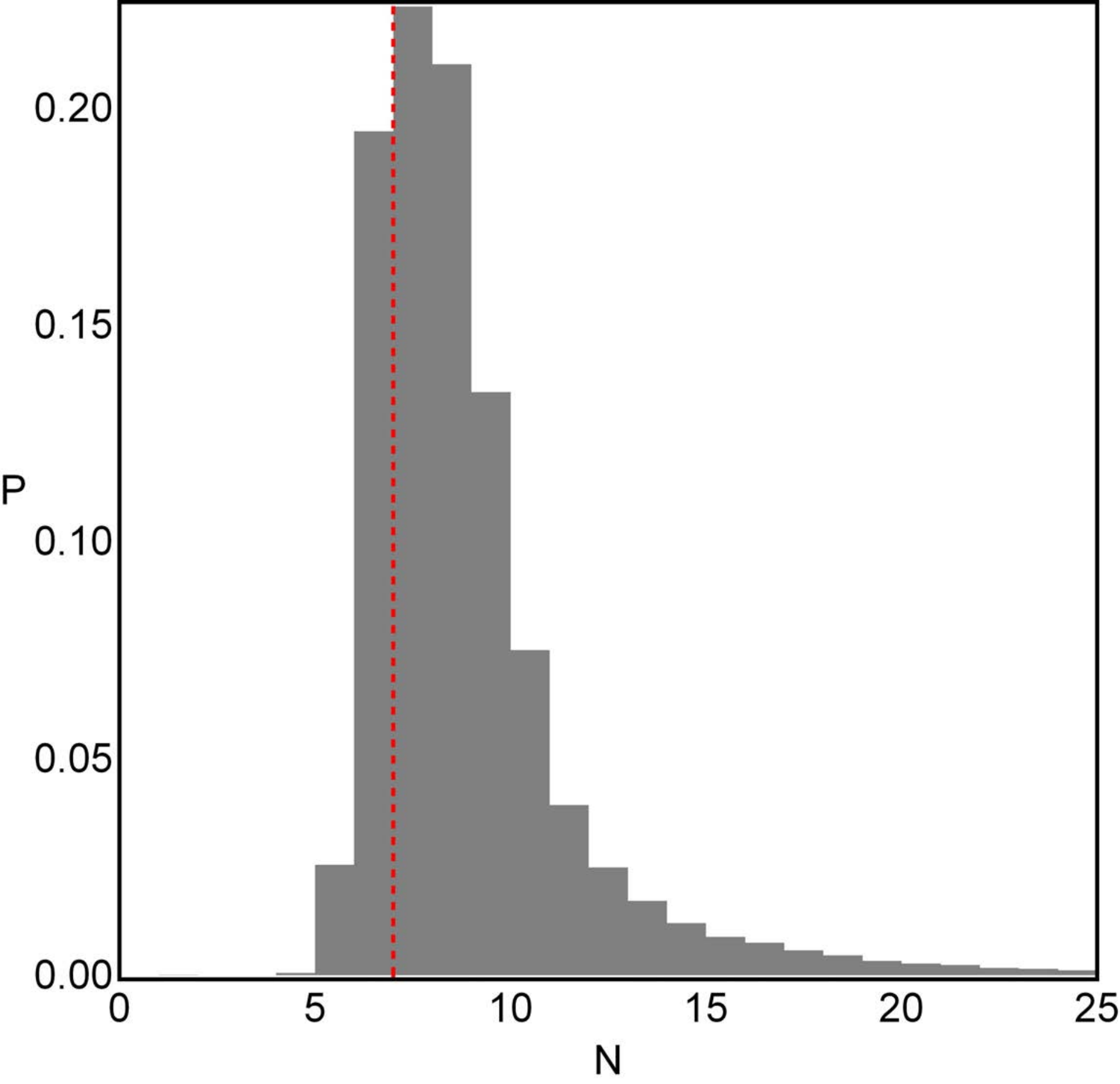}
\caption{The Newton-Raphson basins of attraction on the $(x,y)$ plane
for the case when eleven libration points exist for:  (a) $e=-0.1947$;
(d) $e=-0.1795$; (g) $e=-0.1743$. The color code denoting the attractors
is as follows: $L_4$ (\emph{green}); $L_8$ (\emph{purple}); $L_9$ (\emph{crimson});
 $L_{10}$ (\emph{teal}); $L_{11}$ (\emph{magenta}); $L_{12}$ (\emph{orange});
 $L_{13}$ (\emph{light green}) and non-converging points (\emph{white}).
 (b, e, h: the middle panels)
  The corresponding
  distribution of the number $N$ of required iterations for obtaining the attracting
  regions, (c, f, i: the right panels) the corresponding probability distributions of required number of iterations
  for obtaining the Newton-Raphson basins of convergence, shown in panels-(a, d, g) respectively. The vertical, dashed, red line indicates, in each case,
the most probable number $N^*$ of iterations. The black dots show the position of the libration points. (Color figure online).}
\end{center}
\end{figure*}
\subsection{\emph{Case IV: when thirteen libration points exist}}
\label{sec:404}
The following subsection deals with the case where thirteen libration points exist for $e \in (-0.46671,$
$ -0,457853)$ (case-i): $L_4$ central libration point, $L_{1,2,6,7}$ on $x-$axis, $L_{3,5,8,9}$ on $y-$axis and
$L_{10,11,12,13}$ on $(x, y)$ plane while for $e \in (-0.173395,$ $0)$ (case-ii): $L_4$ central libration point,
$L_{1,2,3,5,6,7}$ on $x-$axis, $L_{8,9}$ on $y-$axis and $L_{10,11,12,13}$ on $(x, y)$ plane. In Fig.
\textcolor[rgb]{1.00,0.00,0.50}{8}(a-f), the Newton-Raphson basins of convergence are illustrated for six values of the parameter $e$. The topology of the configuration $(x,y)$ plane changes in a drastic manner, as the value of the parameter $e$ decreases.

When the parameter $e$ is very close to $0$, there exist eight libration
points in two sets of four originating in the vicinity of each of the
primaries. The basins of convergence associated with the libration points
shown in panel-a, resemble with that of the classical restricted three-body problem.
Two exotic bugs, with many legs and antennas, shown in yellow and cyan color correspond to libration points $L_{1,7}$ respectively. Moreover,
the butterfly wing shaped domains of convergence associated with
the libration points $L_{8,9}$ shown in purple and crimson color
exist. The most notable change due to existence of eight extra libration
points in the presence of a negative value of the parameter is the following four lobe
appear, corresponding to non collinear libration points ($L_{10,11,12,13}$) in the vicinity of the primaries
as well as at the back of both the exotic bugs. Let us denote the region at the
back of exotic bugs corresponding to the libration points $L_1$ (yellow)
and $L_7$ (cyan) by $R_1$ and $R_7$ respectively.\\
As the value of parameter $e$ decreases, the legs and antennas
of the exotic bugs shaped region decrease continuously and they are converted to \textcolor[rgb]{1.00,0.00,0.50}{a}
chaotic region composed of a mixture of initial conditions. This is why,
the regions $R_1$ and $R_7$ increase. In these regions, the lobes corresponding
to $L_{10, 11}$ (in $R_1$) and $L_{12, 13}$ (in $R_7$) also increase.
Moreover, the basins of convergence corresponding to the libration
points $L_3$ (gray colour in $R_1$) and $L_{5}$ (red colour in $R_7$)
increase rapidly with the decrease in $e$. Finally, most of the
regions of the basins of convergence are covered by finite domains of convergence
corresponding to the libration points $L_{3,5}$ (see panel-f). The basins boundaries
inside these regions are highly fractal.

The most important changes for decreasing value of parameter $e$
can be listed as follows:
\begin{itemize}
  \item The domain of the basins of convergence corresponding to
  libration points $L_{1, 7, 8, 9}$ decrease.
  \item The domain of the basins of convergence corresponding to
  libration points $L_{3, 5, 10,...,13}$ increase.
  \item The extent of the basins of convergence associated with the
  central libration point $L_4$ is infinite whereas for other libration
  points, it is finite in all the cases.  Moreover, the area of the basins
  of convergence corresponding to $L_4$ inside central region decreases
  continuously.
  \item The majority of the area of the regions $R_{1,7}$ is covered by a
  domain of convergence associated to the libration points $L_{1,7}$ and
  a prominent part of these regions is also covered by four lobes and the rest
  of the area is composed of highly chaotic mixtures of the initial conditions.
  \end{itemize}
\begin{figure*}\label{Fig:9}
\begin{center}
(a)\includegraphics[scale=.45]{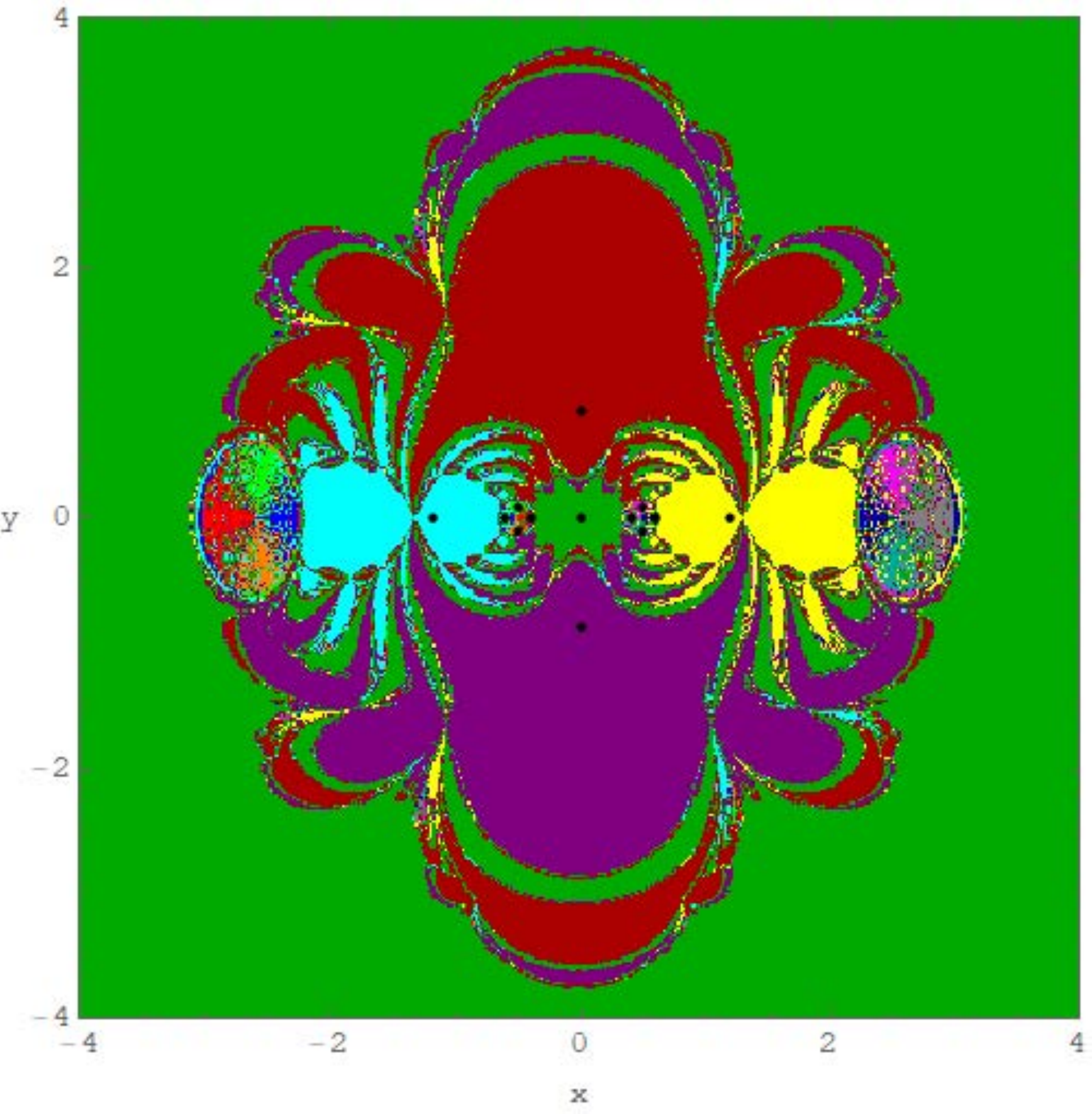}
(b)\includegraphics[scale=.45]{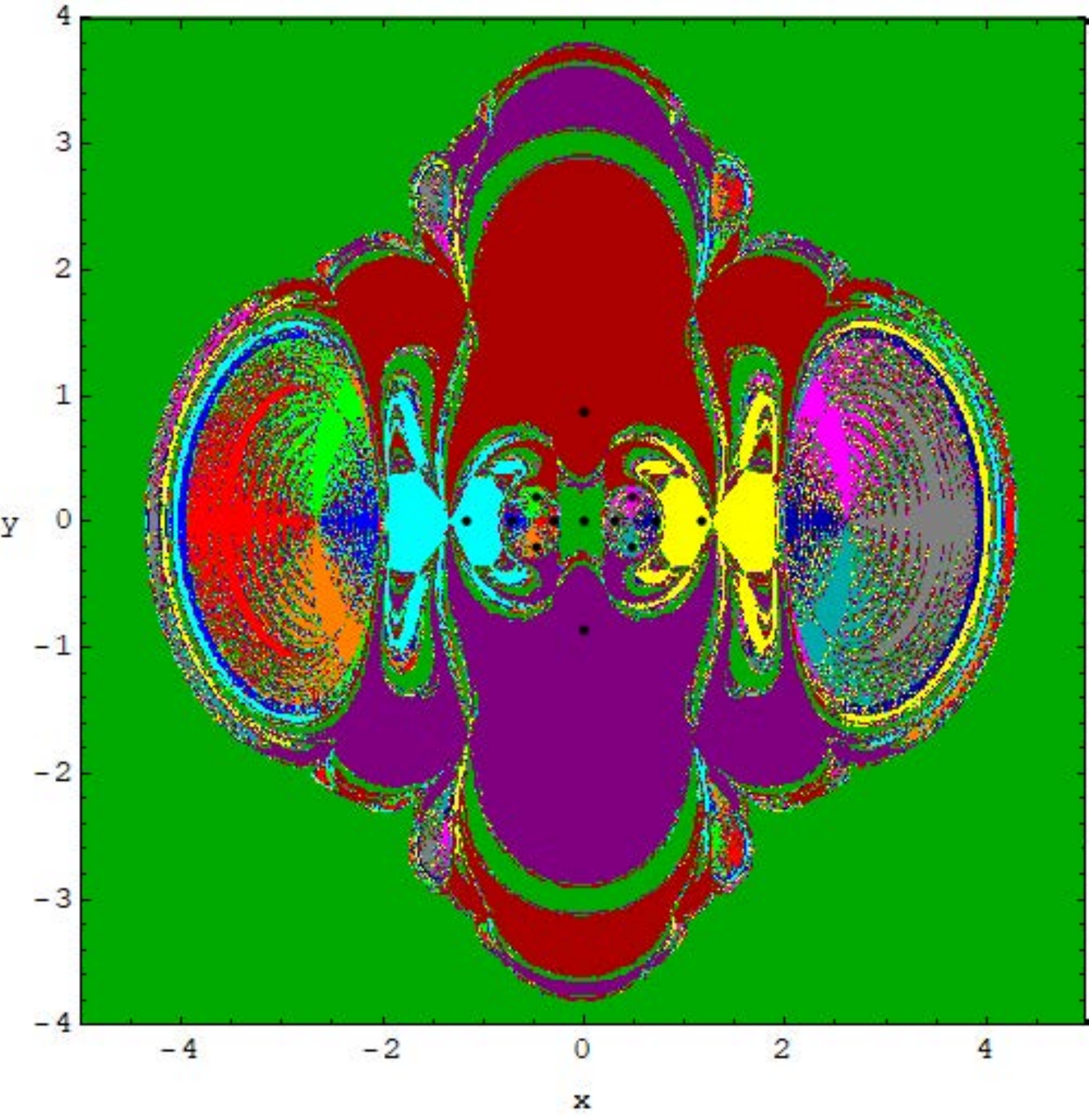}\\
(c)\includegraphics[scale=.45]{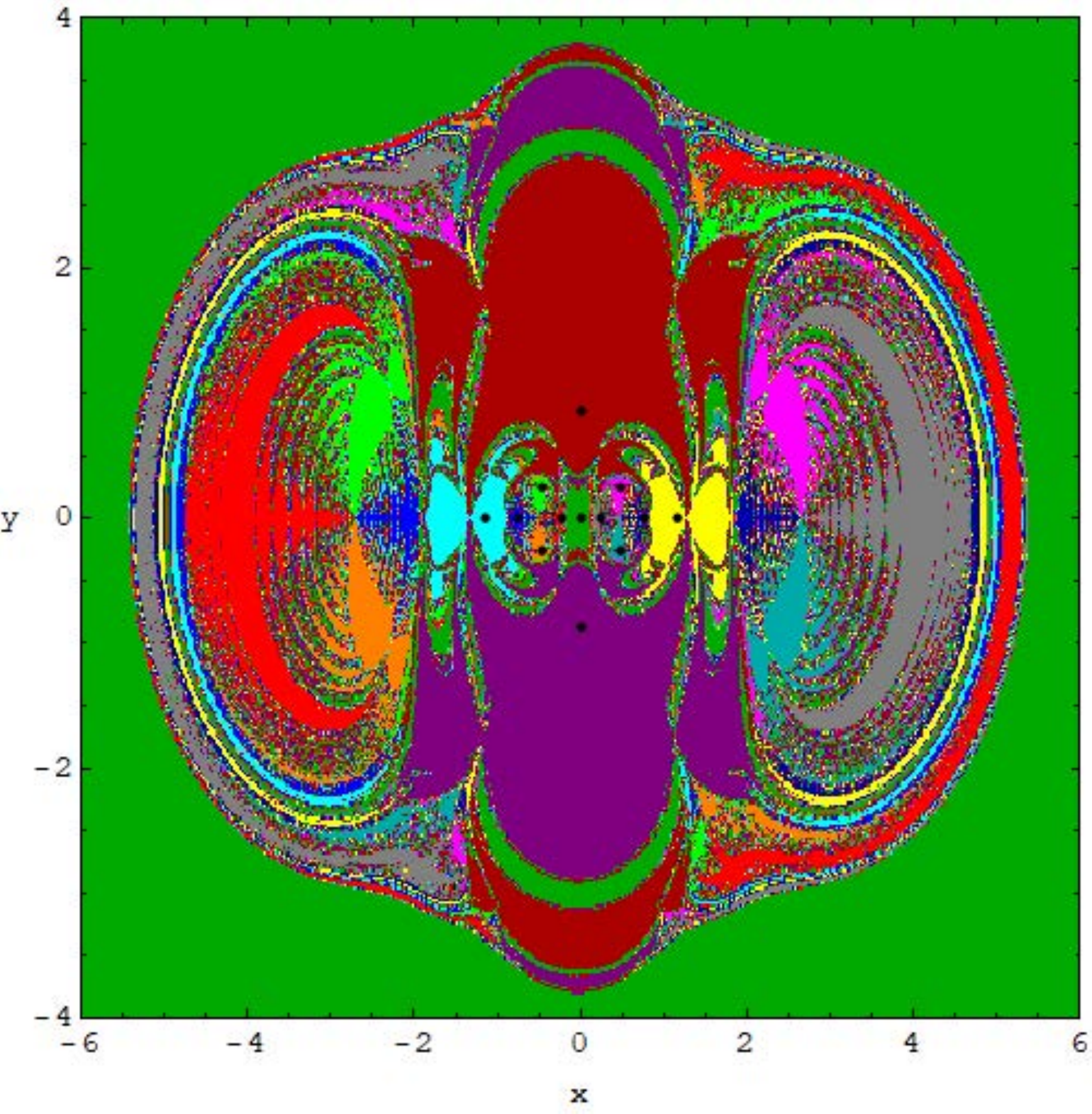}
(d)\includegraphics[scale=.45]{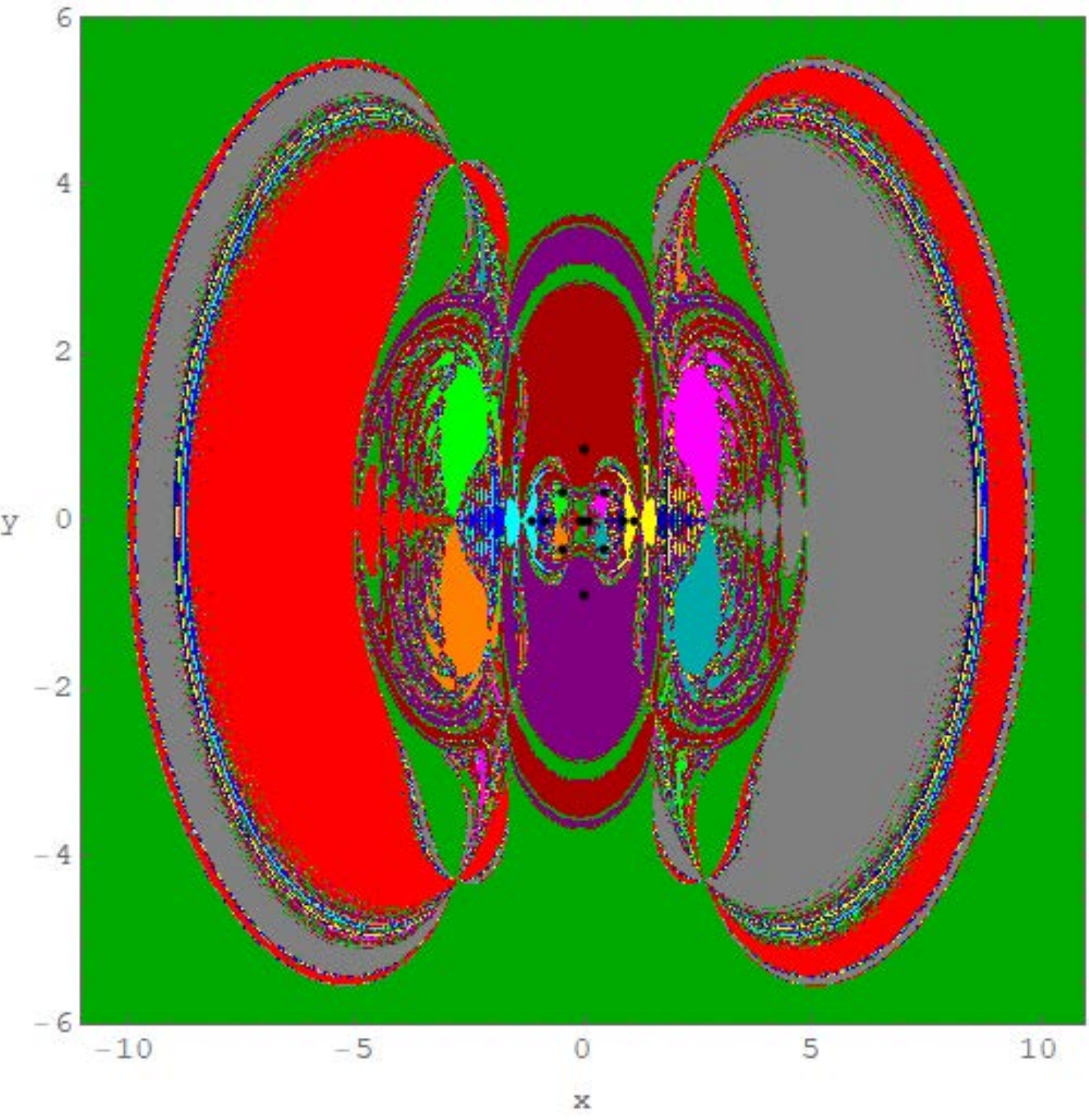}\\
(e)\includegraphics[scale=.45]{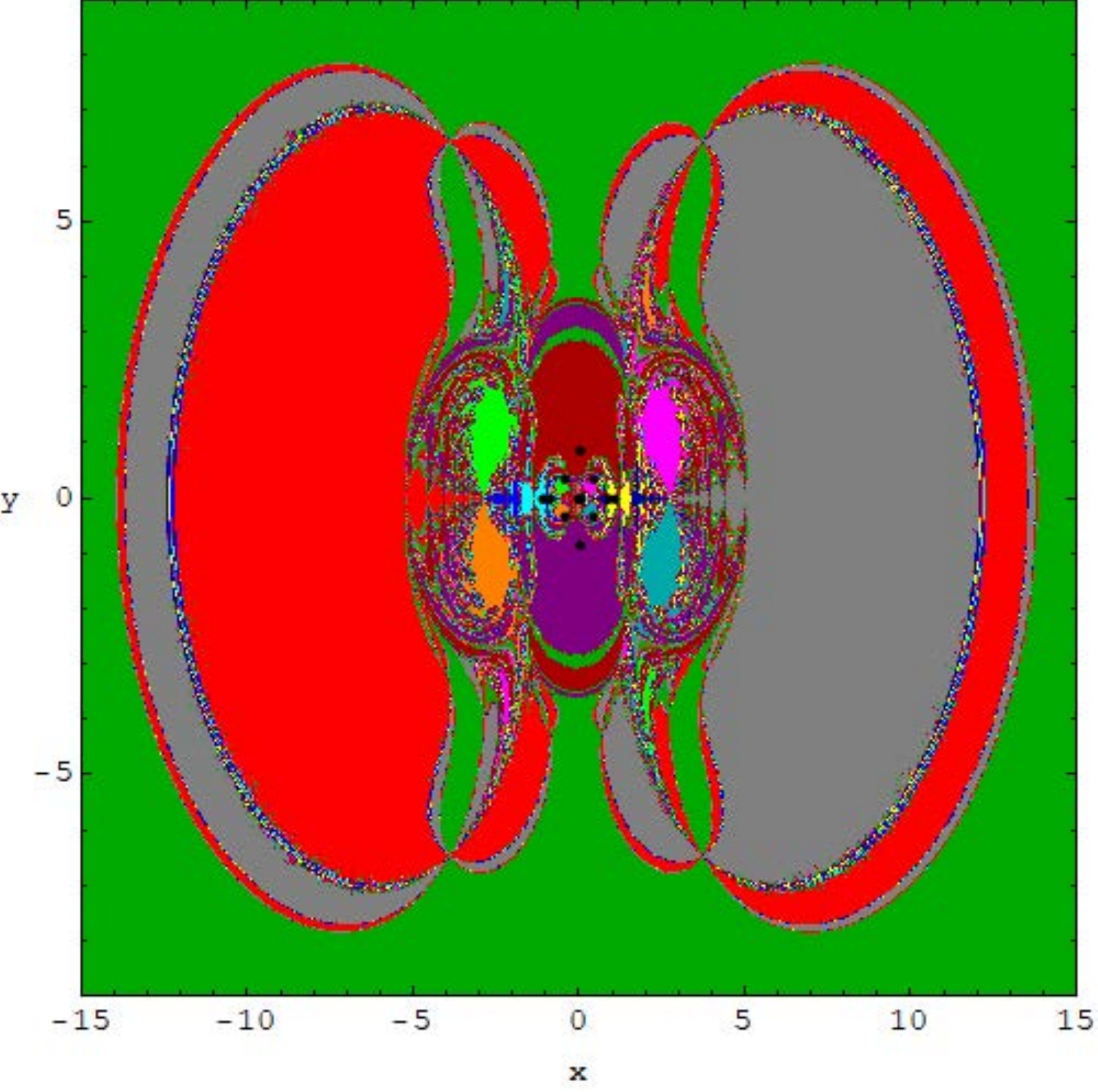}
(f)\includegraphics[scale=.615]{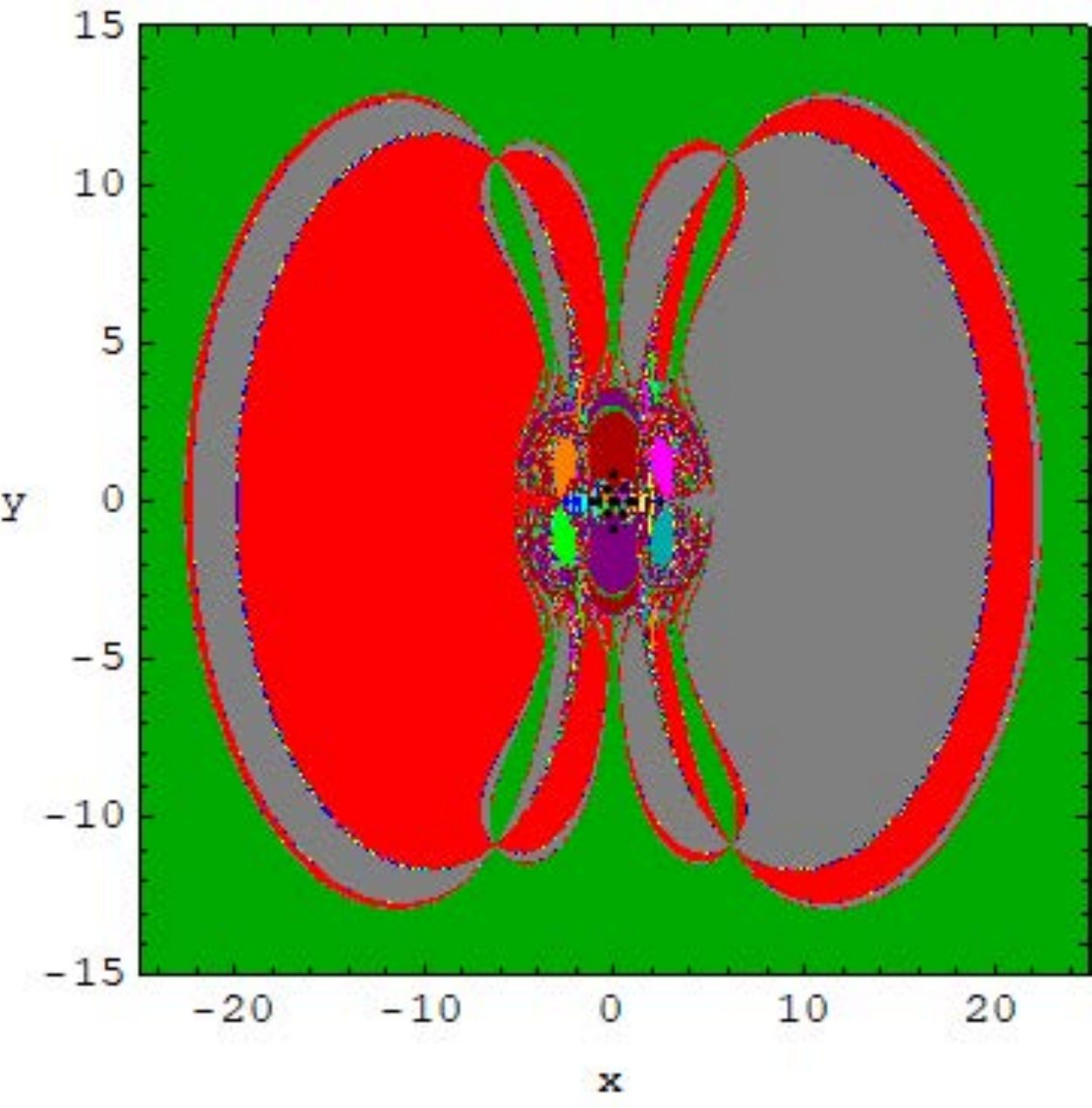}
\caption{The Newton-Raphson basins of attraction on the $xy$-plane for the case when thirteen libration points exist for:  (a) $e=-0.05$; (b) $e=-0.1$; (c) $e=-0.125$; (d) $e=-0.165$ ; (e) $e=-0.171$; (f) $e=-0.1731$. The color code denoting the attractors is as follows: $L_1$ (\emph{yellow}); $L_2$ (\emph{Darker blue}); $L_3$ (\emph{gray}); $L_4$ (\emph{green}); $L_5$ (\emph{red}); $L_6$ (\emph{blue}); $L_7$ (\emph{cyan}); $L_8$ (\emph{purple}); $L_9$ (\emph{crimson}); $L_{10}$ (\emph{teal}); $L_{11}$ (\emph{magenta}); $L_{12}$ (\emph{orange}); $L_{13}$ (\emph{light green}) and non-converging points (\emph{white}). The black dots show the position of the libration points.}
\end{center}
\end{figure*}
\begin{figure*}\label{Fig:10}
\begin{center}
(a)\includegraphics[scale=.36]{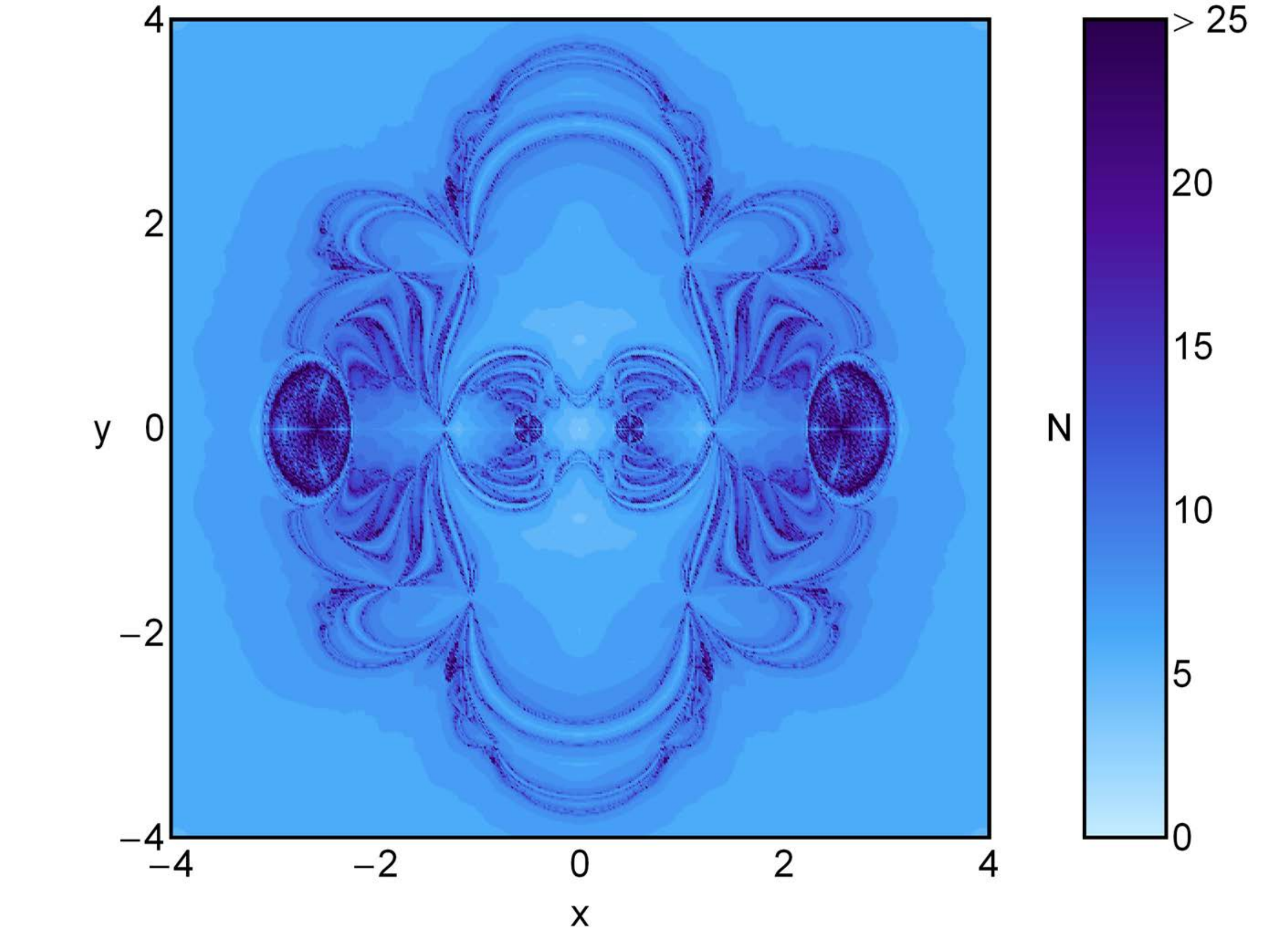}
(b)\includegraphics[scale=.36]{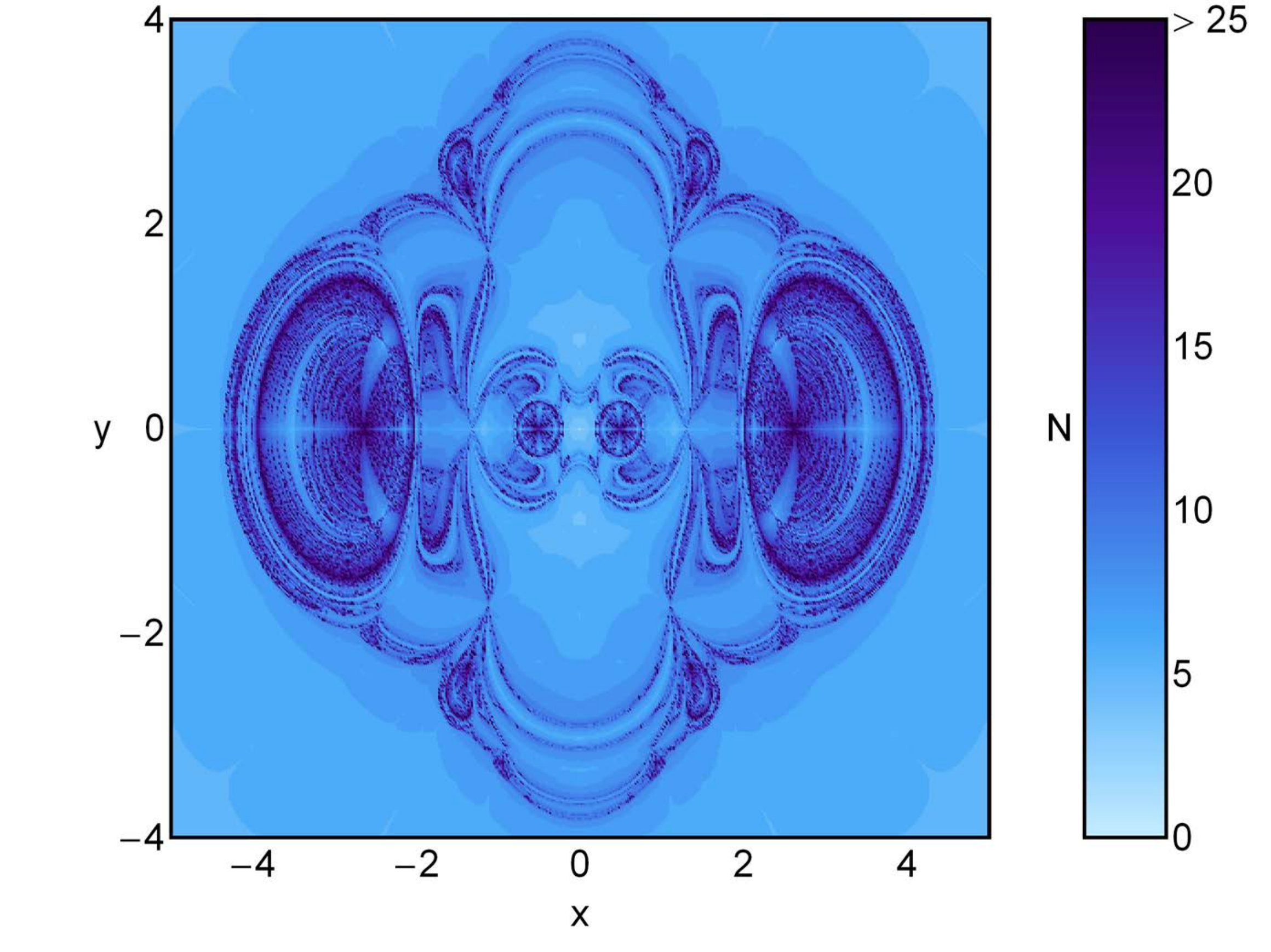}\\
(c)\includegraphics[scale=.36]{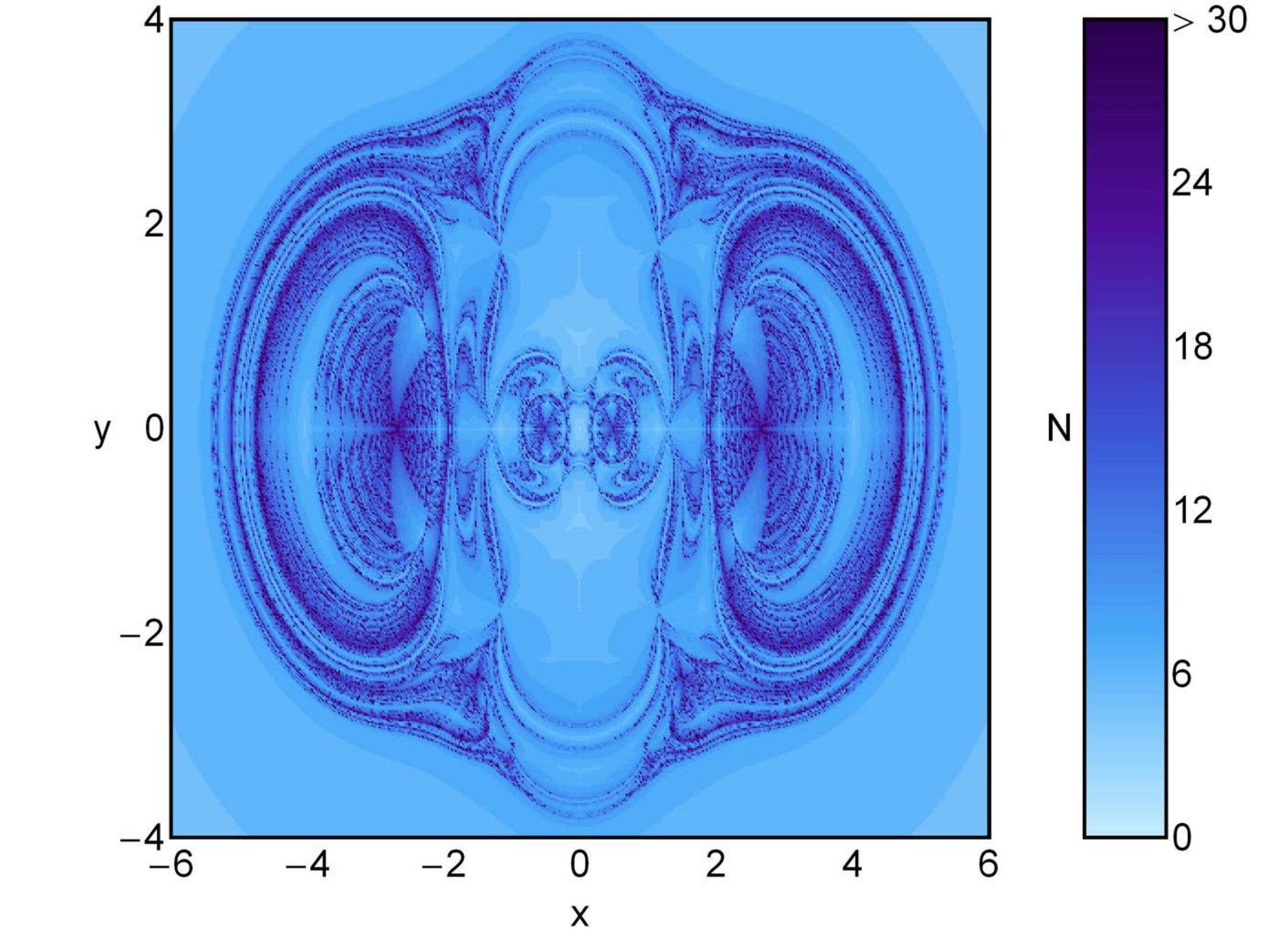}
(d)\includegraphics[scale=.36]{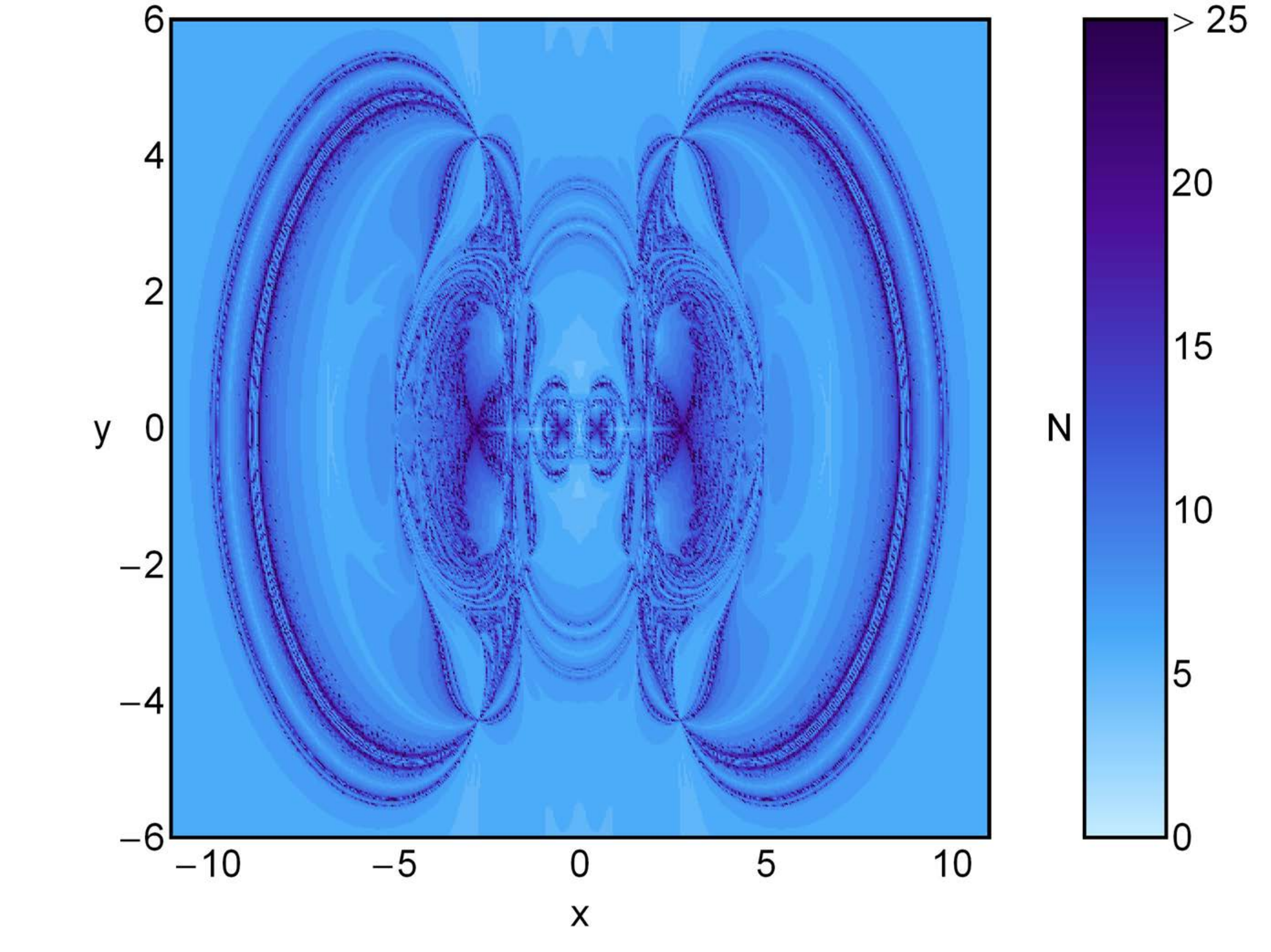}\\
(e)\includegraphics[scale=.36]{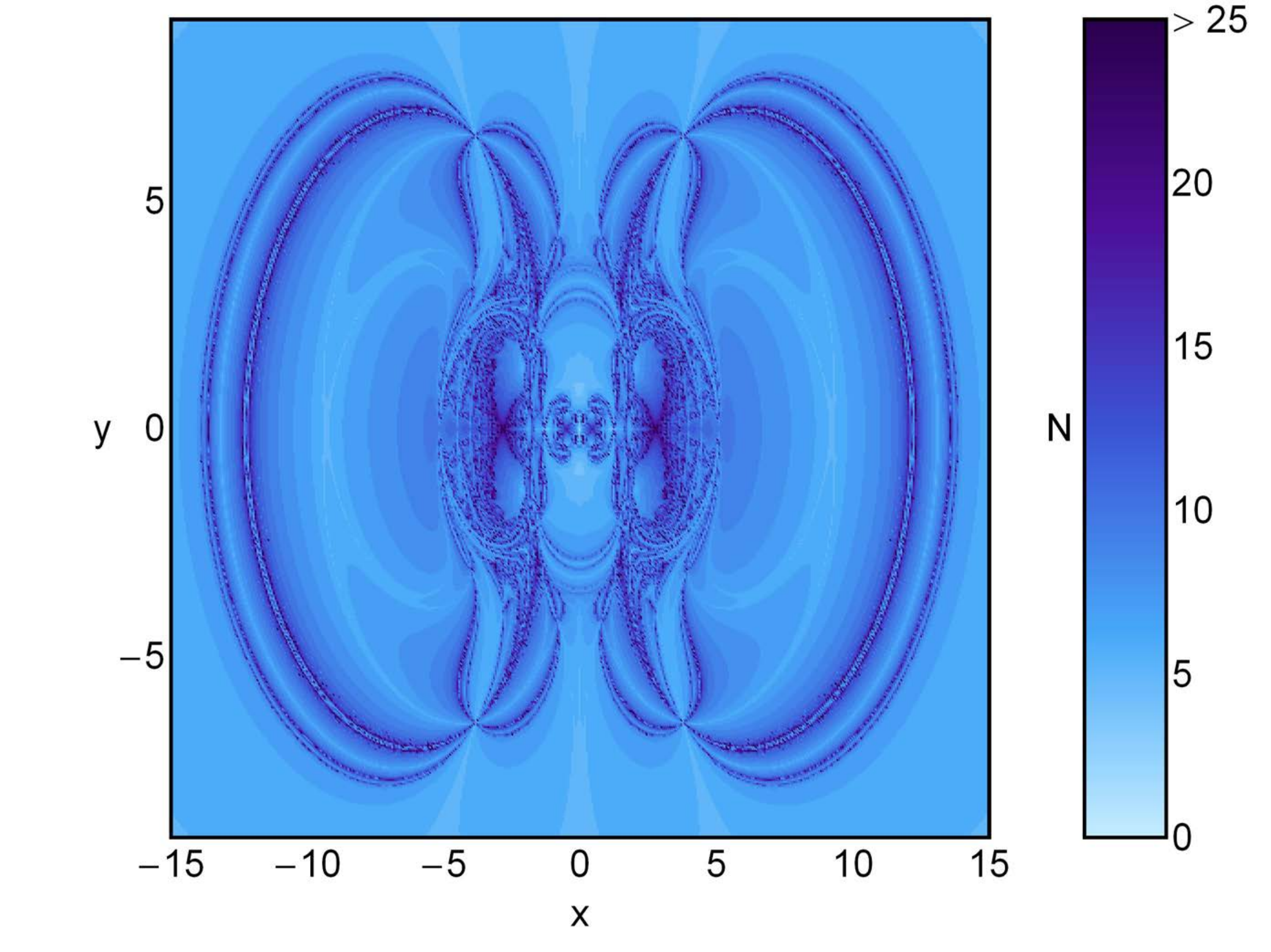}
(f)\includegraphics[scale=.36]{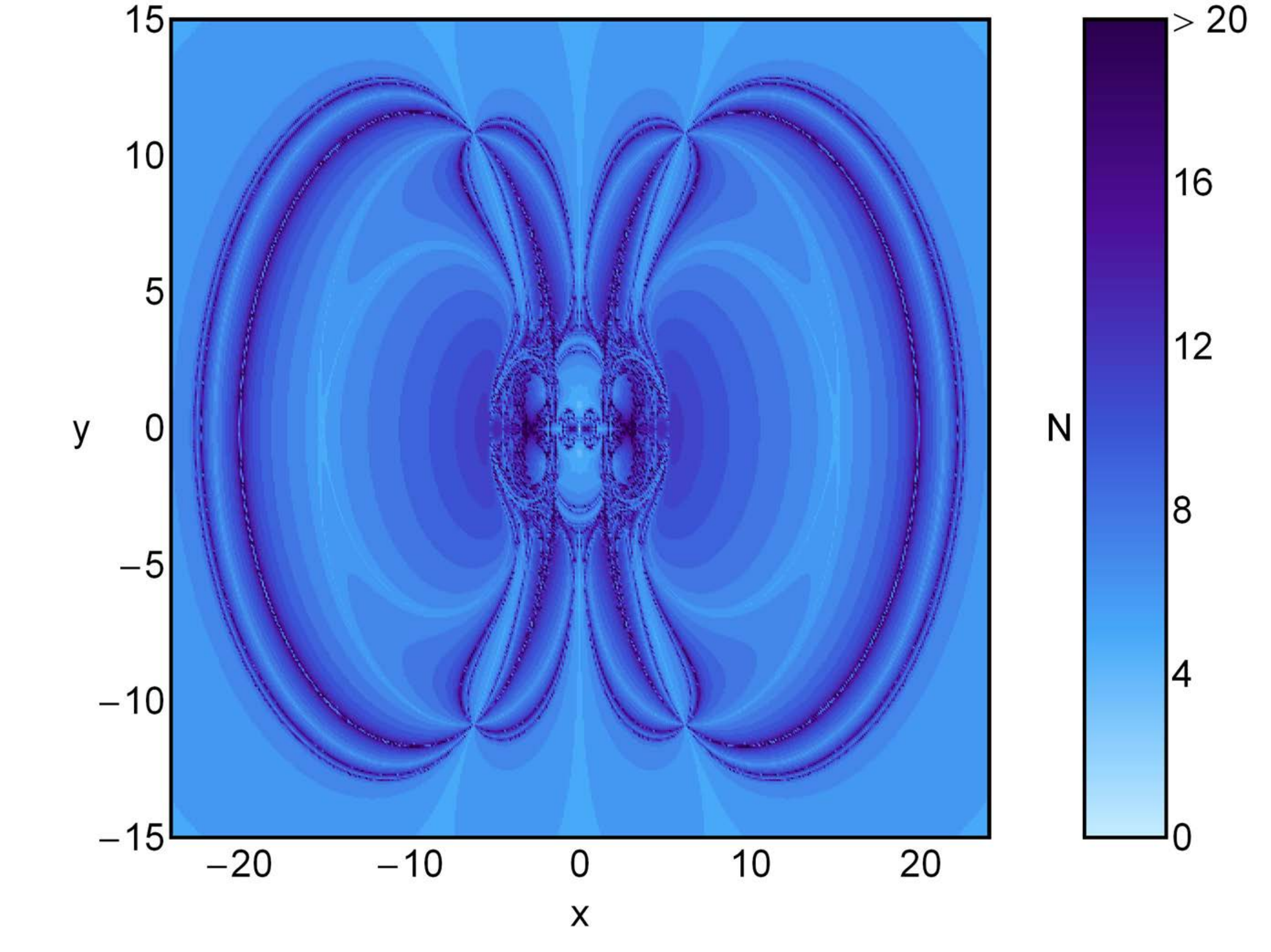}
\caption{The corresponding distributions of number $N$ of the required iterations for obtaining the Newton-Raphson basins of convergence,
shown in Fig. \textcolor[rgb]{1.00,0.00,0.50}{8}(a-f). (Color figure online).}
\end{center}
\end{figure*}
\begin{figure*}\label{Fig:11}
\begin{center}
(a)\includegraphics[scale=.36]{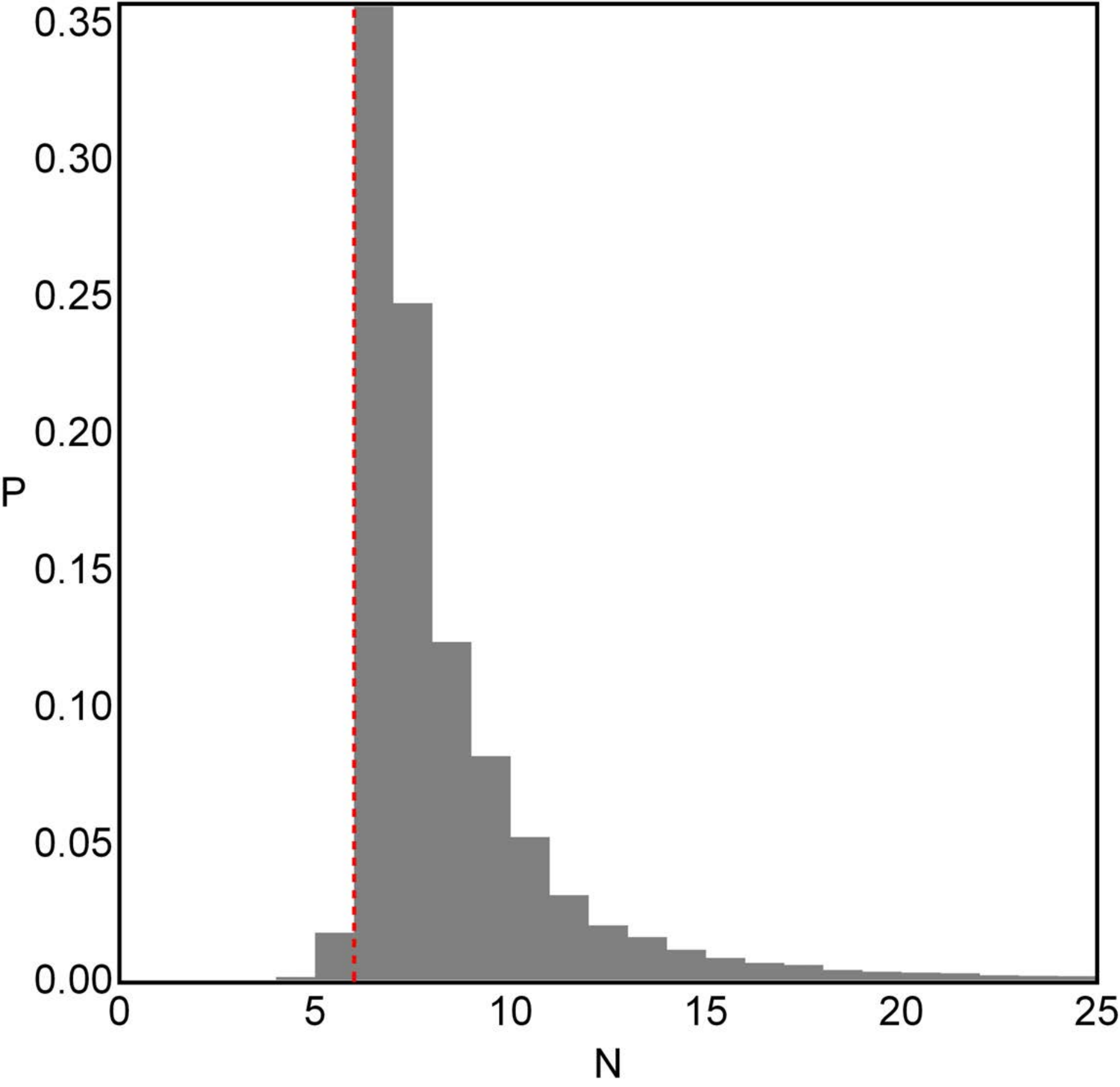}
(b)\includegraphics[scale=.36]{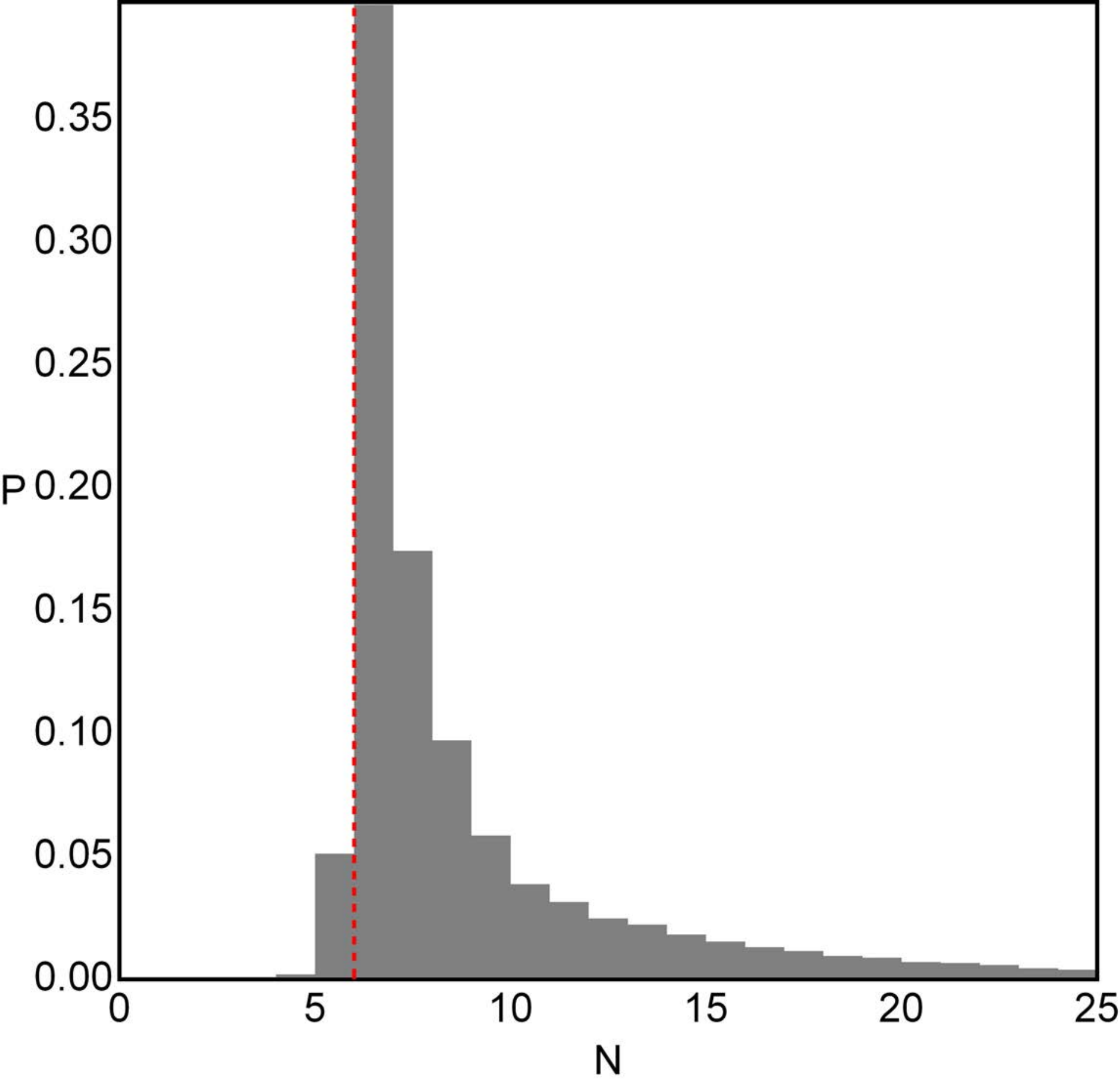}\\
(c)\includegraphics[scale=.36]{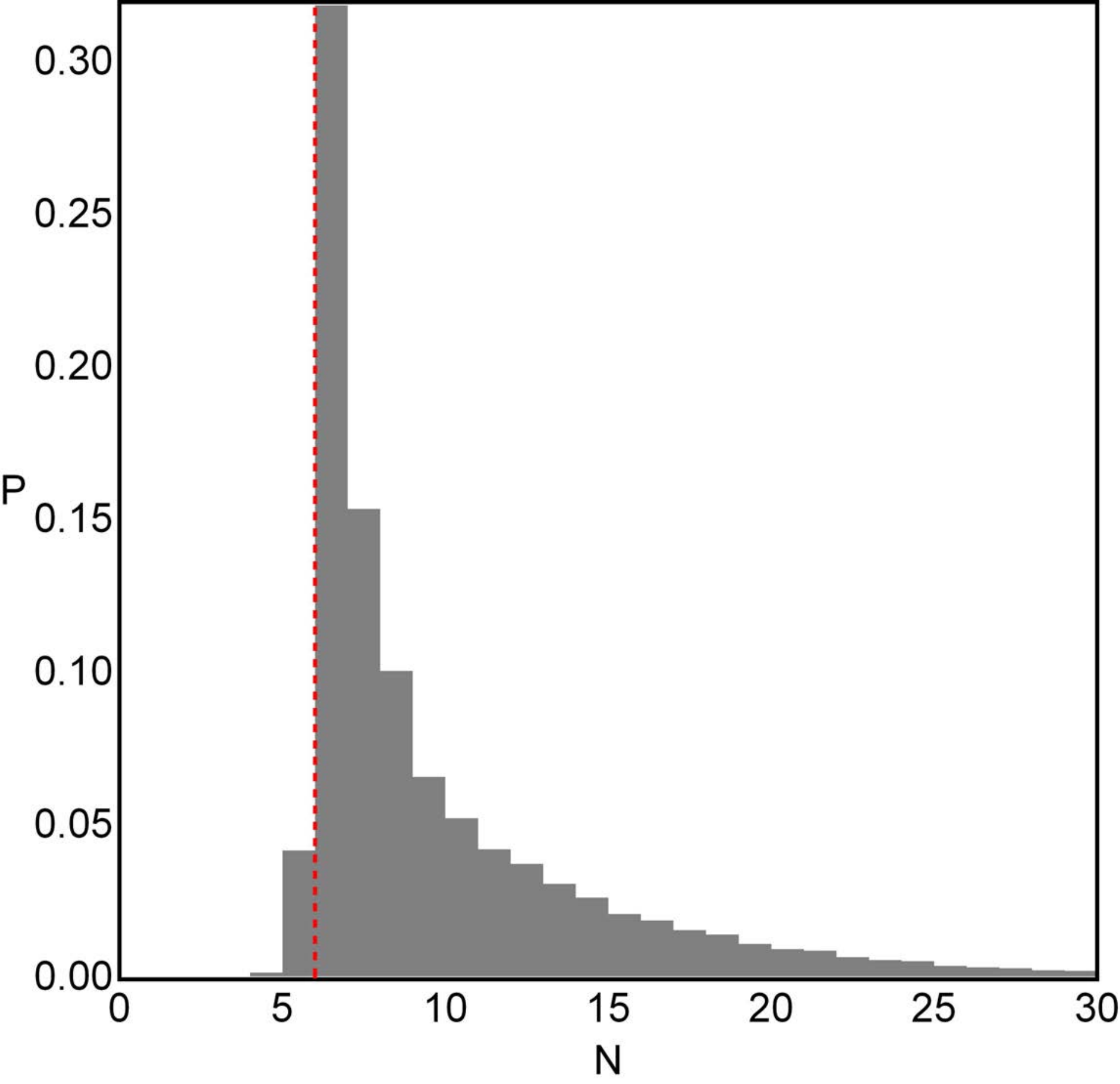}
(d)\includegraphics[scale=.36]{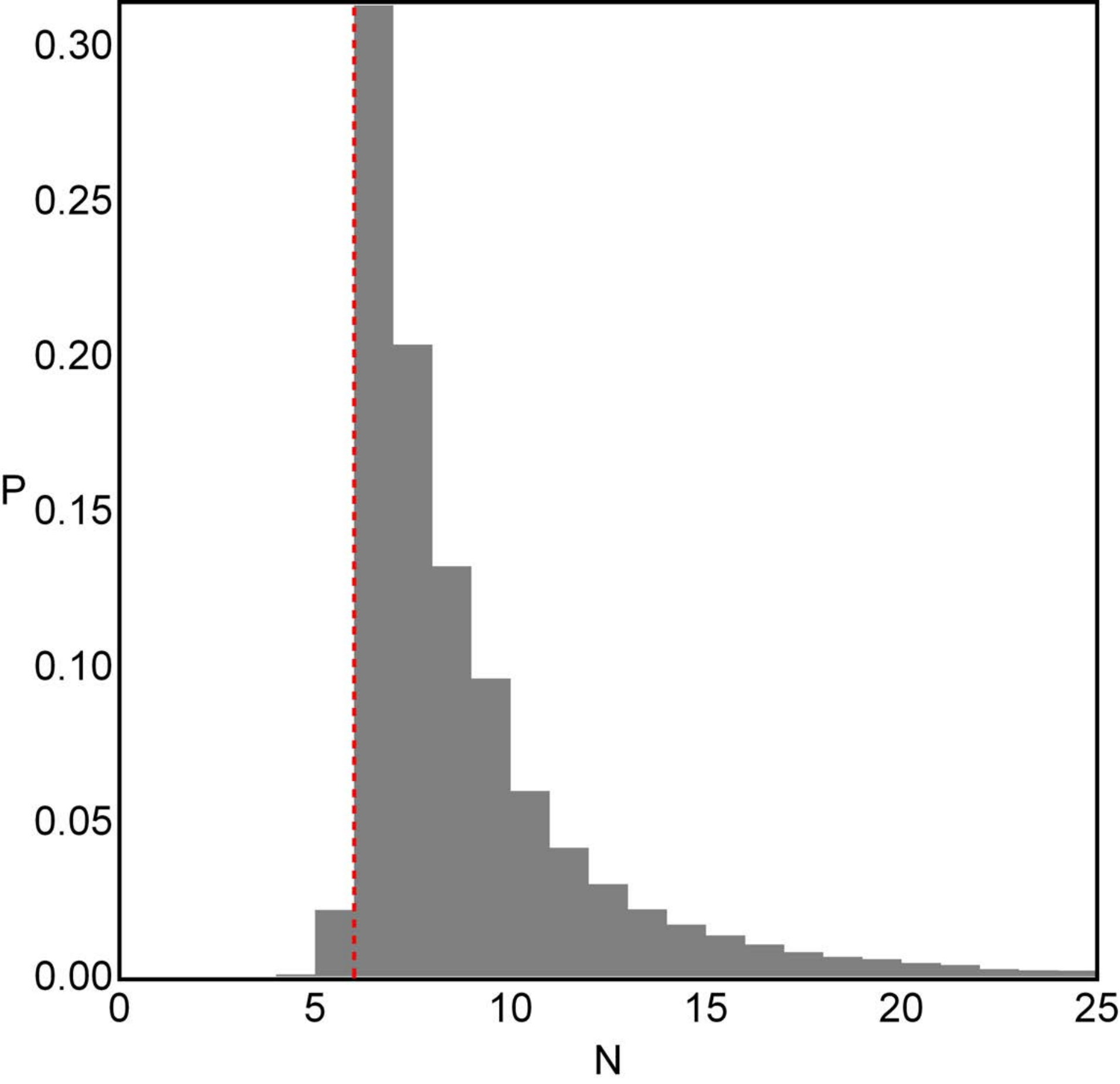}\\
(e)\includegraphics[scale=.36]{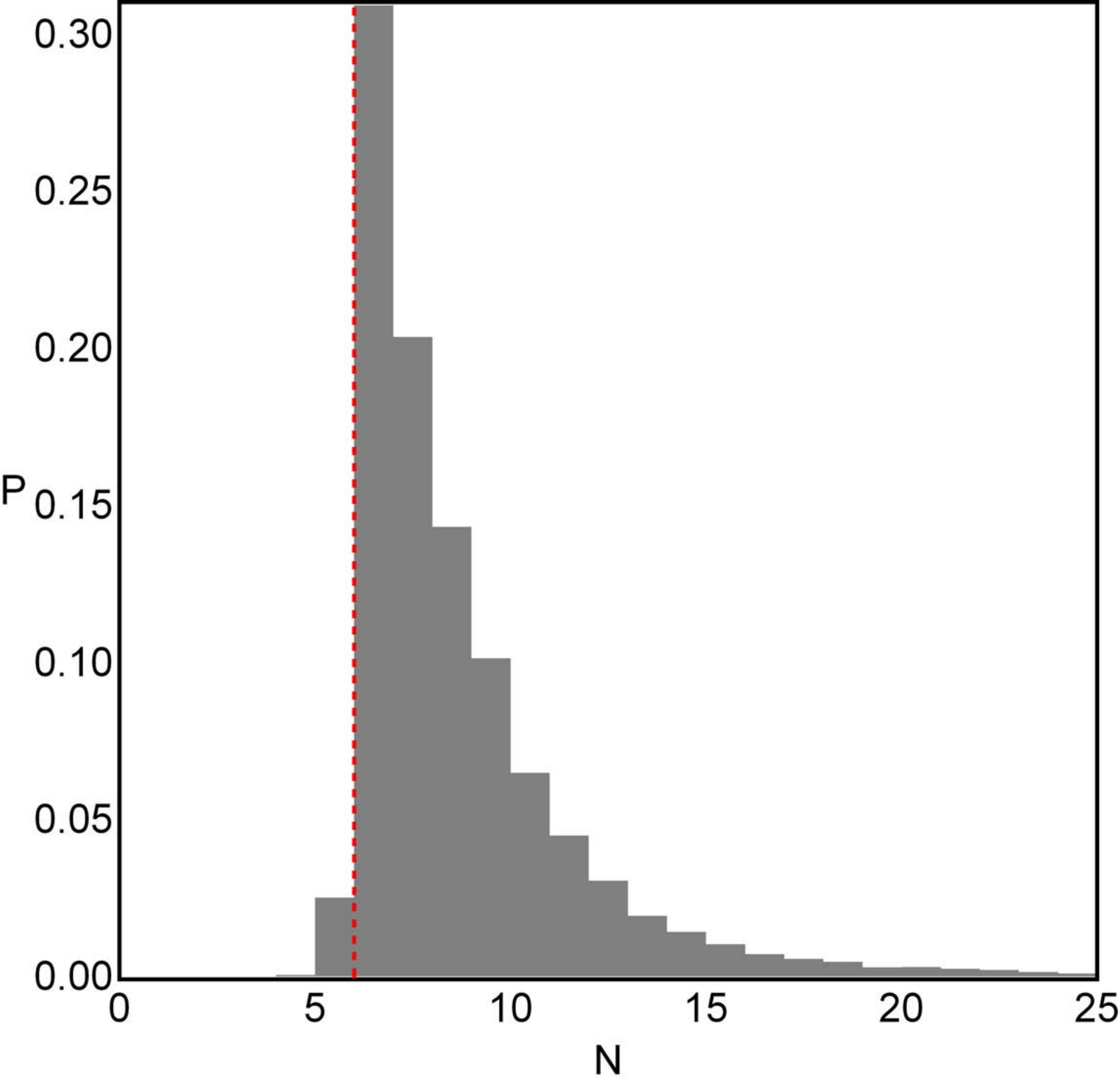}
(f)\includegraphics[scale=.36]{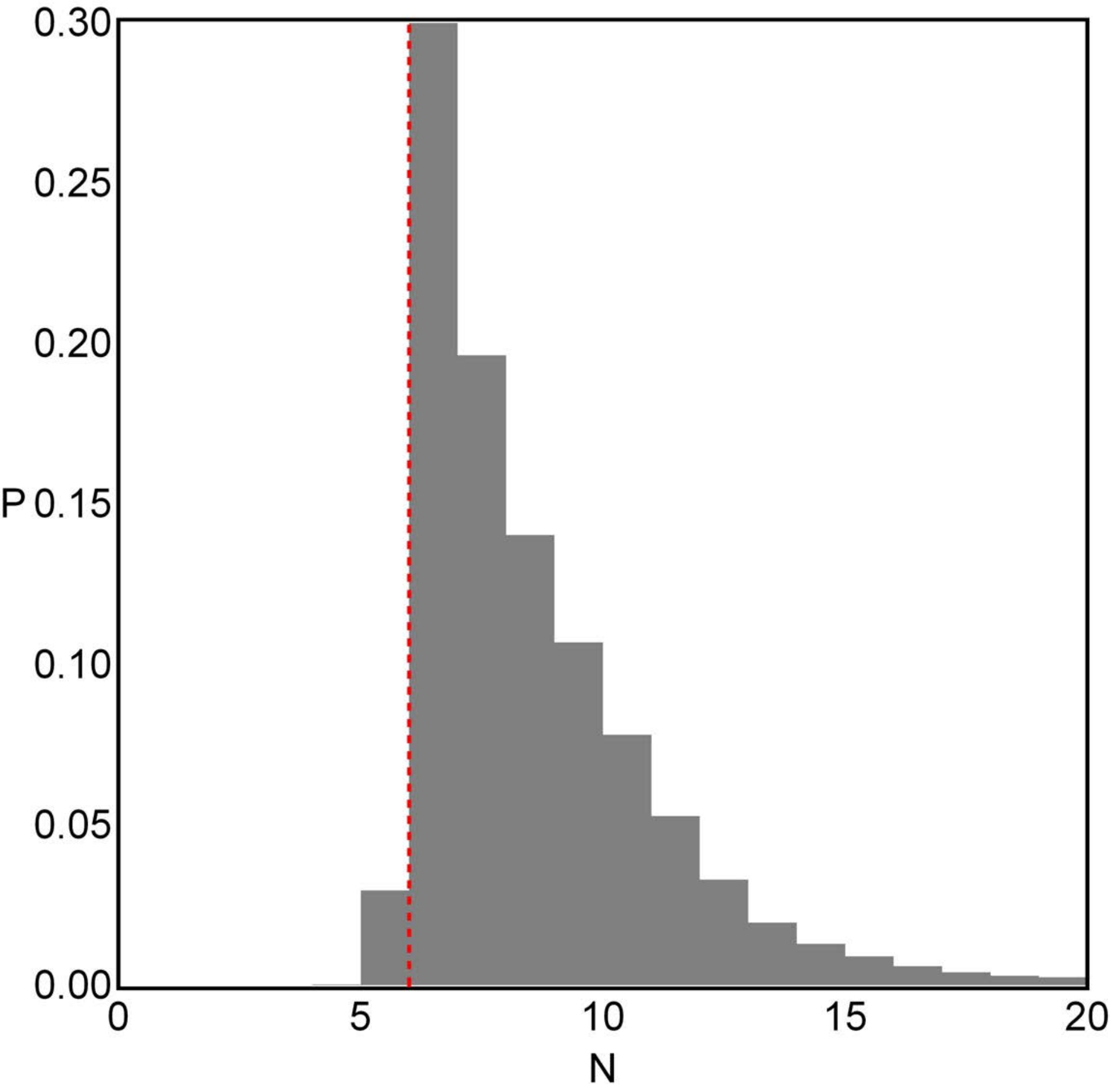}
\caption{The corresponding probability distributions of required number of iterations for obtaining the Newton-Raphson basins of convergence,
shown in Fig. \textcolor[rgb]{1.00,0.00,0.50}{8}(a-f). The vertical, dashed, red line indicates, in each case, the most probable number $N^*$ of iterations. (Color figure online).}
\end{center}
\end{figure*}
\begin{figure*}\label{Fig:12}
\begin{center}
(a)\includegraphics[scale=.26]{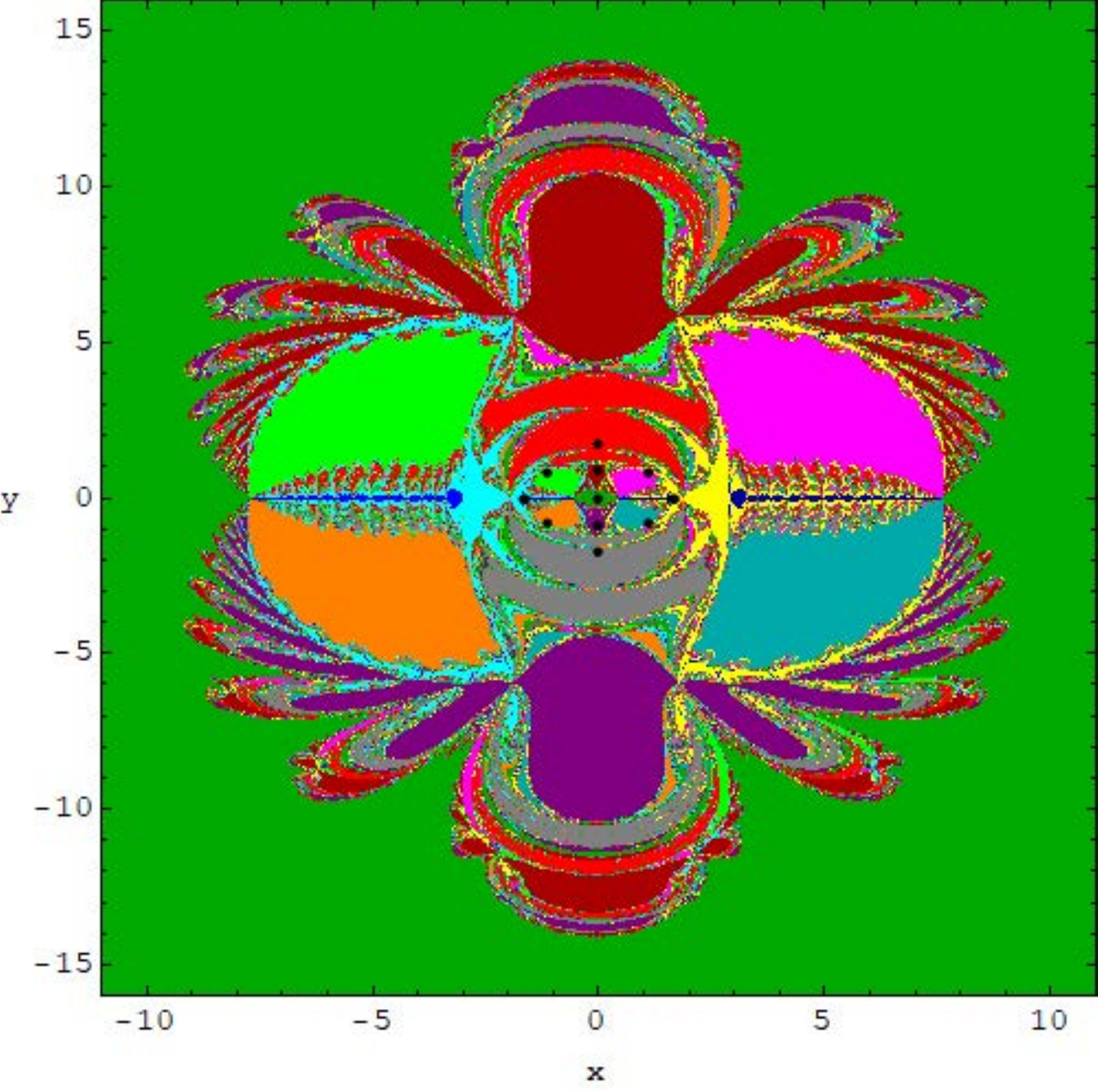}
(b)\includegraphics[scale=.26]{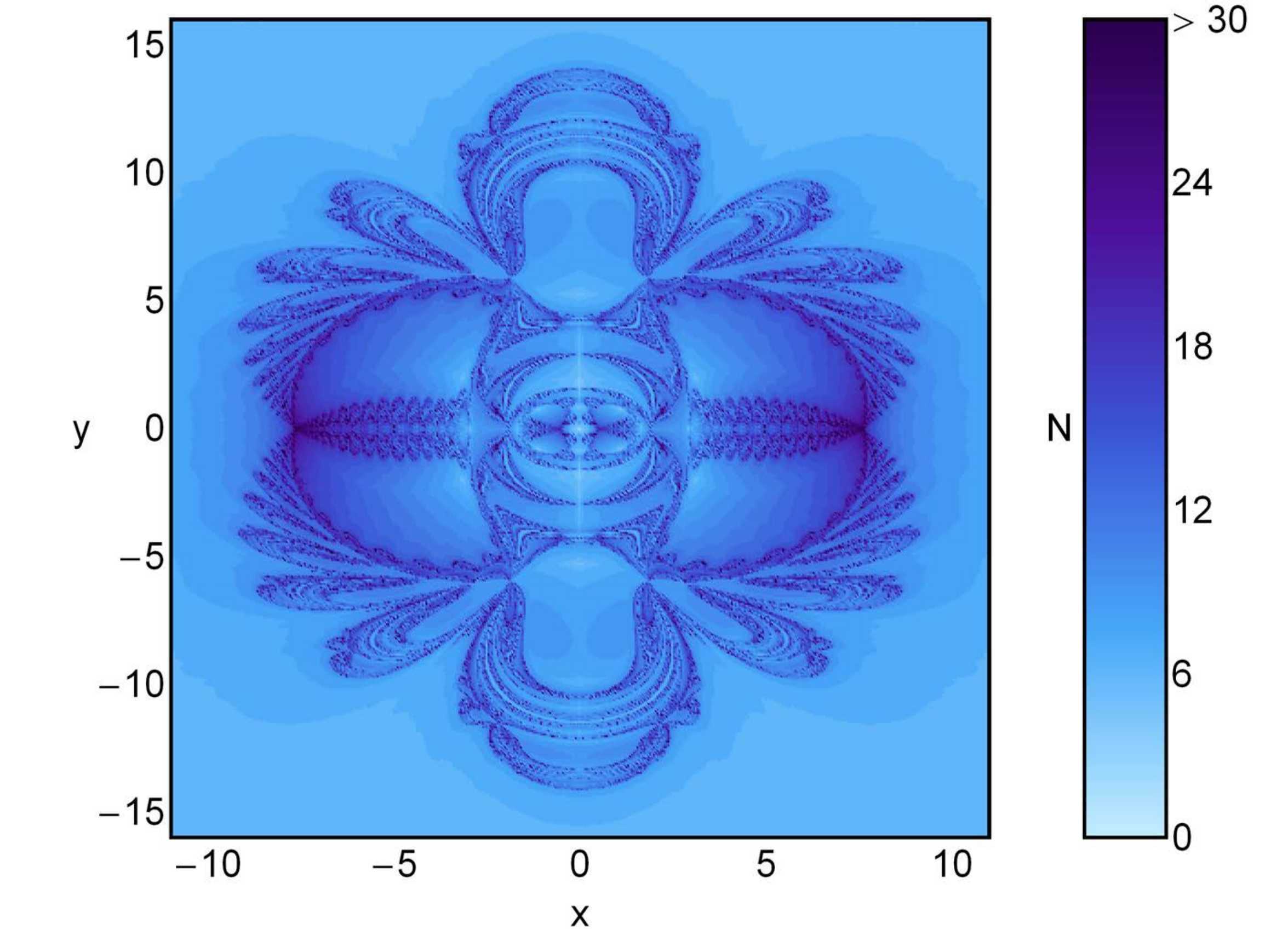}
(c)\includegraphics[scale=.26]{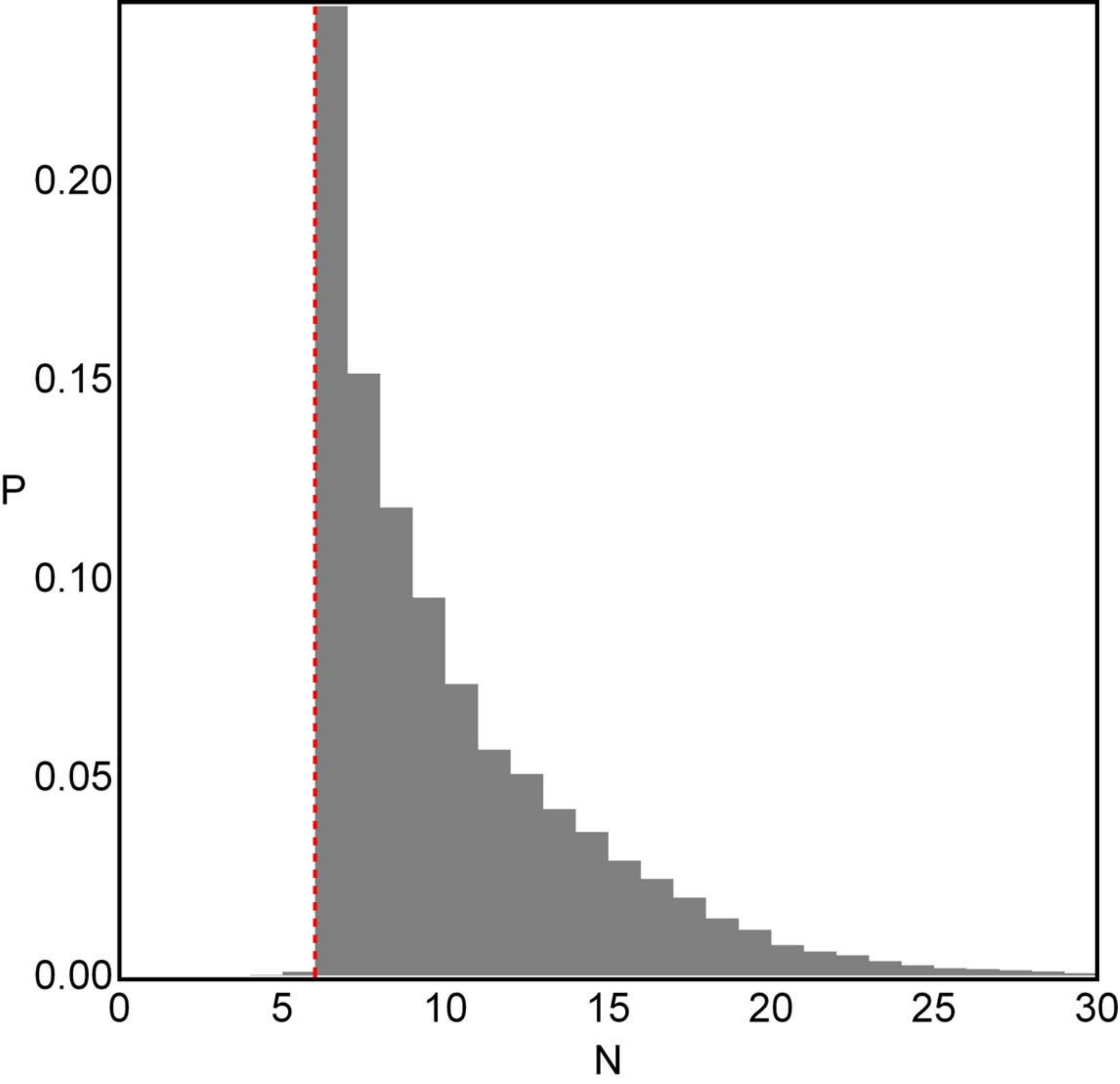}\\
(d)\includegraphics[scale=.26]{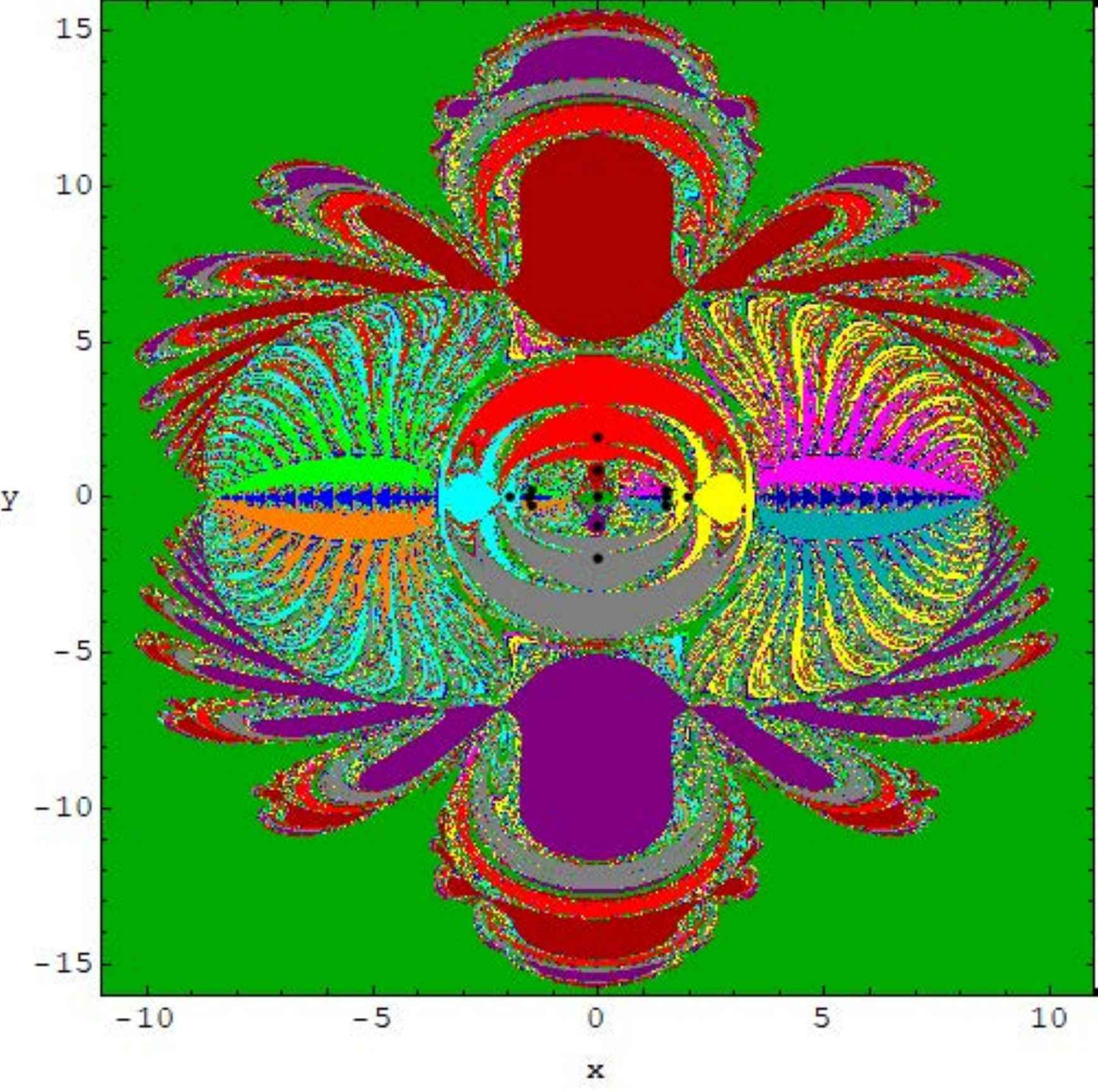}
(e)\includegraphics[scale=.26]{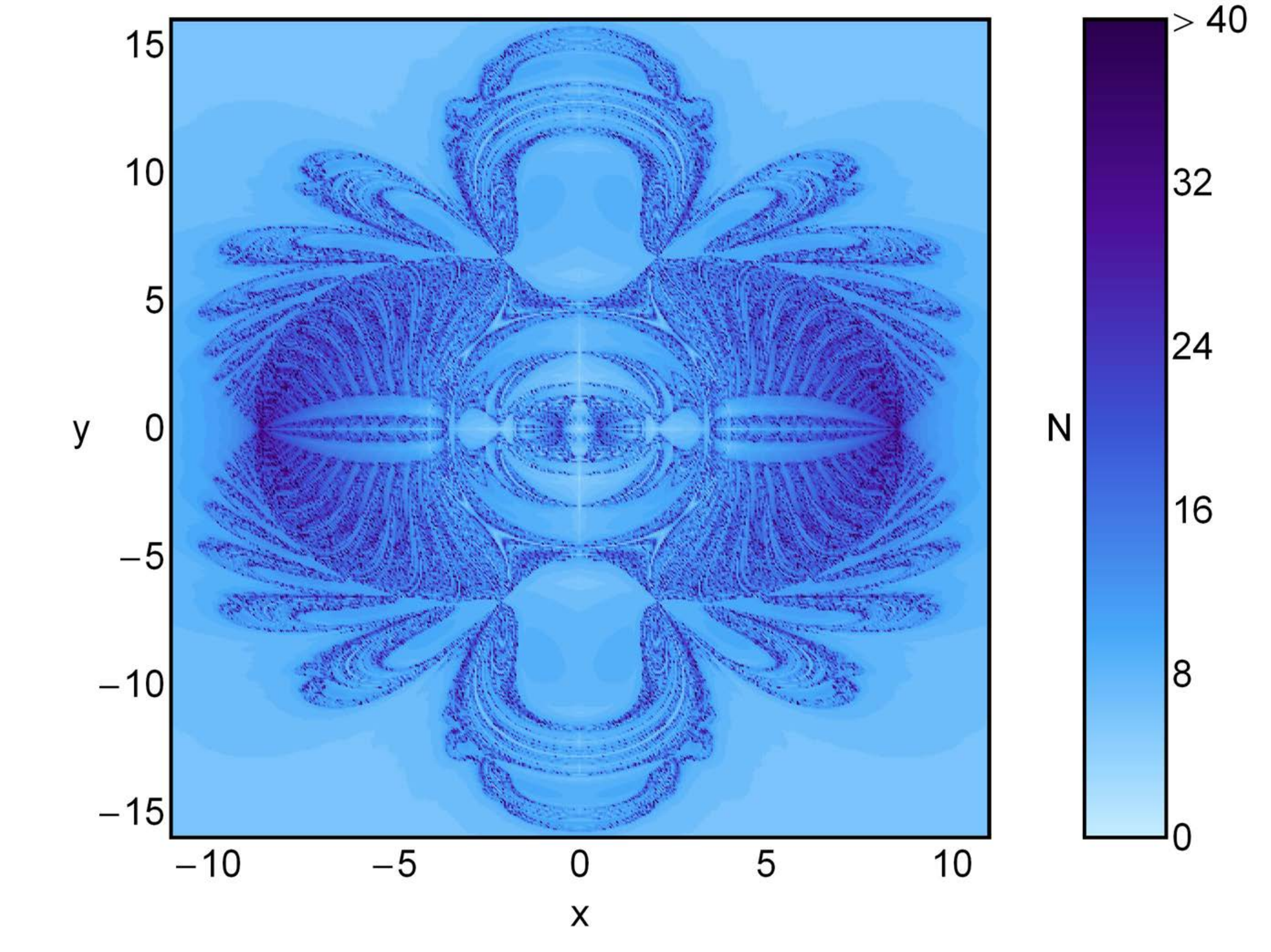}
(f)\includegraphics[scale=.26]{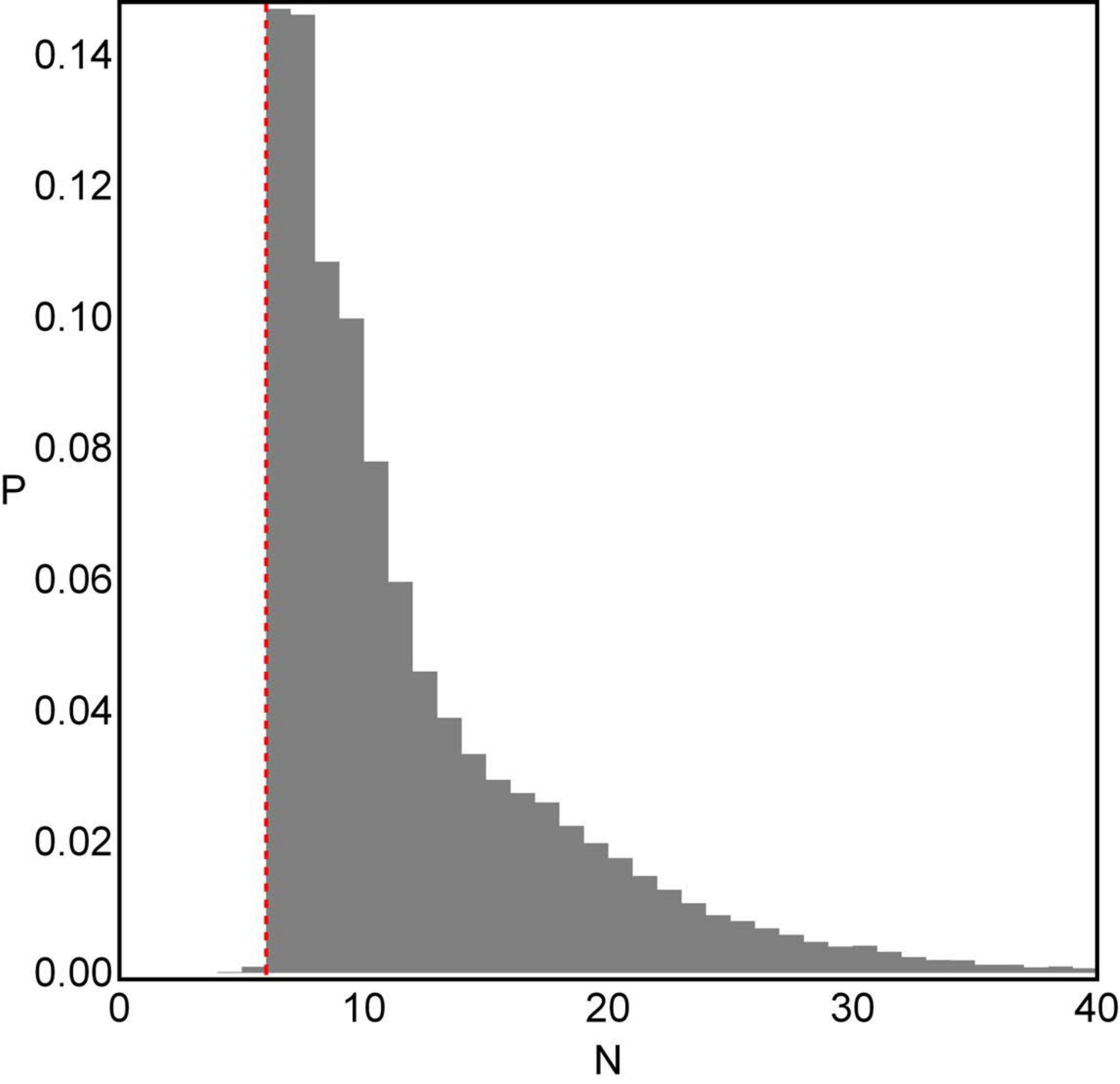}
\caption{The Newton-Raphson basins of attraction on the $xy$-plane for the case when thirteen libration points exist for:  (a) $e=-0.457853$; (d) $e=-0.466$. The color code denoting the attractors is as follows:$L_1$ (\emph{yellow}); $L_2$ (\emph{Darker blue}); $L_3$ (\emph{gray}); $L_4$ (\emph{green}); $L_5$ (\emph{red}); $L_6$ (\emph{blue}); $L_7$ (\emph{cyan}); $L_8$ (\emph{purple}); $L_9$ (\emph{crimson}); $L_{10}$ (\emph{teal}); $L_{11}$ (\emph{magenta}); $L_{12}$ (\emph{orange}); $L_{13}$ (\emph{light green}) and non-converging points (\emph{white}). (b, e: the middle panels)
  The corresponding
  distribution of the number $N$ of required iterations for obtaining the attracting
  regions, (c, f: the right panels) the corresponding probability distributions of required number of iterations
  for obtaining the Newton-Raphson basins of convergence, shown in panels (a, d) respectively. The vertical, dashed, red line indicates, in each case,
the most probable number $N^*$ of iterations. The black dots show the position of the libration points. (colour figure online).}
\end{center}
\end{figure*}
In Fig. \textcolor[rgb]{1.00,0.00,0.50}{9}(a-f), the distribution of the corresponding number $N $
of iterations required for the predefined accuracy is illustrated,
using tones of blue. It is
seen that initial conditions falling inside the attracting regions
converge relatively fast $(N < 10)$, where as the slowest converging
points $(N > 30)$ are those which lie in the vicinity of the basin
boundaries. Figure \textcolor[rgb]{1.00,0.00,0.50}{10}(a-f)
deals with the corresponding probability distributions of the iterations.
The red dashed line is corresponding to the most probable number $N^*$
of iterations, which remains unchanged and equal to 7 throughout
this region of values of parameter $e$.\\
Figure \textcolor[rgb]{1.00,0.00,0.50}{11}(a,d) corresponds to
the case-(ii), and we observe drastic changes in the domain of
the basins of convergence for decreasing value of parameter $e$.
In this case also the basin of convergence corresponding to
central libration point $L_4$ has infinite extent, whereas the
extent of all the other basins of attraction are finite. Observing
carefully at the color coded diagram, we realize that
the domain of the basins of convergence corresponding to libration
points $L_1$ and $L_7$  looks like exotic bugs with many legs and many
antennas, whereas the shape of the domain of convergence associated
with the remaining libration points, except the central libration
point $L_4$, seems like butterfly wings which remain unperturbed.
The most notable changes corresponding to regions $R_1$ and $R_7$ are:
most of the area of the regions $R_{1,7}$ occupied by well formed finite
tadpole shaped region (panel-a) corresponding to libration points $L_{10,...,13}$, which is
converted to exotic bugs shaped region (combination of three well shaped
region) with many legs and antennas. Moreover, these regions are filled
with highly chaotic mixture of initial conditions, except those bugs shaped region, therefore
it is next to impossible for an
initial condition $(x_0, y_0)$ falling inside the chaotic fractal area,
its final state (attractor) is highly sensitive. In panels-(b, e), the distribution of the corresponding number $N$ of iterations, required to obtain the predefined accuracy in our computations is illustrated. We may observe that the distribution of required iterations, corresponding to libration points in the regions $R_{1, 7}$ is very abuzz. Moreover, it is almost impossible to estimate the required number of iterations for all initial conditions falling inside the regions $R_{1,7}$  to converge to any of the attractor. This phenomenon is evident from the panel-(f), where the corresponding probability distribution of iterations is illustrated. It is clear that the corresponding probability distribution extends slightly i.e. up to $N=40$  whereas in panel-(c) it was $N=30$. The most probable number $N^*$ of iteration remains six in both the panels.
\subsection{\emph{Case V: when five libration points exist}}
\label{sec:405}
This subsection corresponds to the case $e > 0$, where five libration points exist in two different classes of the equilibria. These two classes are same as in the classical case i.e. the pure gravitational one where the collinear ($L_1$, $L_4$, $L_7$) and the triangular ($L_8$, $L_9$) libration points occur. In Fig. \textcolor[rgb]{1.00,0.00,0.50}{12} (a), the basins of convergence associated with the libration points are presented for $e=0.4$. It is observed that the basin for central libration point $L_4$ has infinite extent while all the remaining libration points have well formed and finite domains of convergence. In addition, the two exotic bugs with many legs and antennas still exist, corresponding to libration points $L_{1,7}$ and for other libration points the butterfly wings shaped region occur. It is observed that the geometry of the basins of convergence in the present case highly resembles to that of the classical case (see \cite{zot15}).
\begin{figure*}\label{Fig:13}
\begin{center}
(a) \includegraphics[scale=.26]{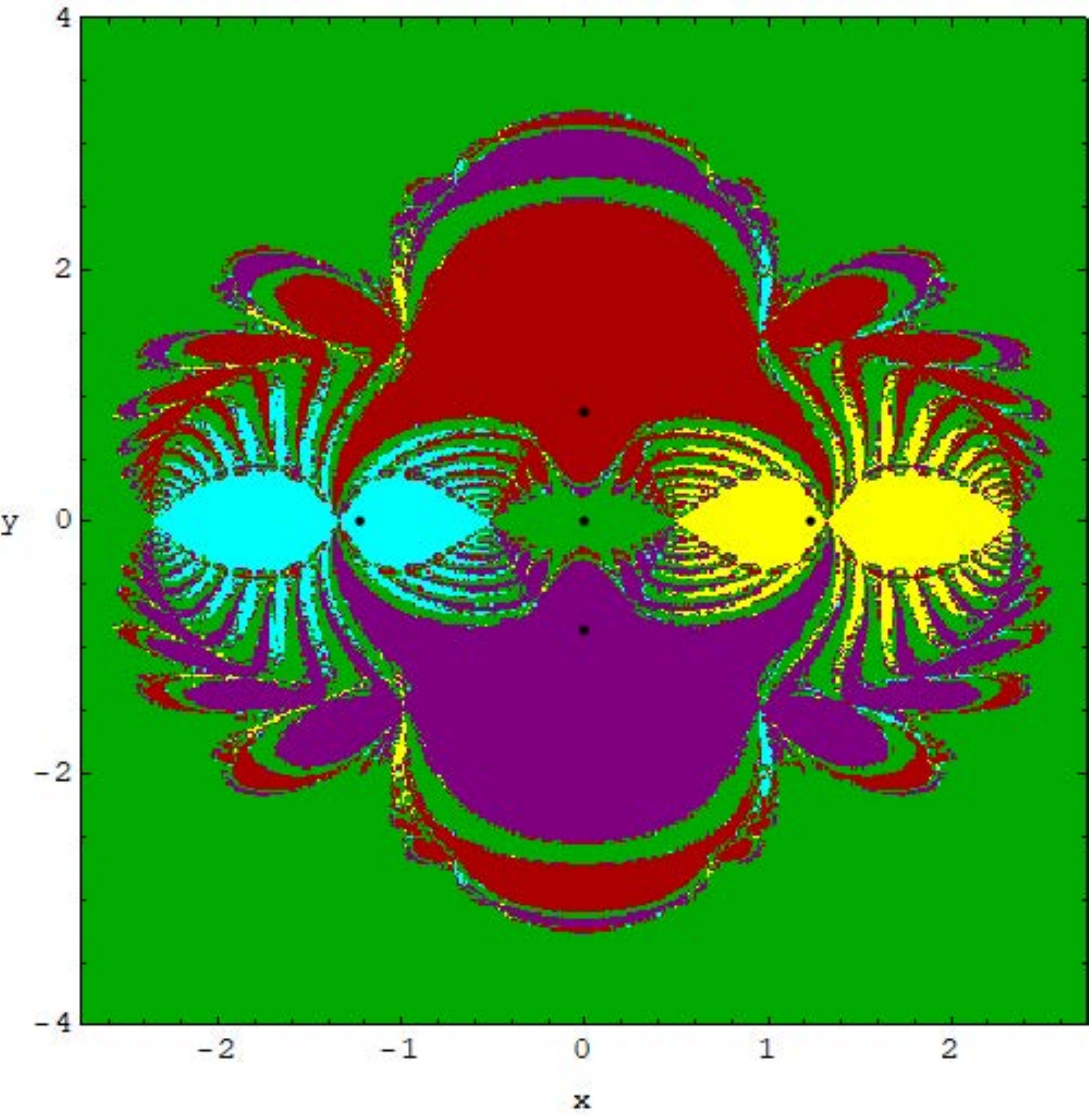}
(b) \includegraphics[scale=.26]{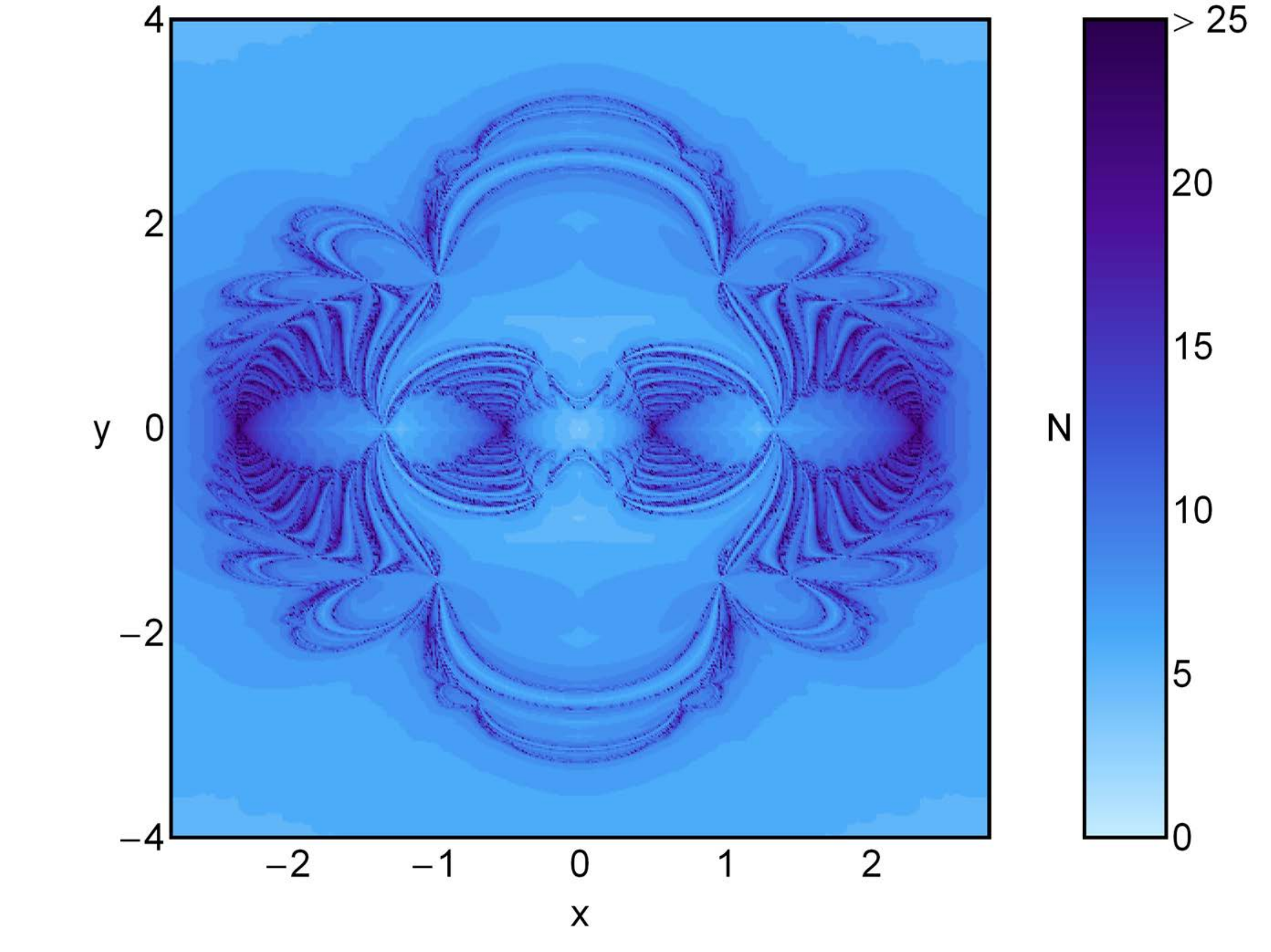}
(c) \includegraphics[scale=.26]{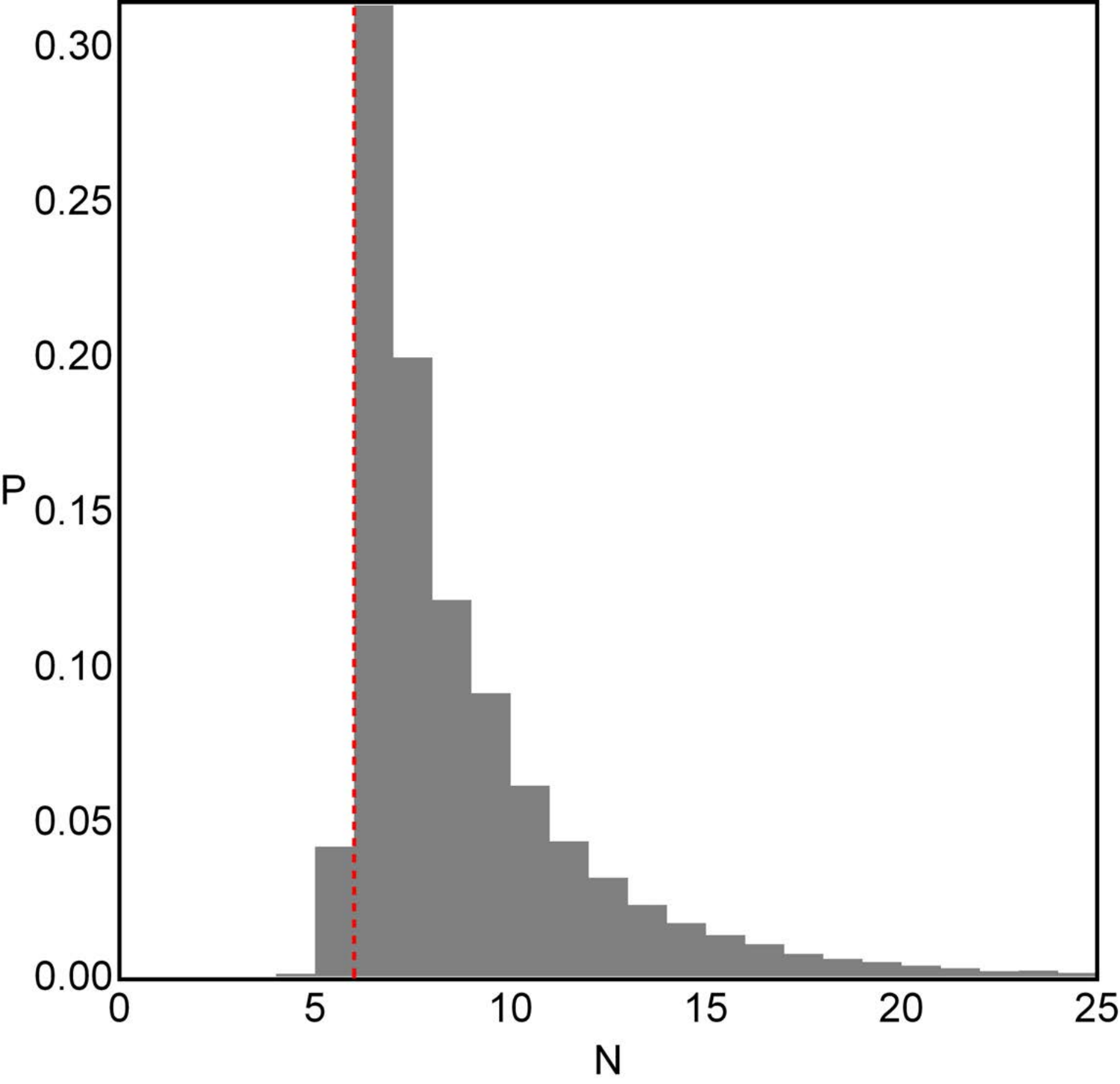}
\caption{The Newton-Raphson basins of attraction on the $(x,y)$ plane for the case when five libration points exist for:  (a) $e=0.4$. The color code denoting the attractors is as in previous figures.  (b: the middle panel)
  The corresponding
  distribution of the number $N$ of required iterations for obtaining the attracting
  regions, (c: the right panel) the corresponding probability distributions of required number of iterations
  for obtaining the Newton-Raphson basins of convergence, shown in panel (a). The vertical, dashed, red line indicates the most probable number $N^*$ of iterations.}
\end{center}
\end{figure*}
\section{Discussion and conclusions}
\label{sec:5}
The present study has aspired to reveal the most intrinsic properties of the Copenhagen problem with a quasi-homogeneous potential by exploring the basins of convergence associated with the coplanar libration points of this system.  Using the multivariate version of the Newton-Raphson iterative scheme, we unveiled the structure of the basins of convergence on the configuration plane. In addition, the relations among the basins of convergence, the corresponding distributions of the number of required iterations, and the probability distributions are illustrated.

The most important results of our study can be recapitulated as follows:
\begin{itemize}
  \item The parameter $e$ has a significant influence on the dynamical properties of the present system. When $e$ varies in the interval $(-0.5, 0)$, it is observed that the total number of the libration points changes in a drastic manner as various points originated, while several points collide with each other and disappear.
  \item The libration points on the $(x,y)$ plane except on the $y-$axis are unstable throughout the interval $(-0.5,0)$. Only the libration points which lie on either $x-$axis or $y-$ axis are stable for specific intervals.
  \item The two-dimensional planes are covered by a complicated mixture of attracting domains composed of initial conditions with highly fractal basins boundaries. In the neighbourhood of the basins boundaries, it is next to impossible to predict the final state of an initial condition since the degree of fractality is very eminent.
  \item In all the examined cases, the domain of convergence corresponding to the central libration point $L_4$ has infinite extent. In addition, the areas of the basins of convergence corresponding to the remaining libration points are finite.
  \item When $e > 0$ the number of the libration points as well as the basins of convergence, associated with them, almost resemble with those of the classical case where the classical Newtonian gravitation is considered.
\end{itemize}
For all the calculation and the graphical illustration presented in the paper we used the latest version 11 of Mathematica$^\circledR$. We believe that the present study and the obtained results may be useful in the field of basins of convergence in dynamical systems. In future, it is worth studying some different types of iterative formulae other than the Newton-Raphson iterative scheme in order to reveal the similarities as well as the differences, regarding the domains of the basin of convergence in the Copenhagen problem with a quasi homogeneous potential.

\section*{Acknowledgments}
\footnotesize

The authors are thankful to Center for Fundamental Research in Space dynamics and Celestial mechanics (CFRSC), New Delhi, India for providing research facilities. We would also like to express our warmest thanks to the two anonymous referees for the careful reading of the manuscript and for all the apt suggestions and comments which allowed us to improve both the quality and the clarity of the paper.\par

\textbf{Compliance with Ethical Standards}
\begin{description}
  \item[-] Funding: The authors state that they have not received any research
grants.
  \item[-] Conflict of interest: The authors declare that they have no conflict of
interest.
\end{description}
\bibliographystyle{aps-nameyear}      
\end{document}